\documentclass[12pt]{article}
\usepackage{amsthm}
\usepackage{amscd}
\usepackage{indentfirst}
\usepackage{bbm}
\usepackage{mathrsfs}
\usepackage{amssymb}
\usepackage{graphics}
\usepackage{graphicx}
\usepackage{amsfonts}
\usepackage{amsmath}
\usepackage{color}
\usepackage{fancyhdr}
\usepackage[numbers,sort&compress]{natbib}
\usepackage{float}
\usepackage{subfigure}
\usepackage{hyperref}
\begin{document}
\date{}

\title{Asymptotic behaviors and dynamics of degenerate and mixed solitons for the coupled Hirota system with strong coherent coupling effects}

\author{Zhong Du$^1$, Mingke Qin$^1$, Lei Liu$^{2}$\thanks{Corresponding
		author, with e-mail address as
		liueli@126.com}
 \\
\\{\em 1. Department of Mathematics and Physics, and}\\
{\em Hebei Key Laboratory of Physics and Energy Technology,}\\
{\em North China Electric Power University, Baoding 071003, China}\\
{\em 2. College of Mathematics and Statistics, and}\\
{\em Key Laboratory of Nonlinear Analysis and its Applications, Ministry of Education,}\\
{\em Chongqing University, Chongqing 401331, China}}

\maketitle \maketitle


\begin{abstract}
In this work, we investigate the asymptotic behaviors and dynamics of degenerate and mixed solitons in the coupled Hirota system with strong coherent coupling effect. Through the binary Darboux transformation, we obtain three types of degenerate solitons and their asymptotic expressions. These degenerate solitons admit time-dependent velocities, and the relative distance between the two asymptotic solitons logarithmically increases with the increase of the higher-order perturbation parameter $|\varepsilon|$. We also reveal four mechanisms of interaction between degenerate solitons and bell-shaped solitons: (1) elastic interaction with a position shift; (2) inelastic interaction for the degenerate soliton but elastic for the bell-shaped one; (3) elastic interaction for the degenerate soliton but inelastic for the bell-shaped one; and (4) elastic interaction based on coherent interaction under specific parameter conditions. Furthermore, we analyze a special degenerate vector soliton with strong coherent coupling effects, and through numerical studies, we investigate the relationship between the soliton's robustness and the parameter $\varepsilon$. The results show that $\varepsilon$ significantly affects the coherence of the solitons, and its robustness decreases as $|\varepsilon|$ increases. Our research results not only provide a new theoretical basis for understanding soliton dynamics, but also offer important guidance for practical applications, such as optical fiber communication and fluid dynamics.
\end{abstract}

\vspace{3mm}


\newpage

\vspace{5mm} \noindent\textbf{~1.~Introduction}\\

Fluid mechanics plays an important role in engineering and science by verifying theories to predict and analyze complex flow phenomena~\cite{White}. This discipline studies fluid motion and static behavior, adheres to Newton's laws, and establishes parametric scaling models through dimensionless analysis, providing a foundation for interdisciplinary research~\cite{Kato,Hamid}. Solitons, which preserve their shapes during the propagation~\cite{Mollenauer,Zabusky}, hold significant importance in fluid mechanics, particularly in the study of shallow water waves, ocean waves, and internal waves in stratified fluids~\cite{Hamid}. Besides, solitons are also extensively studied in nonlinear optics~\cite{Hasegawa1,Hasegawa2,LLPRL}, Bose-Einstein condensates~\cite{Nguyen}, plasma physics~\cite{Bailung}, and molecular biology~\cite{Scott}. Solitons are steady nonlinear waves, while other transient studies focusing on their unsteady forms can be found in Refs.~\cite{ZhaoGong,Ullah}. The focusing nonlinear Schr\"{o}dinger (NLS) equation governs the propagation of optical solitons in the picosecond regime, balancing group velocity dispersion with self-phase modulation~\cite{Agrawal}. Experimental observations and theoretical predictions of optical soliton propagation in optical fibers~\cite{Mollenauer,Hasegawa1,Hasegawa2} have strongly driven the development of both mathematical and physical models for solitons. The focusing NLS equation is known to admit $N$-soliton solutions, which describe the elastic interactions of $N$ solitons in an ideal optical Kerr medium, where the reflection coefficient has $N$ simple poles according to the inverse scattering transform~\cite{AblowitzClarkson,Zakharov}. Namely, an $N$-soliton solution corresponds to $N$ distinct eigenvalues of the linear spectral problem. As the time evolution $t\to\pm \infty$, an $N$-soliton solution can be viewed as a superposition of $N$ individual solitons, which separate linearly with $t$ and exhibit no interaction force~\cite{AblowitzClarkson,Zakharov}. When the eigenvalues have the same real parts but different imaginary parts, the $N$ interacting solitons have the same velocity, thereby forming a bound state called soliton molecule\cite{Agrawal,LiBiondini}. These soliton molecules exhibit molecular dynamic properties, and are proposed for increasing the bit rate in multi-level optical communication applications~\cite{Stratmann,Hause}.

When the eigenvalues tend to be conjugate or same value, the soliton solution exhibits special dynamical behaviors. In this case, the soliton solution will transform from the ordinary multi-soliton solution to a degenerate soliton solution with more complex behaviors, i.e., multi-pole solution in the context of the inverse scattering transform~\cite{Zakharov,Bilman,Schiebold,Olmedilla}. Distinct from the usual $N$-soliton interactions, degenerate solitons, referred to as special soliton molecules, can display intense interactions in the near-field region, where the relative distance between interacting solitons increases logarithmically with $|t|$~\cite{Schiebold,Olmedilla}. Additionally, the initial conditions such as the initial amplitude and phase of the solitons, as well as external disturbances, non-uniform media, or higher-power input, can all trigger degradation phenomena, disrupting the stable equilibrium of the solitons and altering their dynamic behaviors~\cite{Schiebold,Olmedilla}.

Degradation phenomena usually occur under specific conditions. For instance, in the coupled Hirota system, when the higher-order dispersion parameter reaches a certain value, the characteristic values of the solitons may tend to be conjugate, thereby triggering degradation~\cite{Schiebold,Olmedilla}.
The degradation phenomenon has various impacts on the actual system. In fluid mechanics, degenerate solitons may correspond to complex wave interactions, affecting the transmission and distribution of wave energy~\cite{Gagnon,Gordon,Karlsson}. For instance, their strong interactions may lead to energy redistribution, which has potential applications in marine engineering and wave energy utilization. 
Degenerate soliton solutions serve as the useful models for describing the behavior of multiple chirped pulses with the identical amplitudes and group velocities, especially when they are introduced without any phase difference~\cite{Gagnon,Gordon,Karlsson}. Mixed solitons can describe complex interactions among different types of solitons, such as degenerate solitons and bell-shaped solitons, which can reveal new behaviors of solitons in coupled systems~\cite{Gordon,Karlsson}. The study of mixed solitons not only helps to investigate complex interaction dynamics, but also provides a more realistic description of multi-component systems, and can incorporate higher-order effects such as third-order dispersion and self-steeping effects, which is crucial for a comprehensive understanding of soliton behavior. In terms of practical applications, the research results of mixed solitons are also of great significance in fields such as nonlinear optics including optical communication systems and optical switch devices, and fluid mechanics including wave behavior and coastal protection measures~\cite{Gordon,Karlsson}. In recent years, degenerate soliton and mixed soliton solutions have been investigated in coupled NLS systems, revealing some fascinating dynamical properties~\cite{Karlsson,RaoJG1,RaoJG2}.

Coupled NLS systems are highly significant in both theory and practice for describing the behavior of multiple modes, frequencies, and polarizations in optical fibers and related structures~\cite{Karpman,Gerdjikov,Baronio}. These systems enable intensities transfer among additional degrees of freedom and generate a wide variety of vector solutions~\cite{Karpman,Gerdjikov,Baronio}. To describe the propagation of orthogonally polarized optical waves in an isotropic medium, the coupled NLS system with negative coherent coupling has been studied, which exhibits abundant coherent nonlinear phenomena~\cite{Park,SunLiuPRSA}. Additionally, for higher pulse input powers, it is important to incorporate higher-order effects into the basic NLS systems~\cite{Agrawal,Porsezian,Kodama,Ankiewicz}. In nonlinear optics, the third-order dispersion and self-steepening effects are used to describe the propagation of ultrashort pulses in optical fibers~\cite{Agrawal,Porsezian,Kodama,Ankiewicz}. Consequently, researchers have introduced the coupled Hirota system with higher-order effects which can find applications in nonlinear optics~\cite{SunWR,DuZ,DuZPS}
\begin{subequations}
\begin{eqnarray}
&&\hspace{-1.5cm}{\rm i}q_{1,t}+q_{1,xx}+2 \left(|q_1|^2+2 |q_2|^2\right)q_1-2 q_1^*q_2^2+{\rm i} \varepsilon q_{1,xxx}\nonumber\\
&&\hspace{-1cm}+6{\rm i}\varepsilon \left(|q_1|^2+|q_2|^2\right)q_{1,x}+6{\rm i}\varepsilon \left(q_1q_2^*-q_2q_1^*\right)q_{2,x}=0,\\
&&\hspace{-1.5cm}{\rm i}q_{2,t}+q_{2,xx}+2 \left(|q_2|^2+2 |q_1|^2\right)q_2-2 q_2^*q_1^2+{\rm i} \varepsilon q_{2,xxx}\nonumber\\
&&\hspace{-1cm}+6{\rm i}\varepsilon \left(|q_1|^2+|q_2|^2\right)q_{2,x}+6{\rm i}\varepsilon \left(q_2q_1^*-q_1q_2^*\right)q_{1,x}=0,
\end{eqnarray}\label{equations}
\end{subequations}
\hspace{-0.15cm}where $q_1=q_1(x,t)$ and $q_2=q_2(x,t)$ represent the slowly varying complex amplitudes of two interacting optical modes, the functions depend on the propagation variable  $x$ and the time variable $t$, and the asterisk denotes the complex conjugate. In the context of short-pulse propagation through weakly birefringent media exhibiting Kerr-type nonlinearity, the variable $t$ represents the retarded time. The terms $|q_1|^2 q_1$ and $|q_2|^2q_2$ account for self-phase modulation, while the terms $|q_1|^2q_2$ and $|q_2|^2q_1$ describe cross-phase modulation, and the terms $q_1^*q_2^2$ and $q_2^*q_1^2$ represent coherent coupling effects governing the energy exchange between the two modes of the fiber, and $\varepsilon$ is a higher-order perturbation parameter. When $\varepsilon\neq 0$, System~(\ref{equations}) incorporates high-order effects such as third-order dispersion and self-steepening, which facilitate the description of orthogonally polarized optical pulse propagation in an isotropic medium. System~(\ref{equations}) can be symplified into the matrix form of
\begin{eqnarray}
\hspace{0cm}{\rm i}{\bf Q}_t+{\bf Q}_{xx}+2{\bf QQ}^{\dagger}{\bf Q}+{\rm i}\varepsilon{\bf Q}_{xxx}+6{\rm i}\varepsilon {\bf QQ}^{\dagger}{\bf Q}_x={\bf0},\label{matrixHirota}
\end{eqnarray}
where ${\bf Q}=
\begin{pmatrix}
q_1 & q_2 \\
-q_2 & q_1
\end{pmatrix}$ and ``\,$\dagger$\,'' denotes the conjugate transpose, which is also called the matrix Hirota equation. It is noted that System~(\ref{equations}) is unchanged through $x\rightarrow-x$ and $\varepsilon\rightarrow-\varepsilon$. Namely, we only need to consider the absolute value
of higher-order perturbation parameter $\varepsilon$ and always set $\varepsilon>0$ in this paper.

For System~(\ref{equations}), Ref.~\cite{SunWR} has explored one- and two-hump solitons, bright and dark rogue waves, and bright and dark breathers with respect to a single spectral parameter using the Darboux transformation (DT) and generalized DT. Ref.~\cite{DuZ} has studied the interactions of vector breathers with two spectral parameters, while Ref.~\cite{DuZPS} has reported the hybrid structures of rogue waves and breathers by the generalized DT. In this paper, we will extend our previous research by investigating degenerate and mixed solitons, giving asymptotic analysis of their interactions, and performing numerical studies on their robustness, all of which are novel contributions not reported in Ref.~\cite{DuZ,DuZPS}.

In this paper, we focus on the degenerate and mixed solitons for the coupled Hirota system with strong coherent coupling effects. This system can more accurately describe the behavior of solitons with strong coherent coupling effects, especially when higher-order dispersion and self-steepening effects cannot be ignored. Our research results not only can enrich the soliton theory but also provide important guidance for practical applications. For example, in optical fiber communication, understanding the degeneration phenomenon and interaction mechanism of solitons helps to optimize signal transmission schemes, reduce signal distortion and energy loss; in fluid dynamics, a deep understanding of soliton dynamics can provide theoretical basis for the development and utilization of ocean wave energy, and also provide support for predicting and preventing marine disasters. Moreover, the binary Darboux transformation method and asymptotic analysis technique adopted in this paper provide new ideas and methods for solving other complex nonlinear systems, with wide applicability and promotion value.

The aim of this paper is to investigate the asymptotic behaviors and coherent dynamics for the higher-order effects of degenerate and mixed solitons for System~(\ref{equations}). The structure of this paper is outlined as follows: In Section 2, we will construct the $N$th-order binary DT and derive the determinant representation of soliton solutions for System~(\ref{equations}). Section 3 will focus on deriving three types of degenerate solitons and their asymptotic expressions based on the binary DT. In Section 4, we will analyze the asymptotic expressions of mixed solitons before and after interactions, describing the four interaction mechanisms. In Section 5, we will discuss the interaction properties of mixed solitons based on both asymptotic and graphical analyses. In Section 6, we will analyze a special degenerate vector soliton with significant coherence effects and numerically investigate the relationship between the soliton's robustness and parameter $\varepsilon$. Section 7 will conclude the paper.

\vspace{5mm} \noindent\textbf{~2.~The $N$th-order binary DT}\\

In this section, we would like to review the $N$th-order binary DT and obtain the determinant representation of soliton solutions for System~(\ref{equations}).

Based on the Ablowitz-Kaup-Newell-Segur inverse scattering formulation~\cite{Ablowitz}, System~(\ref{equations}) is completely integrable and  admit the $4\times 4$ Lax pair~\cite{SunWR,DuZ,DuZPS}
\begin{eqnarray}
&&\hspace{-2cm}\Psi_{x}=U(\lambda;W)\Psi,\qquad \Psi_{t}=V(\lambda;W)\Psi, \label{lax}
\end{eqnarray}
where $\Psi$ is a $4\times 1$ vector complex
differentiable eigenfunction of $x$, $t$ and the complex spectral parameter $\lambda$, and $U$ and $V$ have the following forms of
\begin{eqnarray*}
	&&\hspace{-1.1cm}U(\lambda;W)={\rm i} \lambda \Omega+W,\\
	&&\hspace{-1.1cm}V(\lambda;W)=4{\rm i}\varepsilon \Omega \lambda^3+2\left(2\varepsilon W+{\rm i} \Omega\right)\lambda^2+2\left({\rm i} \varepsilon W^2\Omega+W+{\rm i} \varepsilon  W_x\Omega\right)\lambda\nonumber\\
	&&\hspace{-0.3cm}+2\varepsilon W^3+{\rm i} W^2 \Omega-\varepsilon W_{xx}+{\rm i}W_x\Omega+\varepsilon \left(W_xW - WW_{x}\right),
\end{eqnarray*}
with
\begin{eqnarray*}
	&&\hspace{-1.1cm}\Omega=\begin{pmatrix}
		-{\bf I}_{2\times2} &  {\bf 0}_{2\times2} \\
		{\bf 0}_{2\times2} & {\bf I}_{2\times2}
	\end{pmatrix},\qquad W=\begin{pmatrix}
		{\bf 0}_{2\times2} & {\bf Q} \\
		-{\bf Q}^{\dag} & {\bf 0}_{2\times2}
	\end{pmatrix},
\end{eqnarray*}
${\bf I}_{2\times2}$ and ${\bf 0}_{2\times2}$ as the $2\times2$ identity matrix and zero matrix, respectively. From Lax Pair (\ref{lax}), we find that the compatibility condition $U_t-V_x+UV-VU=0$ is equivalent to System~(\ref{equations}).

Observing the expression of $W$, we find that $W$ has the symmetry property as follows
\begin{eqnarray*}
 W=S^{-1}WS,\qquad S=\left(
 \begin{array}{cccc}
 	0 & -1 & 0 & 0 \\
 	1 & 0 & 0 & 0 \\
 	0 & 0 & 0 & -1 \\
 	0 & 0 & 1 & 0
 \end{array}
 \right),
\end{eqnarray*}
with the superscript ``\,$-1$\,'' as the inverse of a matrix. Hence, supposing that $\Psi_j=\left(\psi_{j,1},\psi_{j,2},\psi_{j,3},\psi_{j,4}\right)^T$ represents a vector eigenfunction of Lax Pair (\ref{lax}) corresponding to the complex eigenvalue $\lambda_j$ $(j=1,2,3,\cdots,N)$, it follows that $S\Psi_j=(-\psi_{j,2},\psi_{j,1},-\psi_{j,4},\psi_{j,3})^T$ is also a vector eigenfunction of Lax Pair (\ref{lax}) at $\lambda=\lambda_j$, where $N$ is a positive integer, $\psi_{j,k}$'s $(k=1,2,3,4)$ are the complex functions with respect to $x$ and $t$, and the superscript ``\,$T$\,'' denotes the transpose operation on a vector/matrix. Setting that $\mathcal{H}_j=\left(
\begin{array}{cc}
\psi_{j,1} & -\psi_{j,2} \\
\psi_{j,2} & \psi_{j,1}
\end{array}
\right)$, $\mathcal{Y}_j=\left(
\begin{array}{cc}
\psi_{j,3} & -\psi_{j,4} \\
\psi_{j,4} & \psi_{j,3}
\end{array}
\right)$ and $\Gamma_j=\left(\Psi_j,S\Psi_j\right)=\left(
\begin{array}{c}
\mathcal{H}_j  \\
\mathcal{Y}_j
\end{array}
\right)$, we deduce that $\Gamma_j$ is a $4\times 2$ matrix eigenfunction of Lax Pair (\ref{lax}) at $\lambda=\lambda_j$, i.e.,
\begin{eqnarray}
&&\hspace{-1cm}\Gamma_{jx}=U(\lambda_j;W)\Gamma_j,\qquad \Gamma_{jt}=V(\lambda_j;W)\Gamma_j.
\end{eqnarray}

By means of the binary DT construction method~\cite{Nimmo,LingZhao,ZhangCRCNSNS} and Lax Pair (\ref{lax}), we obtain the $N$th-order binary DT for System~(\ref{equations}) as
\begin{eqnarray}
{\bf Q}[N]={\bf Q}-2{\rm i}\varUpsilon_1[N]\varOmega[N]^{-1}\varUpsilon_2[N]^{\dagger},\label{BDT}
\end{eqnarray}
where ${\bf Q}[N]=\left(
\begin{array}{cc}
q_1[N]     & q_2[N]  \\
-q_2[N]   & q_1[N]
\end{array}
\right)$, $[N]$ indicates the $N$th iteration with respect to $q_1$ and $q_2$, $\varUpsilon_1[N]$ and $\varUpsilon_2[N]$ are both $2\times 2N$ matrices, which can be respectively represented as
\begin{eqnarray*}
&&\hspace{-1cm}\varUpsilon_1[N]=\left(
\begin{array}{ccccccc}
\psi_{11} & -\psi_{12} & \psi_{21} & -\psi_{22} & \cdots & \psi_{N,1} & -\psi_{N,2}\\
\psi_{12} & \psi_{11}  & \psi_{22} & \psi_{21}  & \cdots & \psi_{N,2} & \psi_{N,1}
\end{array}
\right)=\left(\mathcal{H}_1,\mathcal{H}_2,\cdots,\mathcal{H}_N\right),\\
&&\hspace{-1cm}\varUpsilon_2[N]=\left(
\begin{array}{ccccccc}
\psi_{13} & -\psi_{14} & \psi_{23} & -\psi_{24} & \cdots & \psi_{N,3} & -\psi_{N,4}\\
\psi_{14} & \psi_{13}  & \psi_{24} & \psi_{23}  & \cdots & \psi_{N,4} & \psi_{N,3}
\end{array}
\right)=\left(\mathcal{Y}_1,\mathcal{Y}_2,\cdots,\mathcal{Y}_N\right),
\end{eqnarray*}
and $\varOmega[N]$ is a $2N\times 2N$ matrix written as
\begin{eqnarray}
&&\hspace{-1cm} \varOmega[N]=\left(
\begin{array}{cccc}
\varOmega\left(\Gamma_1,\Gamma_1\right) & \varOmega\left(\Gamma_1,\Gamma_2\right) & \cdots & \varOmega\left(\Gamma_1,\Gamma_N\right) \\
\varOmega\left(\Gamma_2,\Gamma_1\right) & \varOmega\left(\Gamma_2,\Gamma_2\right) & \cdots & \varOmega\left(\Gamma_2,\Gamma_N\right) \\
\vdots & \vdots & \ddots & \vdots \\
\varOmega\left(\Gamma_N,\Gamma_1\right) & \varOmega\left(\Gamma_N,\Gamma_2\right) & \cdots & \varOmega\left(\Gamma_N,\Gamma_N\right) \\
\end{array}
\right),\label{OmegaN}
\end{eqnarray}
with $\varOmega\left(\Gamma_j,\Gamma_d\right)=\frac{\Gamma_j^{\dagger}\Gamma_d}{\lambda_d-\lambda_j^*}$ $(d=1,2,\cdots,N)$.

\vspace{5mm}\noindent\textbf{~3.~Asymptotic behaviors of the degenerate solitons}\\

In this section, we start the analysis with the seed solutions $q_1=q_2=0$ of System~(\ref{equations}), i.e., ${\bf Q}={\bf 0}_{2\times2}$. Substituting it into Lax Pair (\ref{lax}) at $\lambda=\lambda_j$, we derive that
\begin{subequations}
\begin{eqnarray}
&&\Psi_j=\left(
\begin{array}{c}
\psi_{j,1} \\
\psi_{j,2} \\
\psi_{j,3}  \\
\psi_{j,4}
\end{array}
\right)=\left(
\begin{array}{c}
l_{j,1}e^{-{\rm i}\eta_j} \\
l_{j,2}e^{-{\rm i}\eta_j} \\
l_{j,3}e^{{\rm i}\eta_j} \\
l_{j,4}e^{{\rm i}\eta_j}
\end{array}
\right),\\
&&\mathcal{H}_j=\left(
\begin{array}{cc}
l_{j,1} & -l_{j,2} \\
l_{j,2} & l_{j,1}
\end{array}
\right)e^{-{\rm i}\eta_j}, \quad \mathcal{Y}_j=\left(
\begin{array}{cc}
l_{j,3} & -l_{j,4} \\
l_{j,4} & l_{j,3}
\end{array}\right)e^{{\rm i}\eta_j},
\end{eqnarray}\label{Psij}
\end{subequations}
\hspace{-0.15cm}where $\eta_j=\lambda_j\left[x+2 \lambda_j \left(2 \lambda_j \varepsilon +1\right)t\right]$, $l_{j,k}$'s are all the arbitrary complex parameters. Letting that $H_j=\left(
\begin{array}{cc}
l_{j,1} & -l_{j,2} \\
l_{j,2} & l_{j,1}
\end{array}
\right)$, $Y_j=\left(
\begin{array}{cc}
l_{j,3} & -l_{j,4} \\
l_{j,4} & l_{j,3}
\end{array}\right)$ and inserting $q_1=q_2=0$ and Expressions~(\ref{Psij}) into Binary DT (\ref{BDT}), we derive the $N$-soliton solutions for System~(\ref{equations}) as
\begin{eqnarray}
{\bf Q}[N]=-2{\rm i}\left(H_1e^{-{\rm i}\eta_1},H_2e^{-{\rm i}\eta_2},\cdots,H_Ne^{-{\rm i}\eta_N}\right)\varOmega[N]^{-1}\left(Y_1e^{{\rm i}\eta_1},Y_2e^{{\rm i}\eta_2},\cdots,Y_Ne^{{\rm i}\eta_N}\right)^{\dagger},\label{Nsoliton}
\end{eqnarray}
where $\varOmega[N]$ is given by Expression~(\ref{OmegaN}), and
\begin{eqnarray}
\varOmega\left(\Gamma_j,\Gamma_d\right)=\frac{\Gamma_j^{\dagger}\Gamma_d}{\lambda_d-\lambda_j^*}=\frac{1}{\lambda_d-\lambda_j^*}\left(H_j^{\dagger}H_de^{{\rm i}\left(\eta_j^*-\eta_d\right)}+Y_j^{\dagger}Y_de^{-{\rm i}\left(\eta_j^*-\eta_d\right)}\right).\label{Omegajd}
\end{eqnarray}
Obviously, when $\lambda_d^*=\lambda_j$, $\varOmega\left(\Gamma_j,\Gamma_d\right)$ and $\varOmega\left(\Gamma_d,\Gamma_j\right)$ will not be directly derived from Expression~(\ref{Omegajd}).
However, when the condition of $H_j^{\dagger}H_d+Y_j^{\dagger}Y_d=0$ is satisfied, this problem can be worked out by taking the limit:
\begin{eqnarray*}
&&\hspace{-2cm}\varOmega\left(\Gamma_d,\Gamma_j\right)=-\varOmega\left(\Gamma_j,\Gamma_d\right)^{\dagger}\\
&&\hspace{-0.15cm}=Y_d^{\dagger}Y_j\lim_{\lambda_d^*\to\lambda_j}\frac{e^{{\rm i} \left(\eta_d^*-\eta_j\right)}-e^{-{\rm i} \left(\eta_d^*-\eta_j\right)}}{\lambda_d^*-\lambda_j}-M_j^{\dagger}\\
&&\hspace{-0.15cm}=Y_d^{\dagger}Y_j\left[\frac{2{\rm i}\eta_{j,I}}{\lambda_{j,I}}-8\lambda_{j,I}\left(1+2{\rm i}\lambda_{j,I}\varepsilon+6\lambda_{j,R}\varepsilon\right)t\right]-M_j^{\dagger},
\end{eqnarray*}
where $M_j=\left(
\begin{array}{cc}
m_{j,1} & -m_{j,2} \\
m_{j,2} & m_{j,1}
\end{array}\right)$, $m_{j,1}$'s and $m_{j,2}$'s are the complex constants, and the subscripts ``$R$'' and ``$I$'' respectively indicate the real and imaginary parts of a complex number.

In what follows, we will investigate the properties of two solitons based on Solutions (\ref{Nsoliton}) with $N=2$. To simplify this problem, we choose that $l_{11}=l_{23}=1$, $l_{12}=l_{24}=0$, $l_{21}=-l_{13}^*$, and $l_{22}=l_{14}^*$ to ensure that $H_1$, $H_2$, $Y_1$ and $Y_2$ satisfy $H_2^{\dagger}H_1+Y_2^{\dagger}Y_1=0$. Then, with these values and substituting $\lambda_2^*\to\lambda_1$ into Solutions (\ref{Nsoliton}) with $N=2$, we find that the limits of two-soliton solutions can be reduced to the degenerate soliton solutions expressed as
\begin{subequations}
\begin{eqnarray}
&&\hspace{-1cm}q_1[2]=-2{\rm i}e^{-2{\rm i}\eta_{1,R}}\left(e^{\theta},0,-l_{13}^*,-l_{14}^*\right)\varOmega[2]^{-1}\left(l_{13}e^{-\theta},-l_{14}e^{-\theta},1,0\right)^{\dagger},\\
&&\hspace{-1cm}q_2[2]=-2{\rm i}e^{-2{\rm i}\eta_{1,R}}\left(e^{\theta},0,-l_{13}^*,-l_{14}^*\right)\varOmega[2]^{-1}\left(l_{14}e^{-\theta},l_{13}e^{-\theta},0,1\right)^{\dagger},
\end{eqnarray}\label{desoliton}
\end{subequations}
where
\begin{eqnarray*}
&&\chi=1+2{\rm i}\lambda_{1,I}\varepsilon+6\lambda_{1,R}\varepsilon,\qquad M_1=\left(
\begin{array}{cc}
m_{11} & -m_{12} \\
m_{12} & m_{11}
\end{array}\right),\\
&&\eta_{1,R}=\frac{1}{2}\left(\eta_1^*+\eta_1\right)=\lambda_{1,R} x+2 \lambda_{1,R}^2 \left(1+2 \lambda_{1,R} \varepsilon\right)t-2 \lambda_{1,I}^2 \left(1+6 \lambda_{1,R} \varepsilon\right)t,\\
&&\theta={\rm i}\left(\eta_1^*-\eta_1\right)=2\eta_{1,I}=2\lambda_{1,I} \left[x+4  \left(\lambda_{1,R}-\lambda_{1,I}^2\varepsilon+3 \lambda_{1,R}^2 \varepsilon \right)t\right],\\
&& \varOmega[2]=\left(
\begin{array}{cc}
	-\frac{{\rm i}}{2\lambda_{1,I}}\left({\bf I}_{2\times2}e^{\theta}+Y_1^{\dagger}Y_1e^{-\theta}\right) & -Y_1^{\dagger}\left(-\frac{{\rm i}\theta}{\lambda_{1,I}}-8\lambda_{1,I}\chi^*t\right)+M_1 \\
	Y_1\left(\frac{{\rm i}\theta}{\lambda_{1,I}}-8\lambda_{1,I}\chi t\right)-M_1^{\dagger} & \frac{{\rm i}}{2\lambda_{1,I}}\left({\bf I}_{2\times 2}e^{\theta}+Y_1Y_1^{\dagger}e^{-\theta}\right)
\end{array}\right).
\end{eqnarray*}
It is noted that Solutions~(\ref{desoliton}) are the semi-rational expressions containing exponential and polynomial functions, which can describe the second-order degenerate solitons.

Next, we will perform the asymptotic analysis on Solutions~(\ref{desoliton}) to illustrate the asymptotic behaviors of degenerate solitons.

When $\lambda_{1,R}=\frac{-1+\sqrt{1+12\lambda_{1,I}^2\varepsilon^2}}{6\varepsilon}$, we derive that $\theta=\mathcal{O}(1)$ as $t\to\pm\infty$.
When $t\to \pm\infty$, we expand Solutions~(\ref{desoliton}) along the straight line $\theta=\mathcal{O}(1)$, and obtain the asymptotic expressions of solitons as
\begin{subequations}
\begin{eqnarray}
&&\hspace{-1.5cm}q_1^{\pm}\to\left\{
 \begin{array}{cc}
 0, & l_{13}^2+l_{14}^2\not=0, \\
 {\rm{i}}\lambda_{1,I}e^{-2{\rm i}\eta_{1,R}}\frac{m_{11}-\frac{l_{13}}{l_{14}}m_{12}}{\left|m_{11}-\frac{l_{13}}{l_{14}}m_{12}\right|}{\rm sech}\left(\theta-\frac{\rho}{2}\right), & l_{13}^2+l_{14}^2=0,
 \end{array}\right.\\
&&\hspace{-1.5cm} q_2^{\pm}=-\frac{l_{13}}{l_{14}}q_1^{\pm},
\end{eqnarray}\label{asymptotic1}
\end{subequations}
\hspace{-0.12cm}where the signs ``$\pm$'' in the superscript correspond to asymptotic limits of the soliton when $t\to\pm\infty$, and $e^{\frac{\rho}{2}}=2\lambda_{1,I}\left|m_{11}-\frac{l_{13}}{l_{14}}m_{12}\right|$. From Asymptotic Expressions~(\ref{asymptotic1}), we find that the
asymptotic limits of Solutions~(\ref{desoliton}) are the sech-type functions under the conditions $l_{13}^2+l_{14}^2=0$, which implies that Solutions~(\ref{desoliton}) can generate the bell-shaped soliton branch the straight line $\theta=\mathcal{O}(1)$ as $|t|\to\infty$.

Since Solutions~(\ref{desoliton}) contain two independent variables $\theta$ and $t$, we also have to consider asymptotic solitons that appear on some curves when $e^{\theta}$ and $t$ reach a state of asymptotic balance. As $|t|\to\infty$, in order to illustrate the behaviors of solitons, we derive the asymptotic balance between $e^{\theta}$ and $t$ in Solutions~(\ref{desoliton}), i.e.,
\begin{eqnarray}
&&\hspace{-1cm} t e^{\pm \theta}\sim \mathcal{O}(1).\label{balance}
\end{eqnarray}
To analyze the asymptotic behaviors of solitons along with Expression~(\ref{balance}), we rewrite Solutions~(\ref{desoliton}) in the form of
\begin{eqnarray}
&& \hspace{-1.3cm} q_s= 2{\rm i}e^{-2{\rm i}\eta_{1,R}}\frac{{\rm det}\left(F_s\right)}{{\rm det}\left(\Omega[2]\right)}=2{\rm i}e^{-2{\rm i}\eta_{1,R}}\frac{{\rm det\left(
		\begin{array}{cc}
		\Omega[2] & B^{[s]\dagger} \\
		A & 0
		\end{array}\right)}}{{\rm det}\left(\Omega[2]\right)},\quad (s=1,2),\label{resolitons}
	\end{eqnarray}
where $A=\left(e^{\theta},0,-l_{13}^*,-l_{14}^*\right)$, the superscript $[s]$ represents the $s$th row of a matrix, and $B=\left(Y_1e^{-\theta},{\bf I}_{2\times2}\right)$. To verify the correctness of these results, we will perform numerical simulations (in Section 6) that demonstrate the consistency between the theoretical and numerical solutions.

Based on Expression~(\ref{balance}) and Solutions~(\ref{resolitons}), we can obtain two distinct asymptotic behaviors of solitons:

(1) When $\theta\to +\infty$ with $t\to\pm\infty$, we find that $t e^{-\theta}\sim \mathcal{O}(1)$, and derive the asymptotic expression of the soliton as
\begin{eqnarray}
	&&\hspace{-1.5cm}\left(
	\begin{array}{c}
		q_1 \\
		q_2
	\end{array}
	\right)\to \left(
	\begin{array}{c}
		q_1 \\
		q_2
	\end{array}
	\right)^{(1)}=\frac{2{\rm i}e^{-2{\rm i}\eta_{1,R}}}{e^{4\theta}{\rm det}\left(\tilde{\Omega}[2]^{(1)}\right)}\left(
	\begin{array}{c}
		e^{4\theta}{\rm det}\left(\tilde{F}_1^{(1)}\right) \\
		e^{4\theta}{\rm det}\left(\tilde{F}_2^{(1)}\right)
	\end{array}
	\right)\nonumber\\
	&&\hspace{2.41cm} =\frac{1}{2}\left[\left(\begin{array}{c}
		\varphi_1^{(1)} \\
		-{\rm i} \varphi_1^{(1)}
	\end{array}\right)+\left(\begin{array}{c}
	\varphi_2^{(1)} \\
	{\rm i}\varphi_2^{(1)}
\end{array}\right)\right],\label{asymptotic2}
\end{eqnarray}
with
\begin{eqnarray*}
	&&\hspace{-0.1cm}\tilde{\Omega}[2]^{(1)}=\left(
	\begin{array}{cc}
		-\frac{{\rm i}}{2\lambda_{1,I}}{\bf I}_{2\times2} & 8\lambda_{1,I}\chi^*Y_1^{\dagger}\frac{t}{e^{\theta}} \\
		-8\lambda_{1,I}\chi Y_1\frac{t}{e^{\theta}} & \frac{{\rm i}}{2\lambda_{1,I}}{\bf I}_{2\times 2}
	\end{array}\right),\\
&&\hspace{-0.1cm} \tilde{F}_s^{(1)}=\left(
\begin{array}{ccc}
-\frac{{\rm i}}{2\lambda_{1,I}}{\bf I}_{2\times2} & 8\lambda_{1,I}\chi^*Y_1^{\dagger}\frac{t}{e^{\theta}} & {\bf 0}_{2\times 1} \\
-8\lambda_{1,I}\chi Y_1\frac{t}{e^{\theta}} & \frac{{\rm i}}{2\lambda_{1,I}}{\bf I}_{2\times 2} & \left({\bf I}_{2\times 2}^{[s]}\right)^{\dagger} \\
{\bf I}_{2\times 2}^{[1]} & {\bf 0}_{1\times2} & 0
\end{array}\right),\\
&&\hspace{-0.1cm} \varphi_1^{(1)}=2{\rm i}\lambda_{1,I}e^{-2{\rm i}\eta_{1,R}}\frac{\left(l_{13}-{\rm i}l_{14}\right)^{*}\chi^*}{\left|\left(l_{13}-{\rm i}l_{14}\right)\chi\right|}{\rm sech}\left(\xi+\mu_1\right),\\
&&\hspace{-0.1cm} \varphi_2^{(1)}=2{\rm i}\lambda_{1,I}e^{-2{\rm i}\eta_{1,R}}\frac{\left(l_{13}+{\rm i}l_{14}\right)^{*}\chi^*}{\left|\left(l_{13}+{\rm i}l_{14}\right)\chi\right|}{\rm sech}\left(\xi+\nu_1\right),
	\end{eqnarray*}
where ${\bf 0}_{2\times1}$ is a $2\times1$ zero matrix, ${\bf 0}_{1\times2}$ denotes a $1\times2$ zero matrix, $e^{\xi}=\frac{8 \lambda_{1,I}t}{e^{\theta}}$, $e^{\mu_1}=2\lambda_{1,I}\left|\left(l_{13}-{\rm i}l_{14}\right)\chi\right|$ and $e^{\nu_1}=2\lambda_{1,I}\left|\left(l_{13}+{\rm i}l_{14}\right)\chi\right|$.
Obviously, Asymptotic Expression~(\ref{asymptotic2}) can be viewed as the linear superposition of two sech-type functions, which implies that Asymptotic Expression~(\ref{asymptotic2}) can describe the superposition of two bell-shaped solitons. The soliton $\left(\begin{array}{c}
q_1^{(1)} \\
q_2^{(1)}
\end{array}
\right)$ propagates at the velocity
\begin{eqnarray}
v_{(1)}(t)=-4\left(\lambda_{1,R}+3\lambda_{1,R}^2\varepsilon-\lambda_{1,I}^2\varepsilon\right)+\frac{1}{2\lambda_{1,I}t},\label{velocity1}
\end{eqnarray}
and keeps the amplitude unchanged.

(2) When $\theta\to -\infty$ with $t\to\pm\infty$, we note that $t e^{\theta}\sim \mathcal{O}(1)$, and obtain the asymptotic state of the soliton as
\begin{eqnarray}
&&\hspace{-1.5cm}\left(
\begin{array}{c}
q_1 \\
q_2
\end{array}
\right)\to \left(
\begin{array}{c}
q_1 \\
q_2
\end{array}
\right)^{(2)}=\frac{2{\rm i}e^{-2{\rm i}\eta_{1,R}}}{e^{-4\theta}{\rm det}\left(\tilde{\Omega}[2]^{(2)}\right)}\left(
\begin{array}{c}
e^{-4\theta}{\rm det}\left(\tilde{F}_1^{(2)}\right) \\
e^{-4\theta}{\rm det}\left(\tilde{F}_2^{(2)}\right)
\end{array}
\right)\nonumber\\
&&\hspace{2.41cm} =\frac{1}{2}\left[\left(\begin{array}{c}
\varphi_1^{(2)} \\
-{\rm i} \varphi_1^{(2)}
\end{array}\right)+\left(\begin{array}{c}
\varphi_2^{(2)} \\
{\rm i}\varphi_2^{(2)}
\end{array}\right)\right],\label{asymptotic3}
\end{eqnarray}
with
\begin{eqnarray*}
	&&\hspace{-0.1cm}\tilde{\Omega}[2]^{(2)}=\left(
	\begin{array}{cc}
		-\frac{{\rm i}}{2\lambda_{1,I}}Y_1^{\dagger}Y_1 & 8\lambda_{1,I}\chi^*Y_1^{\dagger}te^{\theta} \\
		-8\lambda_{1,I}\chi Y_1te^{\theta} & \frac{{\rm i}}{2\lambda_{1,I}}Y_1Y_1^{\dagger}
	\end{array}\right),\\
	&&\hspace{-0.1cm} \tilde{F}_s^{(2)}=\left(
	\begin{array}{ccc}
		-\frac{{\rm i}}{2\lambda_{1,I}}Y_1^{\dagger}Y_1 & 8\lambda_{1,I}\chi^* Y_1^{\dagger}te^{\theta} & \left(Y_1^{[s]}\right)^{\dagger} \\
		-8\lambda_{1,I}\chi Y_1te^{\theta} & \frac{{\rm i}}{2\lambda_{1,I}}Y_1Y_1^{\dagger} & {\bf 0}_{2\times 1} \\
		{\bf 0}_{1\times 2} & -\left(Y_1^{\dagger}\right)^{[1]}  & 0
	\end{array}\right),\\
	&&\hspace{-0.1cm} \varphi_1^{(2)}=2{\rm i}\lambda_{1,I}e^{-2{\rm i}\eta_{1,R}}\frac{\left(l_{13}-{\rm i}l_{14}\right)^{*}\chi}{\left|\left(l_{13}-{\rm i}l_{14}\right)\chi\right|}{\rm sech}\left(\zeta+\mu_2\right),\\
	&&\hspace{-0.1cm} \varphi_2^{(2)}=2{\rm i}\lambda_{1,I}e^{-2{\rm i}\eta_{1,R}}\frac{\left(l_{13}+{\rm i}l_{14}\right)^{*}\chi}{\left|\left(l_{13}+{\rm i}l_{14}\right)\chi\right|}{\rm sech}\left(\zeta+\nu_2\right),
\end{eqnarray*}
where $e^{\zeta}=8 \lambda_{1,I}te^{\theta}$, $e^{\mu_2}=\frac{2\lambda_{1,I}\left|\chi\right|}{\left|l_{13}-{\rm i}l_{14}\right|}$ and $e^{\nu_2}=\frac{2\lambda_{1,I}\left|\chi\right|}{\left|l_{13}+{\rm i}l_{14}\right|}$.
It is worth noting that Asymptotic Expression~(\ref{asymptotic3}) can also depict the superposition of two bell-shaped solitons. Through the computation, we find that the velocity of $\left(\begin{array}{c}
q_1^{(2)} \\
q_2^{(2)}
\end{array}
\right)$ is given by
\begin{eqnarray}
v_{(2)}(t)=-4\left(\lambda_{1,R}+3\lambda_{1,R}^2\varepsilon-\lambda_{1,I}^2\varepsilon\right)-\frac{1}{2\lambda_{1,I}t}.\label{velocity2}
\end{eqnarray}

By means of Expressions~(\ref{velocity1}) and (\ref{velocity2}), we derive that $v_{(1)}(t)=v_{(2)}(-t)$,
which implies the transformations of $q_s^{(1)-}$ to $q_s^{(2)+}$, and $q_s^{(2)-}$ to $q_s^{(1)+}$ in the interaction process between the two solitons $q_s^{(1)}$ and $q_s^{(2)}$. Therefore, $\left(\begin{array}{c}
q_1^{(1)-} \\
q_2^{(1)-}
\end{array}
\right)$ and $\left(\begin{array}{c}
q_1^{(2)+} \\
q_2^{(2)+}
\end{array}
\right)$ form the one soliton, represented as $S^1$, while $\left(\begin{array}{c}
q_1^{(2)-} \\
q_2^{(2)-}
\end{array}
\right)$ and $\left(\begin{array}{c}
q_1^{(1)+} \\
q_2^{(1)+}
\end{array}
\right)$ compose the other soliton, expressed as $S^2$, where the superscript ``1'' or ``2'' denotes the first or second soliton. As $t\to 0$, the absolute difference between the velocities of $S^1$ and $S^2$ $|v_{(1)}-v_{(2)}|$ increases, indicating that the attractive effect between $S^1$ and $S^2$ becomes stronger near the interaction region (around $t=0$). As $t\to \infty$, both $v_{(2)}(t)$ and $v_{(1)}(t)$ tend to  an identical value, i.e., $-4\left(\lambda_{1,R}+3\lambda_{1,R}^2\varepsilon-\lambda_{1,I}^2\varepsilon\right)$, meaning $S^1$ and $S^2$ to become almost parallel to each other over evolution time. Based on the above discussions, treating $S^1$ and $S^2$ as a whole, we find three cases of the degenerate solitons for System~(\ref{equations}), as discussed in what follows.

{\bf Case \textrm{I}.} As $l_{13}^2+l_{14}^2\neq0$, according to Asymptotic Expressions~(\ref{asymptotic2}) and (\ref{asymptotic3}), the asymptotic expressions of $S^1$ and $S^2$ depicted by Solutions~(\ref{desoliton}) can be given as
\begin{subequations}
\begin{eqnarray}
&&\hspace{-1cm}S^{1-}=\left(\begin{array}{c}
q_1^{(1)-} \\
q_2^{(1)-}
\end{array}
\right),\quad S^{1+}=\left(\begin{array}{c}
q_1^{(2)+} \\
q_2^{(2)+}
\end{array}
\right),\\
&&\hspace{-1cm}S^{2-}=\left(\begin{array}{c}
q_1^{(2)-} \\
q_2^{(2)-}
\end{array}
\right), \quad S^{2+}=\left(\begin{array}{c}
q_1^{(1)+} \\
q_2^{(1)+}
\end{array}
\right),
\end{eqnarray}\label{asymptotic4}
\end{subequations}
with
\begin{eqnarray*}
&& q_1^{(1)}=\frac{1}{2}\left(\varphi_1^{(1)}+\varphi_2^{(1)}\right),\qquad q_2^{(1)}=\frac{1}{2}\left(-{\rm i}\varphi_1^{(1)}+{\rm i}\varphi_2^{(1)}\right),\\
&& q_1^{(2)}=\frac{1}{2}\left(\varphi_1^{(2)}+\varphi_2^{(2)}\right),\qquad q_2^{(1)}=\frac{1}{2}\left(-{\rm i}\varphi_1^{(2)}+{\rm i}\varphi_2^{(2)}\right).
\end{eqnarray*}
Analyzing Asymptotic Expressions~(\ref{asymptotic4}), we see that $\left|q_1\right|$ and  $\left|q_2\right|$ are not proportional in the two vector asymptotic solitons $S^1$ and $S^2$, which reflects that $S^1$ and $S^2$ admit the distinct intensity profiles in the $q_1$ and $q_2$ components as seen in Figs.~1.
\begin{center} \vspace{0cm}
	\hspace{0.0cm}{\includegraphics[scale =0.36]{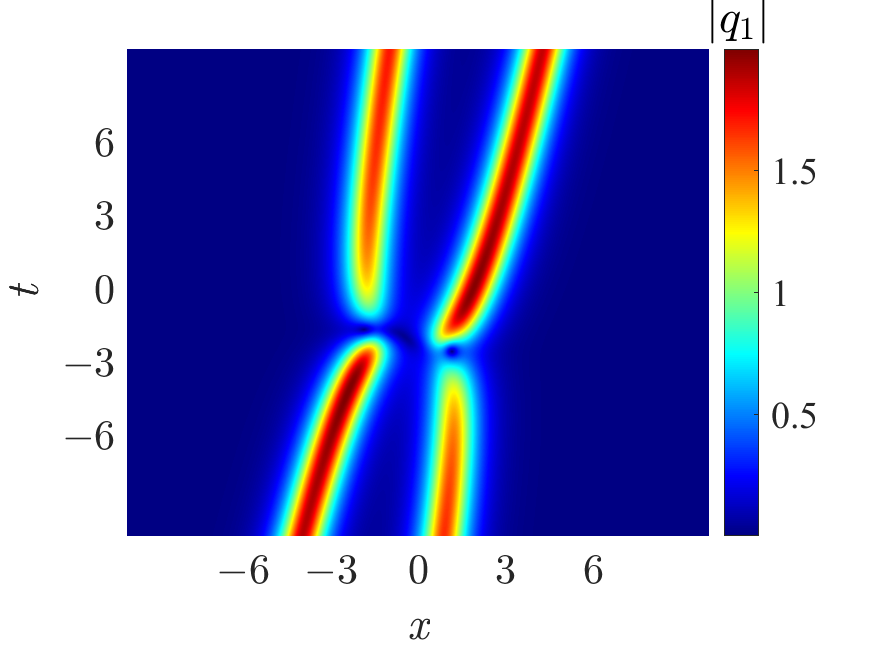}}
	\hspace{0.5cm}{\includegraphics[scale =0.36]{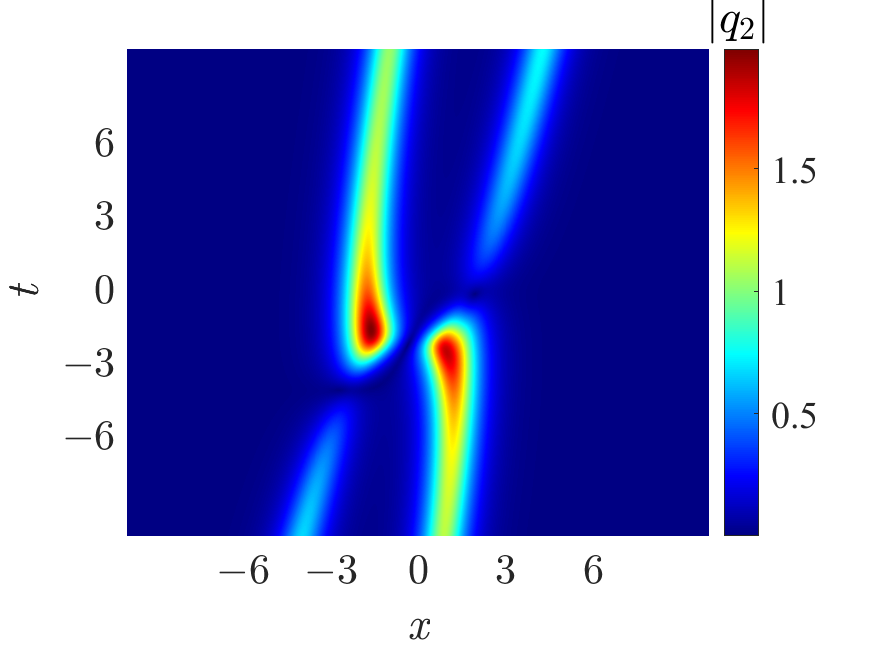}}\\
	\vspace{-0.2cm}{\footnotesize\hspace{-0.2cm}(a)\hspace{5.3cm}(b)}\\
	\vspace{-0.3cm}\flushleft{\footnotesize{\bf Figs.}~1.\,
		The degenerate solitons via Solutions~(\ref{resolitons}) with $\varepsilon=\frac{1}{25}$, $l_{13}=1$, $l_{14}=\frac{1}{2}$, $\lambda_{1,R}=0$, $\lambda_{1,I}=1$, $m_{11}=e^3$, $m_{12}=0$.}
\end{center}

Based on expressions of $\mu_1$, $\mu_2$, $\nu_1$ and $\nu_2$, we find that $\left|\mu_1-\nu_1\right|=\left|\mu_2-\nu_2\right|$, which indicates that the profiles of $S^1$ and $S^2$ are the same and remain invariant both before and after the interactions. For instance, Figs.~2 show the interaction between two-soliton branches: Figs.~2(a$_1$-a$_2$)
depict the interaction between two single-hump solitons in the $q_1$ component, while
Figs.~2(b$_1$-b$_2$) display that two double-hump solitons interact with each other in the $q_2$ component. Both $q_1$ and $q_2$ components exhibit peak profiles in their interaction regions.

\begin{center} \vspace{0cm}
	\hspace{-1cm}\\
\hspace{0.0cm}{\includegraphics[scale =0.36]{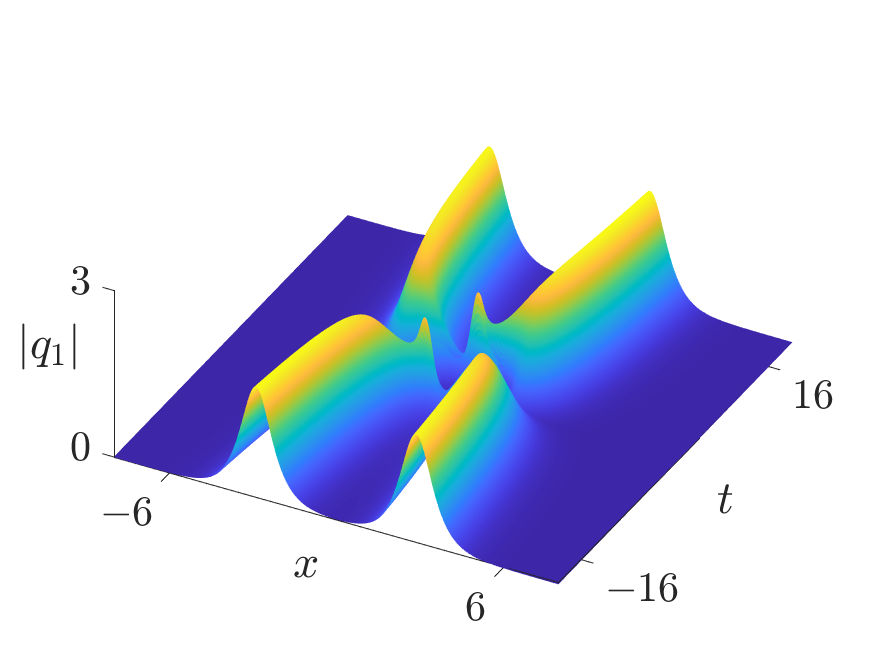}}
\hspace{0.5cm}{\includegraphics[scale =0.36]{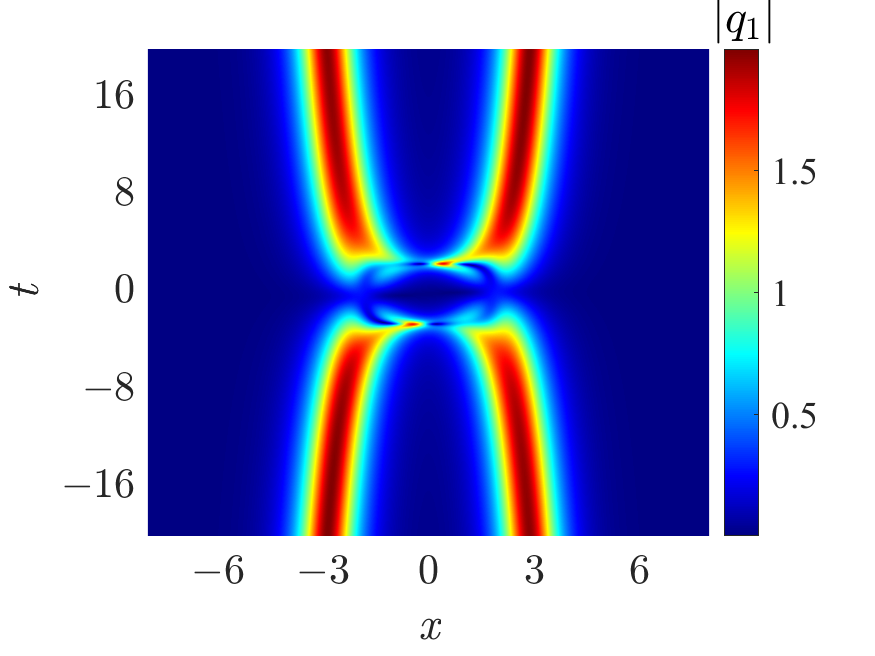}}\\
\vspace{-0.2cm}{\footnotesize\hspace{0.4cm}(a$_1$)\hspace{5cm}(a$_2$)}\\
\hspace{-1cm}\\
\hspace{0.0cm}{\includegraphics[scale =0.36]{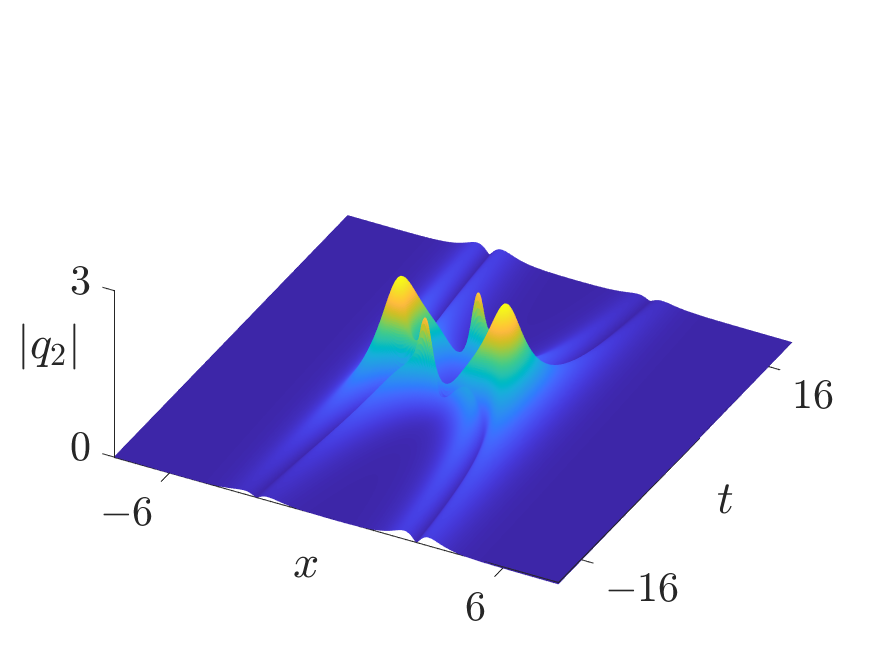}}
\hspace{0.5cm}{\includegraphics[scale =0.36]{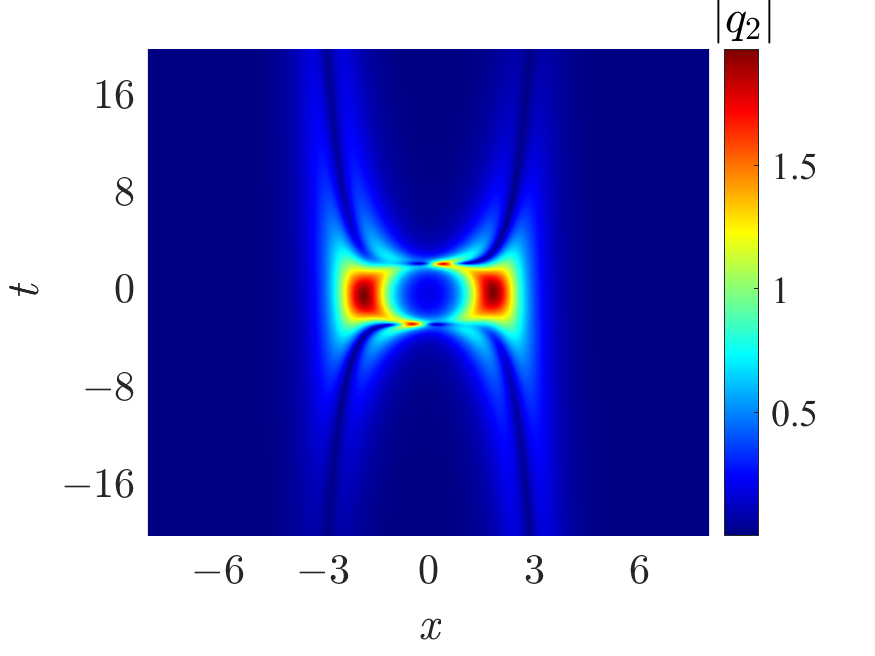}}\\
\vspace{-0.2cm}{\footnotesize\hspace{0.4cm}(b$_1$)\hspace{5cm}(b$_2$)}\\
	\hspace{-1cm}\\
	\vspace{-0.3cm}\flushleft{\footnotesize{\bf Figs.}~2\,
		(a$_1$-a$_2$) and (b$_1$-b$_2$) The degenerate solitons via Solutions~(\ref{resolitons}) with $\varepsilon=\frac{1}{25}$, $l_{13}={\rm i}$, $l_{14}=\frac{1}{20}$, $\lambda_{1,R}=\frac{-1+\sqrt{1+12\lambda_{1,I}^2\varepsilon^2}}{6\varepsilon}$, $\lambda_{1,I}=1$, $m_{11}=0$, $m_{12}=20$.}
\end{center}

{\bf Case \textrm{II}.} As $l_{13}^2+l_{14}^2=0$ and $m_{11}-\frac{l_{13}}{l_{14}}m_{12}\neq0$, we find that Solutions~(\ref{desoliton}) can display the interactions three solitons including $S^1$, $S^2$ and the line soliton $S^{\rm line}$ defined by Expressions~(\ref{asymptotic1}). By virtue of Expressions~(\ref{asymptotic1}), (\ref{asymptotic2}) and (\ref{asymptotic3}), the asymptotic behaviors $S^1$, $S^2$ and $S^{\rm line}$ are expressed as
\begin{subequations}
\begin{eqnarray}
&&\hspace{-1cm}S^{\rm line-}=S^{\rm line+}=\left(\begin{array}{c}
q_1^{\rm line} \\
q_2^{\rm line}
\end{array}
\right),\\
&&\hspace{-1cm}S^{1-}=\left(\begin{array}{c}
q_1^{(1)-} \\
q_2^{(1)-}
\end{array}
\right),\quad S^{1+}=\left(\begin{array}{c}
q_1^{(2)+} \\
q_2^{(2)+}
\end{array}
\right),\\
&&\hspace{-1cm}S^{2-}=\left(\begin{array}{c}
q_1^{(2)-} \\
q_2^{(2)-}
\end{array}
\right), \quad S^{2+}=\left(\begin{array}{c}
q_1^{(1)+} \\
q_2^{(1)+}
\end{array}
\right),
\end{eqnarray}\label{asymptotic5}
\end{subequations}
with
\begin{eqnarray*}
&&\hspace{-1cm}q_1^{(1)}={\rm i}\lambda_{1,I}e^{-2{\rm i}\eta_{1,R}}\frac{l_{13}^{*}\chi^*}{\left|l_{13}\chi\right|}{\rm sech}\left(\xi+\mu\right),\quad q_2^{(1)}=\frac{l_{13}}{l_{14}}q_1^{(1)},\\
&&\hspace{-1cm}q_1^{(2)}={\rm i}\lambda_{1,I}e^{-2{\rm i}\eta_{1,R}}\frac{l_{13}^{*}\chi}{\left|l_{13}\chi\right|}{\rm sech}\left(\zeta+\nu\right),\quad q_2^{(2)}=\frac{l_{13}}{l_{14}}q_1^{(2)},\\
&&\hspace{-1cm}q_1^{\rm line}={\rm{i}}\lambda_{1,I}e^{-2{\rm i}\eta_{1,R}}\frac{m_{11}-\frac{l_{13}}{l_{14}}m_{12}}{\left|m_{11}-\frac{l_{13}}{l_{14}}m_{12}\right|}{\rm sech}\left(\theta-\frac{\rho}{2}\right),\quad q_2^{\rm line}=-\frac{l_{13}}{l_{14}}q_1^{\rm line},
\end{eqnarray*}
where $e^{\mu}=4\lambda_{1,I}\left|l_{13}\chi\right|$ and $e^{\nu}=\frac{\lambda_{1,I}\left|\chi\right|}{\left|l_{13}\right|}$. Through Asymptotic Expressions~(\ref{asymptotic5}), we observe that $\left|q_1\right|$ and $\left|q_2\right|$ are  proportional in the three vector asymptotic solitons $S^1$, $S^2$ and $S^{\rm line}$. This proportionality indicates that $S^1$, $S^2$ and $S^{\rm line}$ exhibit the same intensity profiles in both the $q_1$ and $q_2$ components. In addition, the expressions of $S^1$, $S^2$ and $S^{\rm line}$ are all the sech-type functions which can depict the single-hump bell solitons with the identical amplitudes, i.e., $\left|\lambda_{1,I}\right|$. Both before and after the interactions, the profiles of $S^1$, $S^2$ and $S^{\rm line}$ keep invariant.

The limiting values of $q_s^{(1)-}$ and $q_s^{(2)+}$ indicate that the center trajectory of the asymptotic soliton $S^1$ lies on the curve as follows:
\begin{eqnarray}
&&\left\{\begin{array}{lc}
\mathcal{C}^{-}_{(1)}: \left|e^{\xi}\right|=\frac{8\left|\lambda_{1,I}t\right|}{e^{\theta}}=e^{-\mu}, & t<0,\\
\mathcal{C}^{+}_{(2)}: \left|e^{\zeta}\right|=8\left|\lambda_{1,I}t\right|e^{\theta}=e^{-\nu}, & t>0.
\end{array}\right.
\end{eqnarray}
The limiting values of $q_s^{(2)-}$ and $q_s^{(1)+}$ imply that the central trajectory of the asymptotic soliton $S^2$ is located on the curve as follows:
\begin{eqnarray}
&&\left\{\begin{array}{lc}
\mathcal{C}^{-}_{(2)}: \left|e^{\zeta}\right|=8\left|\lambda_{1,I}t\right|e^{\theta}=e^{-\nu}, & t<0,\\
\mathcal{C}^{+}_{(1)}: \left|e^{\xi}\right|=\frac{8\left|\lambda_{1,I}t\right|}{e^{\theta}}=e^{-\mu}, & t>0.
\end{array}\right.
\end{eqnarray}
In addition, the central trajectory of the asymptotic soliton $S^{\rm line}$ is aligned with the straight line $\theta=\frac{\rho}{2}$.

Figs.~3(a$_1$-a$_2$) and 3(b$_1$-b$_2$) display the elastic interaction among the three vector solitons $S^1$, $S^2$ and $S^{\rm line}$. Obviously,
the intensity profiles of  $S^1$, $S^2$ and $S^{\rm line}$ in both the $q_1$ and $q_2$ components exhibit similarity due to their proportionality. In addition, during the process of the interaction, the profile of the line soliton $S^{\rm line}$ remains unchanged, with no observed phase shifts.

\begin{center} \vspace{0cm}
	\hspace{-1cm}\\
	\hspace{0.0cm}{\includegraphics[scale =0.45]{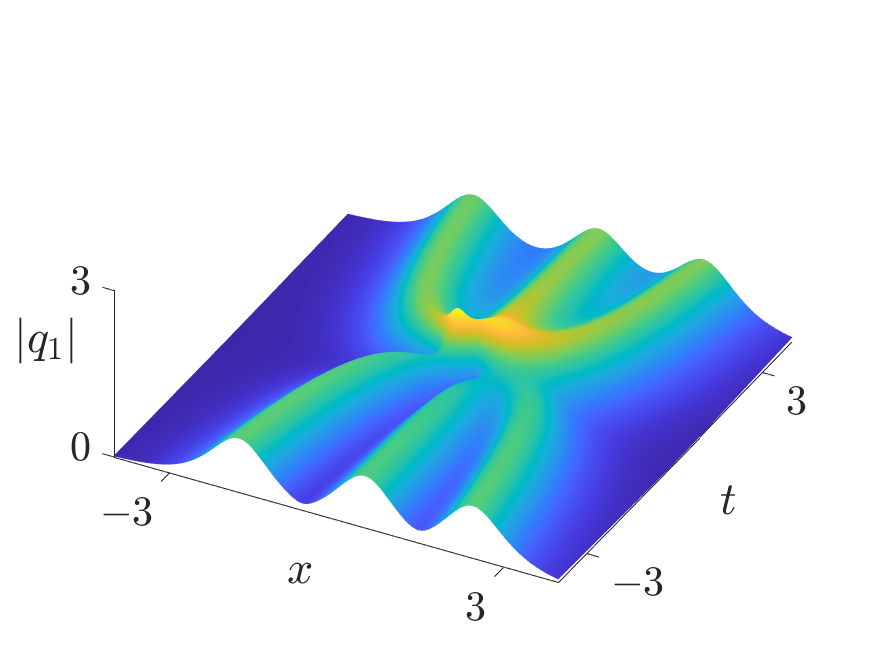}}
	\hspace{0.5cm}{\includegraphics[scale =0.45]{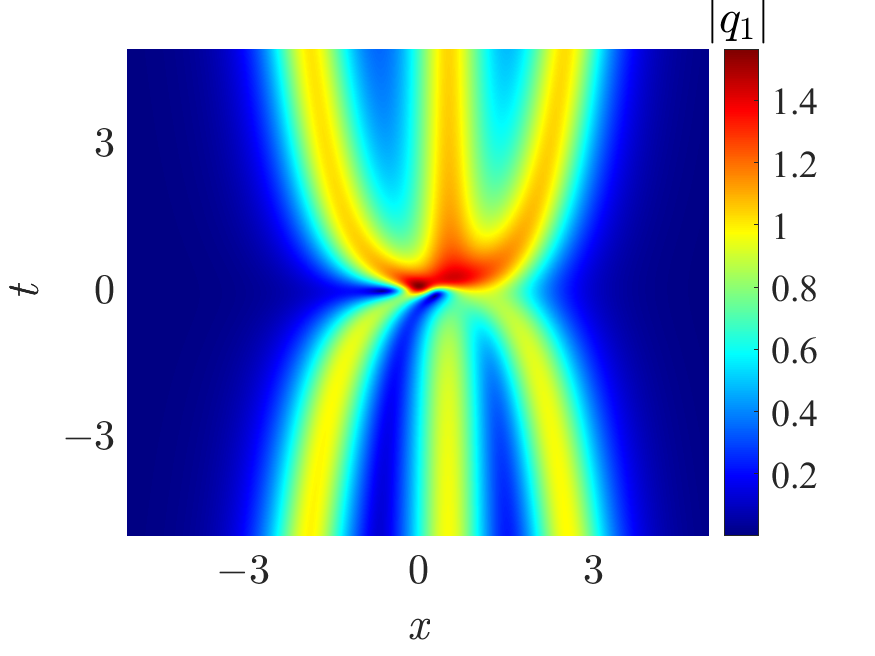}}\\
	\vspace{-0.2cm}{\footnotesize\hspace{0.4cm}(a$_1$)\hspace{5cm}(a$_2$)}\\
	\hspace{-1cm}\\
	\hspace{0.0cm}{\includegraphics[scale =0.45]{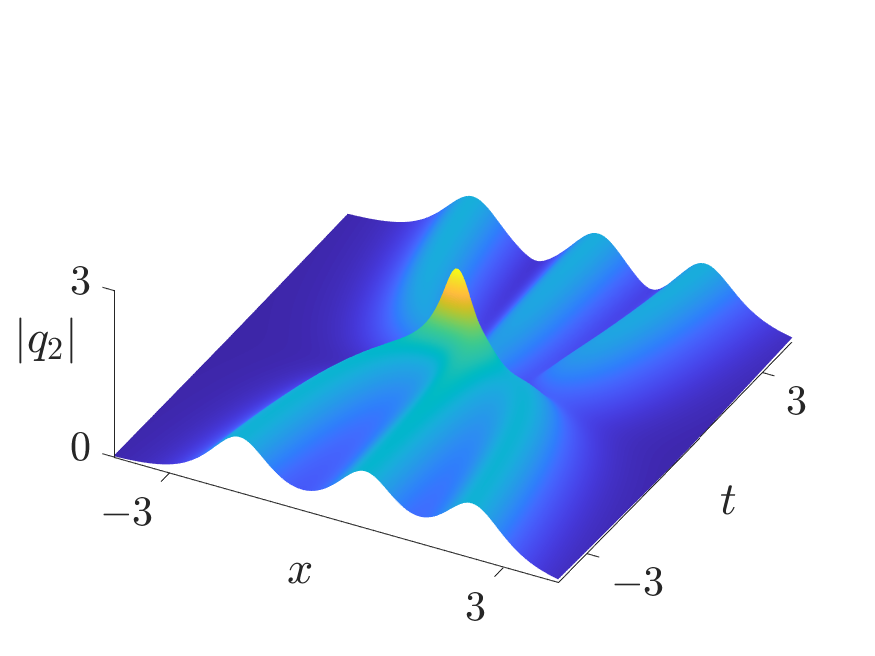}}
	\hspace{0.5cm}{\includegraphics[scale =0.45]{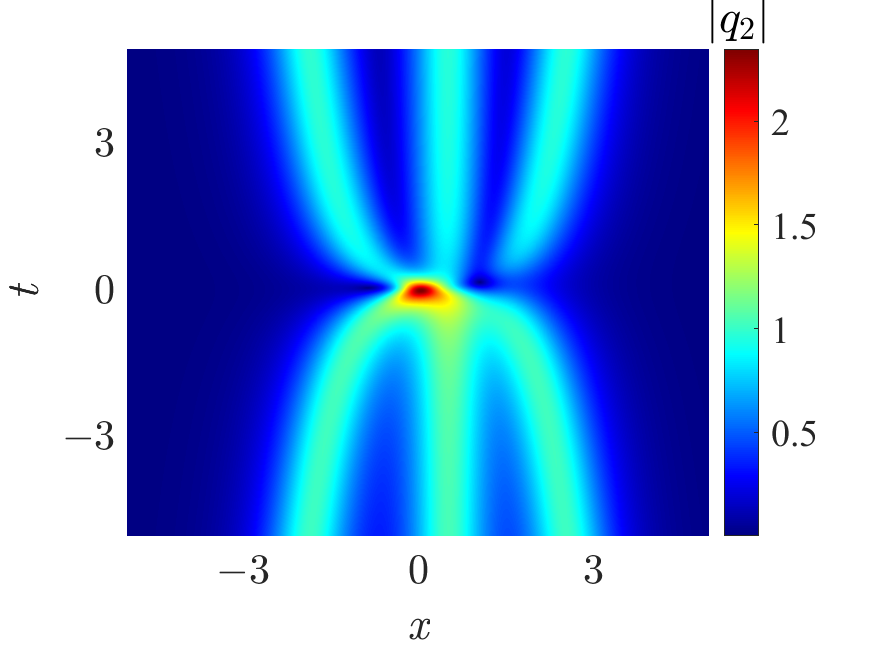}}\\
	\vspace{-0.2cm}{\footnotesize\hspace{0.4cm}(b$_1$)\hspace{5cm}(b$_2$)}\\
	\hspace{-1cm}\\
	\vspace{-0.3cm}\flushleft{\footnotesize{\bf Figs.}~3\,
		(a$_1$-a$_2$) The elastic interaction among $S^1$, $S^2$ and $S^{\rm line}$ via Solutions~(\ref{resolitons}) with $\varepsilon=\frac{1}{25}$, $l_{13}={\rm i}$, $l_{14}=1$,  $\lambda_{1,R}=\frac{-1+\sqrt{1+12\lambda_{1,I}^2\varepsilon^2}}{6\varepsilon}$, $\lambda_{1,I}=1$, $m_{11}=1$, $m_{12}=1$. (b$_1$-b$_2$) The density plot of (a$_1$-a$_2$)}.
\end{center}

{\bf Case \textrm{III}.} When $l_{13}^2+l_{14}^2=0$ and $m_{11}-\frac{l_{13}}{l_{14}}m_{12}=0$, the solitons $S^1$ and $S^2$, as described by Asymptotic Expressions~(\ref{asymptotic5}), possess the same intensity profiles and velocities in both the $q_1$ and $q_2$ components, whereas the line soliton $S^{\rm line}$ disappears under the conditions, as displayed in Figs.~4. With $l_{13}={\rm i}$ and $l_{14}=1$, we derive that the relative distance between the two asymptotic solitons $S^1$ and $S^2$ can be described by
\begin{eqnarray}
\mathcal{D}=\frac{\ln \left(256 \lambda_{1,I}^4 t^2+4096 \lambda_{1,I}^6 t^2 \varepsilon ^2\right)}{2 |\lambda_{1,I}|}.\label{distance}
\end{eqnarray}
Based on Expression~(\ref{distance}), when the values of $t$ and $\lambda_{1,I}$ are fixed, the relative distance between the two asymptotic solitons $S^1$ and $S^2$ increases logarithmically with the higher-order perturbation parameter $|\varepsilon|$, as shown in Figs.~4. It is noted that the relative distance between such solitons is always unchanged for the scalar NLS equation and the coupled NLS systems when the eigenvalue $\lambda_{1}$ is fixed.

\begin{center} \vspace{0cm}
\hspace{0.0cm}{\includegraphics[scale =0.34]{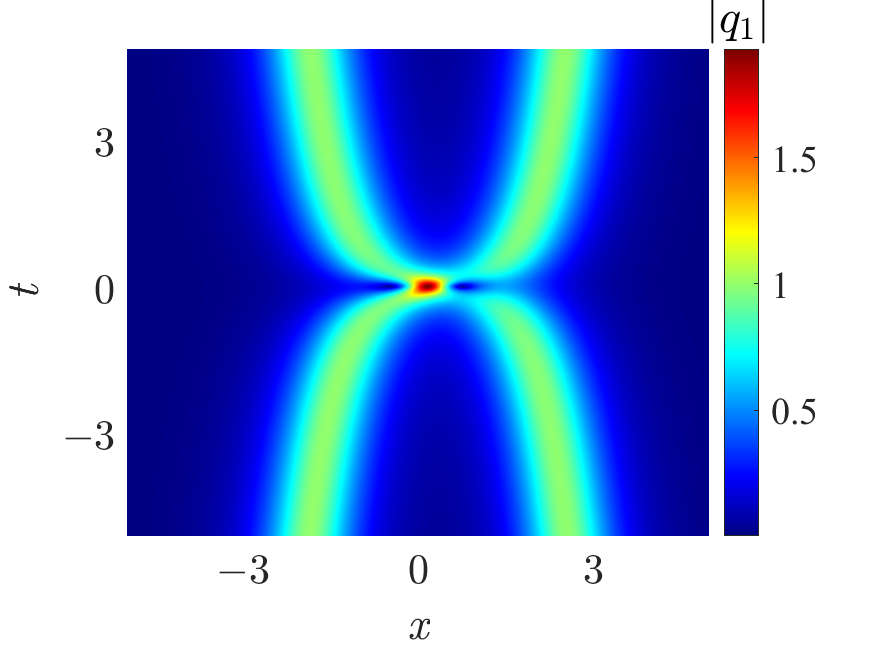}}
\hspace{0.2cm}{\includegraphics[scale =0.34]{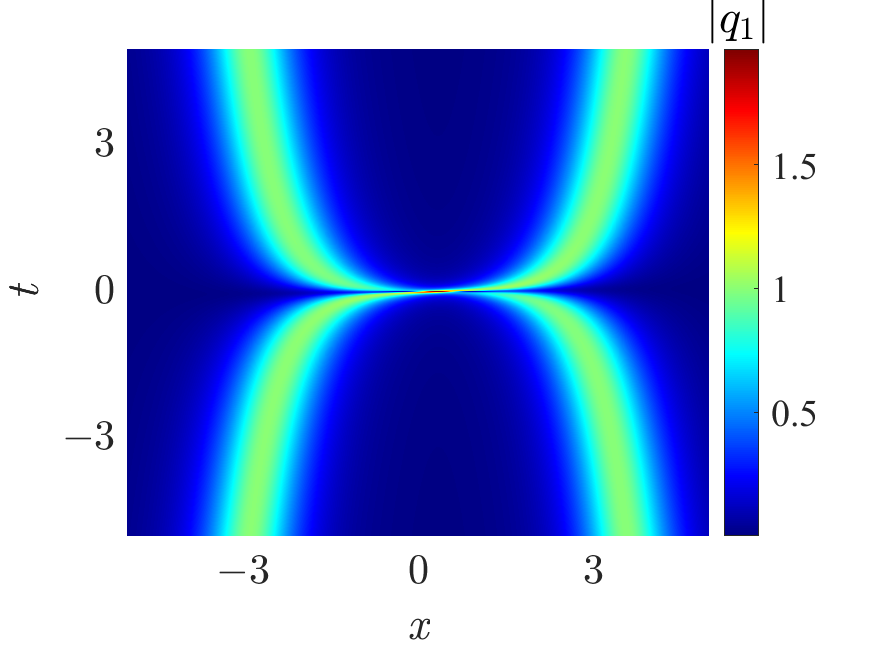}}
\hspace{0.2cm}{\includegraphics[scale =0.34]{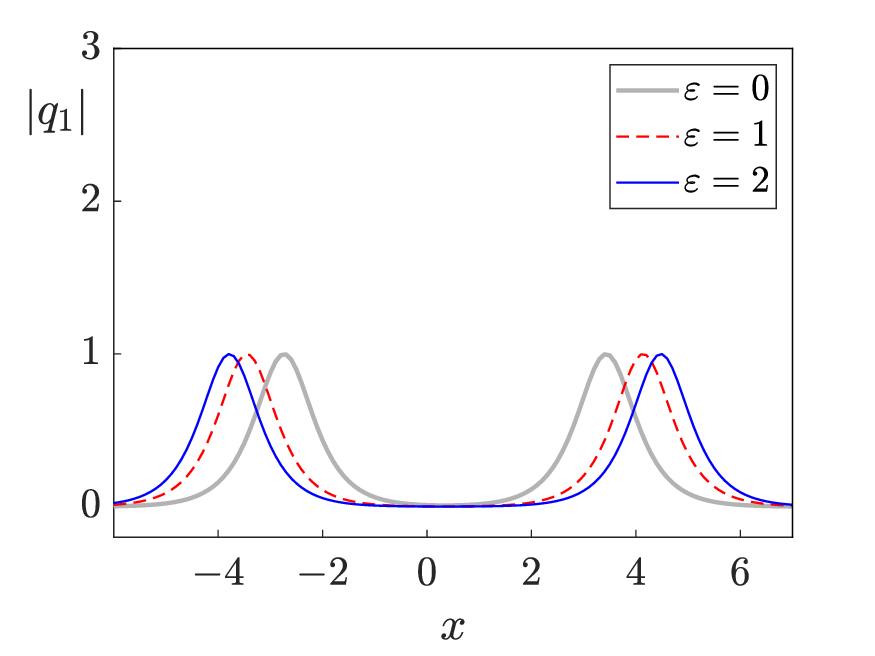}}\\
\vspace{-0.2cm}{\footnotesize\hspace{0.4cm}(a$_1$) \hspace{4.3cm}(b$_1$) \hspace{5cm}(c$_1$)}\\
\hspace{-1cm}\\
\hspace{0.0cm}{\includegraphics[scale =0.34]{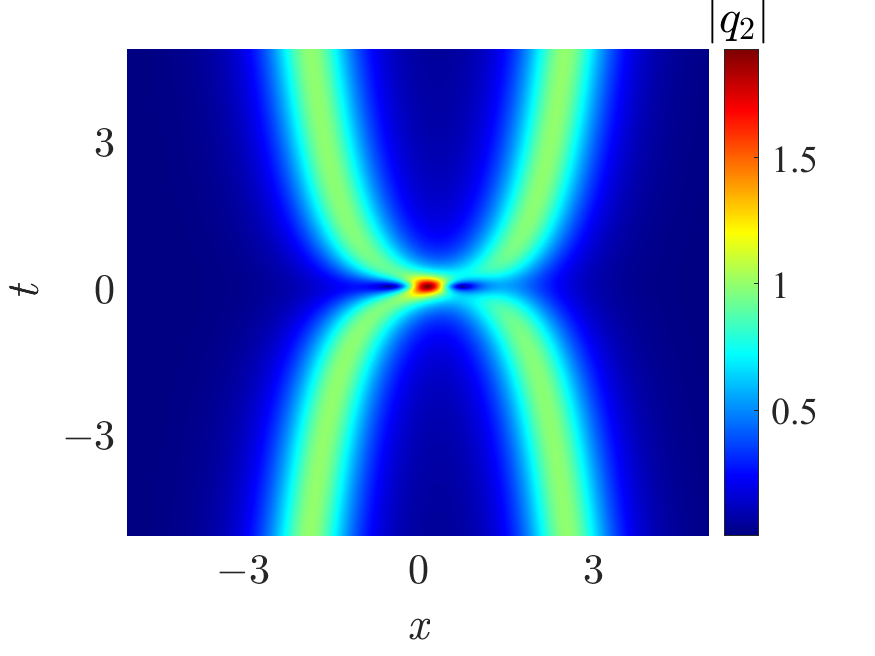}}
\hspace{0.2cm}{\includegraphics[scale =0.34]{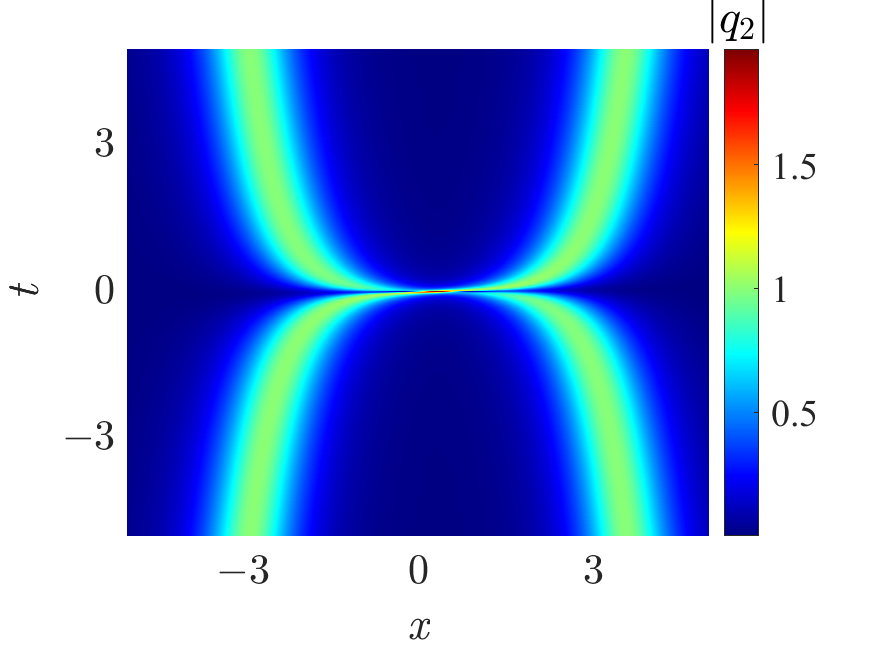}}
\hspace{0.2cm}{\includegraphics[scale =0.34]{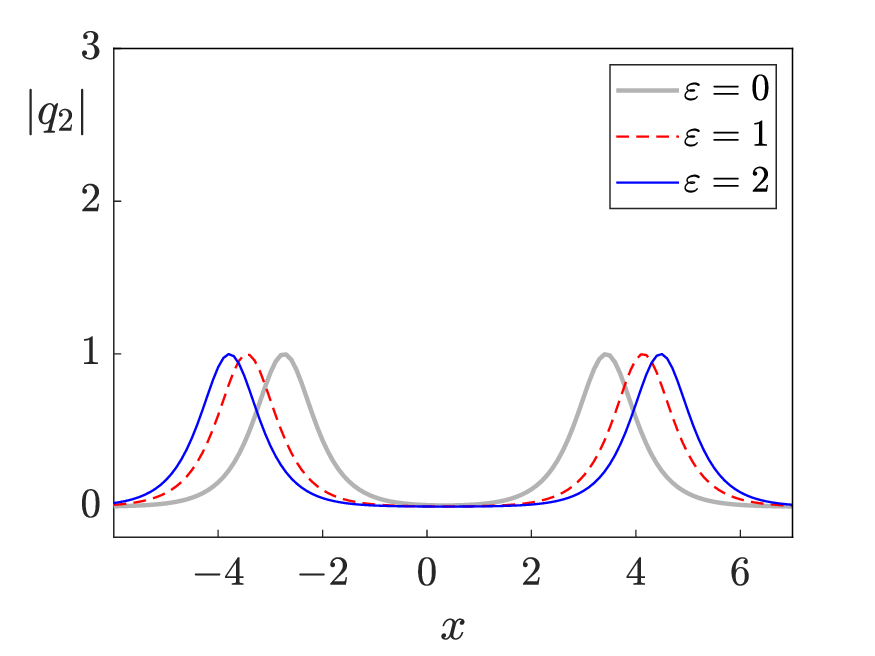}}\\
\vspace{-0.2cm}{\footnotesize\hspace{0.4cm}(a$_2$)\hspace{4.3cm}(b$_2$) \hspace{5cm}(c$_2$)}\\
\hspace{-1cm}\\
	\vspace{-0.3cm}\flushleft{\footnotesize{\bf Figs.}~4\,
		 The elastic interaction between $S^1$ and $S^2$ via Solutions~(\ref{resolitons}) with $l_{13}={\rm i}$, $l_{14}=1$, $m_{11}={\rm i}$, $m_{12}=1$, $\lambda_{1,R}=\frac{-1+\sqrt{1+12\lambda_{1,I}^2\varepsilon^2}}{6\varepsilon}$, $\lambda_{1,I}=1$, (a$_1$-a$_2$) $\varepsilon=\frac{1}{25}$; (b$_1$-b$_2$) $\varepsilon=2$; (c$_1$-c$_2$) Intensity profiles of degenerate solitons at $t=30$ for three different $\varepsilon$ settings.}
\end{center}

In the three cases discussed above, the asymptotic solitons can be localized along the straight line or certain algebraic curves according to Solutions~(\ref{resolitons}), all of which can be approached by the exact solutions when $|t|$ tends to infinity. In what follows, we will graphically demonstrate the validity of the asymptotic analysis. In Figs.~5-7, we compare the degenerate solitons described by Solutions~(\ref{resolitons}) with the asymptotic solitons given in Asymptotic Expressions~(\ref{asymptotic4}) and (\ref{asymptotic5}) for a large value of $|t|$.

In Fig.~5(a), the asymptotic soliton agrees with the degenerate soliton at $t = 30$ in the $q_1$ component. Fig.~5(b) shows a small asymptotic error between the degenerate soliton and asymptotic solitons in the $q_2$ component at $t=30$.
\begin{center} \vspace{-0.4cm}
	\hspace{-1cm}\\
	\hspace{0.0cm}{\includegraphics[scale =0.36]{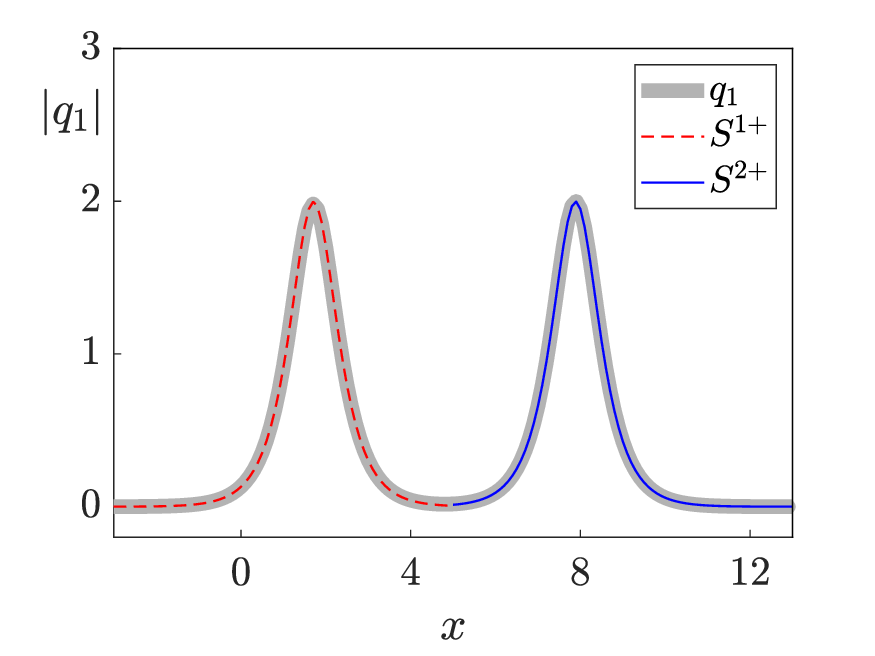}}
	\hspace{0.4cm}{\includegraphics[scale =0.36]{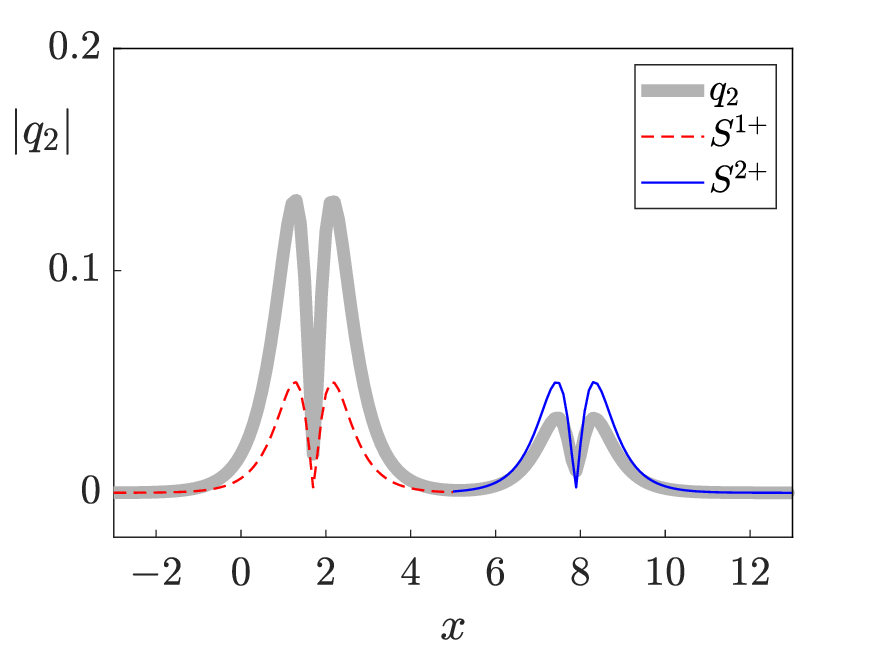}}\\
	\vspace{-0.0cm}{\footnotesize\hspace{0.2cm}(a) \hspace{5.2cm}(b)}\\
	\hspace{-1cm}\\
	\vspace{-0.3cm}\flushleft{\footnotesize{\bf Figs.}~5\,
		(a-b) Comparison of the asymptotic soliton branches $S^{1+}$ and $S^{2+}$ with the analytic degenerate soliton (gray line) in {\bf Case \textrm{I}}. With $t=30$, $S^{1+}$ and $S^{2+}$ are given by Asymptotic Expressions~(\ref{asymptotic4}), while the analytic degenerate soliton is descibed by Solutions~(\ref{resolitons}), using the same parameters as those in Figs.~2(a$_1$-a$_2$) and 2(b$_1$-b$_2$).}
\end{center}

Figs.~6-7 show that all the asymptotic solitons closely match the degenerate solitons when $t=30$, which confirms the accuracy of our asymptotic analysis in describing soliton behaviors.

\begin{center} \vspace{0cm}
	\hspace{-1cm}\\
	\hspace{0.0cm}{\includegraphics[scale =0.45]{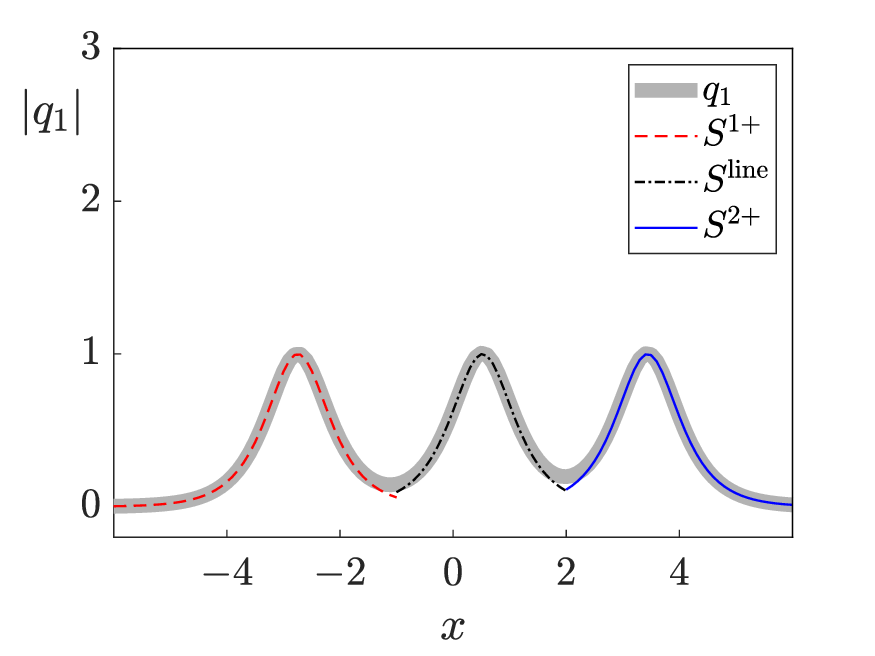}}
	\hspace{0.4cm}{\includegraphics[scale =0.45]{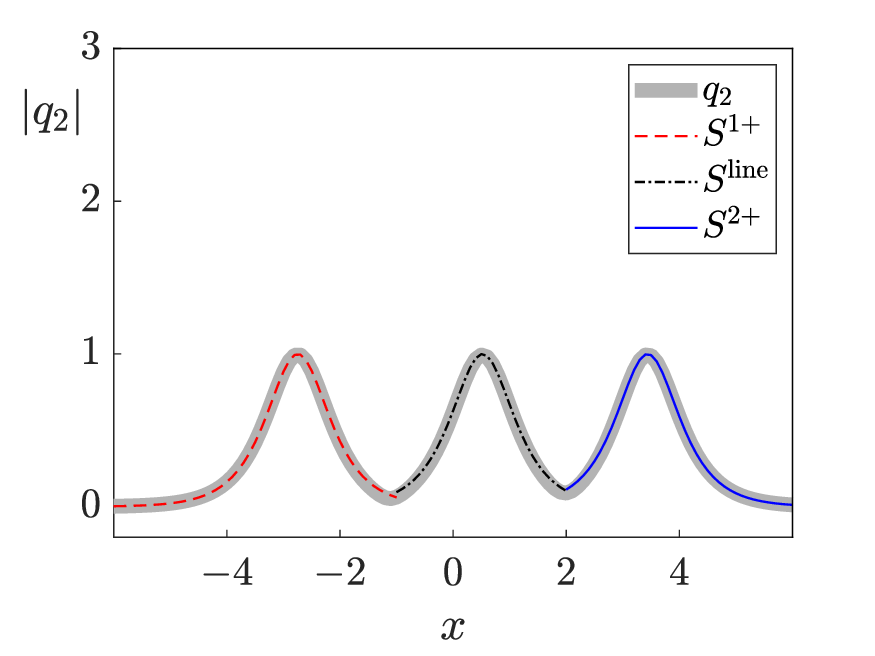}}\\
	\vspace{-0.0cm}{\footnotesize\hspace{0.2cm}(a) \hspace{5.2cm}(b)}\\
	\hspace{-1cm}\\
	\vspace{-0.3cm}\flushleft{\footnotesize{\bf Figs.}~6\,
		(a-b) Comparison of the asymptotic soliton branches $S^{1+}$, $S^{\rm line}$ and $S^{2+}$ with the analytic degenerate soliton (gray line) in {\bf Case \textrm{II}}. With $t=30$, $S^{1+}$, $S^{\rm line}$ and $S^{2+}$ are given by Asymptotic Expressions~(\ref{asymptotic5}), while the analytic degenerate soliton is described by Solutions~(\ref{resolitons}), using the same parameters as those in Figs.~3(a$_1$-a$_2$) and 3(b$_1$-b$_2$).}
\end{center}

\begin{center} \vspace{0cm}
	\hspace{-1cm}\\
	\hspace{0.0cm}{\includegraphics[scale =0.36]{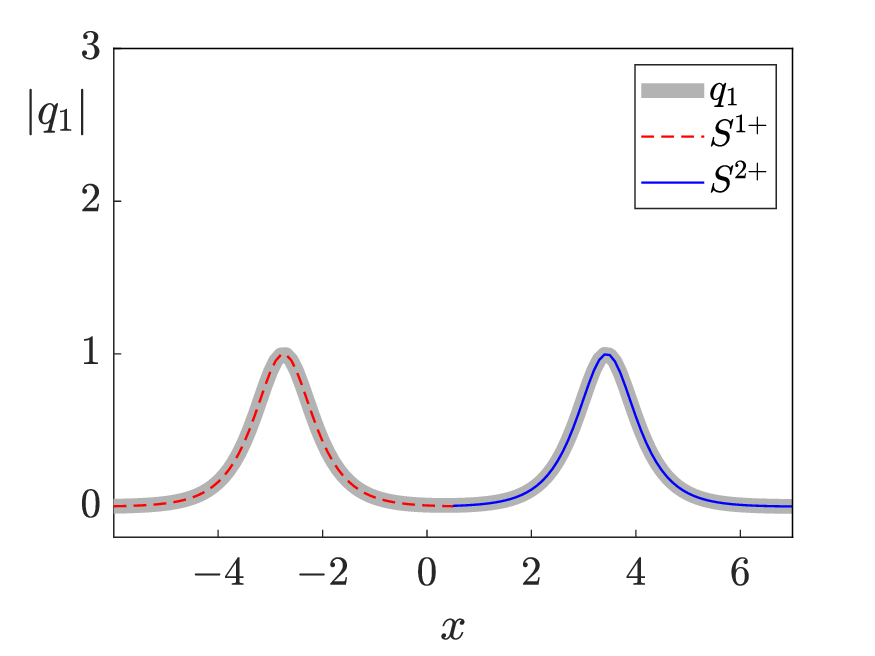}}
	\hspace{0.4cm}{\includegraphics[scale =0.36]{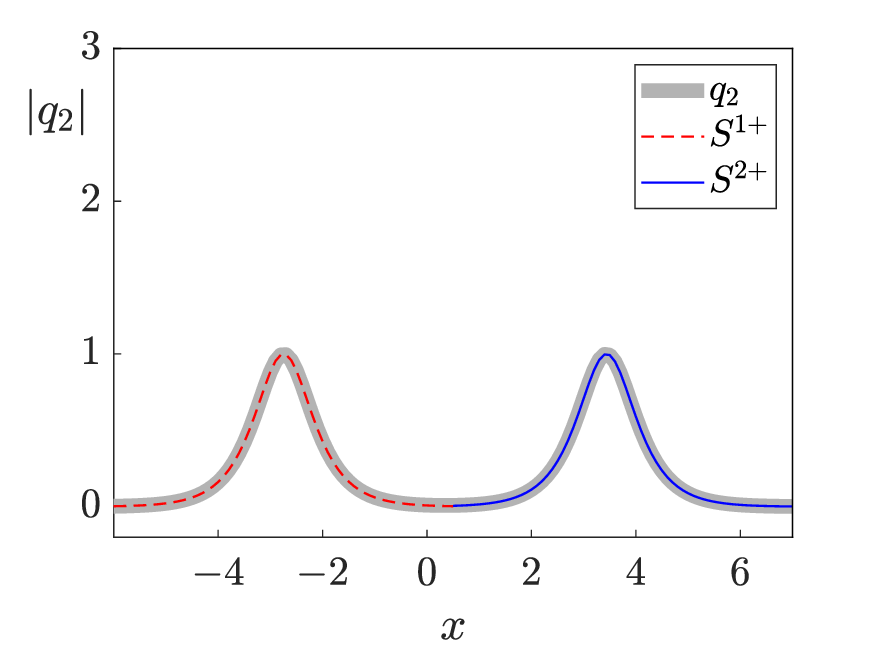}}\\
	\vspace{-0.0cm}{\footnotesize\hspace{0.2cm}(a) \hspace{5.2cm}(b)}\\
	\hspace{-1cm}\\
	\vspace{-0.3cm}\flushleft{\footnotesize{\bf Figs.}~7\,
		(a-b) Comparison of the asymptotic soliton branches $S^{1+}$ and $S^{2+}$ with the analytic degenerate soliton (gray line) in {\bf Case \textrm{III}}. With $t=30$, $S^{1+}$ and $S^{2+}$ are given by Asymptotic Expressions~(\ref{asymptotic5}), while the analytic degenerate soliton is described by Solutions~(\ref{resolitons}), using the same parameters as those in Figs.~4(a$_1$-a$_2$).}
\end{center}

In the above discussion, to simplify the mathematical analysis and focus on certain physical phenomena such as the coherent effect and robustness of solitons, we make certain specific choices and restrictions on the parameters including $l_{13}$, $l_{14}$, $\lambda_1$, and $\varepsilon$ in this paper. These choices enable us to derive the degenerate soliton solutions in a manageable form, but also limit the scope of the solution space. A broader solution space may include other parameter values, which could lead to different types of soliton behaviors or more complex interactions. Our future research will relax these restrictions and explore a broader solution space to reveal new types of solitons and interaction mechanisms, thereby providing a more comprehensive understanding of the physical behavior of the system.

\vspace{5mm}\noindent\textbf{~4.~Asymptotic behaviors of the mixed solitons}\\

As discussed above, the degenerate solitons for System~(\ref{equations}) are composed of the bell-shaped solitons. In this section, we will focus on the interactions between degenerate solitons and other bell-shaped solitons. By means of Solutions (\ref{Nsoliton}) with $N=3$ and $\lambda_2^*\to\lambda_1$, we derive the mixed soliton solutions that describe the interactions between the degenerate solitons and other bell-shaped solitons.

Specifically, as performed in Section~3, we set $l_{11}=l_{23}=l_{31}=1$, $l_{12}=l_{24}=l_{32}=0$, $l_{21}=-l_{13}^*$, and $l_{22}=l_{14}^*$ so that $H_1$, $H_2$, $Y_1$ and $Y_2$ satisfy $H_2^{\dagger}H_1+Y_2^{\dagger}Y_1=0$ and $H_3={\bf I}_{2\times 2}$. With these values, by substituting $\lambda_2^*\to\lambda_1$ into Solutions (\ref{Nsoliton}) with $N=3$, we derive the mixed soliton solutions as follows:
\begin{subequations}
	\begin{eqnarray}
	&&\hspace{-1.7cm}q_1[3]=-2{\rm i}e^{-2{\rm i}\eta_{1,R}}\left(
	\begin{array}{cccccc}
	e^{\theta} & 0 & -l_{13}^* & -l_{14}^* & e^{-{\rm i}\left(\eta_3-\eta_1^*\right)} & 0 \end{array}
	\right)
	\varOmega[3]^{-1}\left(\begin{array}{c}
	l_{13}^*e^{-\theta}\\
	-l_{14}^*e^{-\theta}\\
	1\\
	0\\
	l_{33}^*e^{-{\rm i}\left(\eta_3^*-\eta_1\right)}\\
	-l_{34}^*e^{-{\rm i}\left(\eta_3^*-\eta_1\right)}
	\end{array}\right),\\
	&&\hspace{-1.7cm}q_2[3]=-2{\rm i}e^{-2{\rm i}\eta_{1,R}}\left(
	\begin{array}{cccccc}
	e^{\theta} & 0 & -l_{13}^* & -l_{14}^* & e^{-{\rm i}\left(\eta_3-\eta_1^*\right)} & 0 \end{array}
	\right)\varOmega[3]^{-1}\left(\begin{array}{c}
		l_{14}^*e^{-\theta}\\
	l_{13}^*e^{-\theta}\\
		0\\
		1\\
	l_{34}^*e^{-{\rm i}\left(\eta_3^*-\eta_1\right)}\\
	l_{33}^*e^{-{\rm i}\left(\eta_3^*-\eta_1\right)}
	\end{array}\right),
	\end{eqnarray}\label{mixedsoliton}
\end{subequations}
with
\begin{eqnarray*}
	\varOmega[3]=\left(
	\begin{array}{ccc}
		-\frac{{\rm i}}{2\lambda_{1,I}}\left({\bf I}_{2\times2}e^{\theta}+Y_1^{\dagger}Y_1e^{-\theta}\right) & -Y_1^{\dagger}\left(-\frac{{\rm i}\theta}{\lambda_{1,I}}-8\lambda_{1,I}\chi^*t\right)+M_1 & \varOmega\left(\Gamma_1,\Gamma_3\right)\\
		Y_1\left(\frac{{\rm i}\theta}{\lambda_{1,I}}-8\lambda_{1,I}\chi t\right)-M_1^{\dagger} & \frac{{\rm i}}{2\lambda_{1,I}}\left({\bf I}_{2\times 2}e^{\theta}+Y_1Y_1^{\dagger}e^{-\theta}\right) & \varOmega\left(\Gamma_2,\Gamma_3\right)\\
		\varOmega\left(\Gamma_3,\Gamma_1\right) & \varOmega\left(\Gamma_3,\Gamma_2\right) & \varOmega\left(\Gamma_3,\Gamma_3\right)
	\end{array}\right),
\end{eqnarray*}
where $\varOmega\left(\Gamma_3,\Gamma_s\right)=-\varOmega\left(\Gamma_s,\Gamma_3\right)^{\dagger}=\frac{1}{\lambda_s-\lambda_3^*}\left(H_3^{\dagger}H_se^{{\rm i}\left(\eta_3^*-\eta_s\right)}+Y_3^{\dagger}Y_se^{-{\rm i}\left(\eta_3^*-\eta_s\right)}\right)$ $(s=1,2,3)$.

Considering the influences of dominant terms $te^{-\theta}$, $te^{\theta}$, $e^{\theta}$ and ${\rm i}\left(\eta_3-\eta_3^*\right)$ on mixed soliton behaviors along the different directions, respectively, we will perform the asymptotic analysis on Solutions~(\ref{mixedsoliton}). Through this asymptotic analysis, we will explore changes in the phase shifts, intensity redistributions and velocities of the degenerate solitons and other bell-shaped solitons during their interactions based on the relative values of $l_{13}$, $l_{14}$, $l_{33}$ and $l_{34}$. Furthermore, we will classify the interacting mechanisms for System~(\ref{equations}) in the following four situations.

\vspace{5mm}\noindent{\em ~{\bf A}.~Case 1: $l_{13}^2+l_{14}^2\neq0$ and $l_{33}^2+l_{34}^2\neq0$}\\

In this case, the asymptotic behaviors of mixed solitons are as follows:

(1) {\em Before the interaction} ($t\to-\infty$)

\hspace{0.65cm}{\em Soliton} $S^{1-}$ $\left(\theta\to+\infty, te^{-\theta}\sim\mathcal{O}(1), {\rm i}(\eta_3-\eta_3^*)\to-\infty \right)$
\begin{eqnarray}
\left(\begin{array}{c}
q_1\\
q_2
\end{array}\right)\to\left(\begin{array}{c}
q_1^{(1)-}\\
q_2^{(1)-}
\end{array}\right)=\frac{1}{2}\left[\left(\begin{array}{c}
\phi_1^{(1)-} \\
-{\rm i} \phi_1^{(1)-}
\end{array}\right)+\left(\begin{array}{c}
\phi_2^{(1)-} \\
{\rm i}\phi_2^{(1)-}
\end{array}\right)\right],\label{AS1n}
\end{eqnarray}
with
\begin{eqnarray*}
&&\hspace{-1cm} \phi_1^{(1)-}=2{\rm i}\lambda_{1,I}e^{-2{\rm i}\eta_{1,R}}\frac{(\lambda_1-\lambda_3)(\lambda_1^*-\lambda_3)\left(l_{13}-{\rm i}l_{14}\right)^{*}\chi^*}{\left|(\lambda_1-\lambda_3)(\lambda_1^*-\lambda_3)\left(l_{13}-{\rm i}l_{14}\right)\chi\right|}{\rm sech}\left(\xi+a_1^-\right),\\
&&\hspace{-1cm} \phi_2^{(1)-}=2{\rm i}\lambda_{1,I}e^{-2{\rm i}\eta_{1,R}}\frac{(\lambda_1-\lambda_3)(\lambda_1^*-\lambda_3)\left(l_{13}+{\rm i}l_{14}\right)^{*}\chi^*}{\left|(\lambda_1-\lambda_3)(\lambda_1^*-\lambda_3)\left(l_{13}+{\rm i}l_{14}\right)\chi\right|}{\rm sech}\left(\xi+b_1^-\right),
\end{eqnarray*}
where $e^{a_1^{-}}=2\lambda_{1,I}\left|\frac{\left(l_{13}-{\rm i}l_{14}\right)\chi(\lambda_1^*-\lambda_3)}{\lambda_1-\lambda_3}\right|$ and $e^{b_1^{-}}=2\lambda_{1,I}\left|\frac{\left(l_{13}+{\rm i}l_{14}\right)\chi(\lambda_1^*-\lambda_3)}{\lambda_1-\lambda_3}\right|$.

\hspace{0.65cm}{\em Soliton} $S^{2-}$ $\left(\theta\to -\infty, te^{\theta}\sim\mathcal{O}(1), {\rm i}(\eta_3-\eta_3^*)\to-\infty \right)$
\begin{eqnarray}
\left(\begin{array}{c}
q_1\\
q_2
\end{array}\right)\to\left(\begin{array}{c}
q_1^{(2)-}\\
q_2^{(2)-}
\end{array}\right)=\frac{1}{2}\left[\left(\begin{array}{c}
\phi_1^{(2)-} \\
-{\rm i} \phi_1^{(2)-}
\end{array}\right)+\left(\begin{array}{c}
\phi_2^{(2)-} \\
{\rm i}\phi_2^{(2)-}
\end{array}\right)\right],\label{AS2n}
\end{eqnarray}
with
\begin{eqnarray*}
	&&\hspace{-1cm} \phi_1^{(2)-}=2{\rm i}\lambda_{1,I}e^{-2{\rm i}\eta_{1,R}}\frac{(\lambda_1-\lambda_3)(\lambda_1^*-\lambda_3)\left(l_{13}-{\rm i}l_{14}\right)^{*}\chi}{\left|(\lambda_1-\lambda_3)(\lambda_1^*-\lambda_3)\left(l_{13}-{\rm i}l_{14}\right)\chi\right|}{\rm sech}\left(\zeta+a_2^{-}\right),\\
	&&\hspace{-1cm} \phi_2^{(2)-}=2{\rm i}\lambda_{1,I}e^{-2{\rm i}\eta_{1,R}}\frac{(\lambda_1-\lambda_3)(\lambda_1^*-\lambda_3)\left(l_{13}+{\rm i}l_{14}\right)^{*}\chi}{\left|(\lambda_1-\lambda_3)(\lambda_1^*-\lambda_3)\left(l_{13}+{\rm i}l_{14}\right)\chi\right|}{\rm sech}\left(\zeta+b_2^{-}\right),
\end{eqnarray*}
where $e^{a_2^{-}}=2\lambda_{1,I}\left|\frac{(\lambda_1-\lambda_3)\chi}{\left(l_{13}-{\rm i}l_{14}\right)(\lambda_1^*-\lambda_3)}\right|$ and $e^{b_2^{-}}=2\lambda_{1,I}\left|\frac{(\lambda_1-\lambda_3)\chi}{\left(l_{13}+{\rm i}l_{14}\right)(\lambda_1^*-\lambda_3)}\right|$.

\hspace{0.65cm}{\em Soliton} $S^{3-}$ $\left( {\rm i}(\eta_3-\eta_3^*)\sim\mathcal{O}(1), \theta\to +\infty \right)$
\begin{eqnarray}
\left(\begin{array}{c}
q_1\\
q_2
\end{array}\right)\to\left(\begin{array}{c}
q_1^{(3)-}\\
q_2^{(3)-}
\end{array}\right)=\frac{1}{2}\left[\left(\begin{array}{c}
\phi_1^{(3)-} \\
-{\rm i} \phi_1^{(3)-}
\end{array}\right)+\left(\begin{array}{c}
\phi_2^{(3)-} \\
{\rm i}\phi_2^{(3)-}
\end{array}\right)\right],\label{AS3n}
\end{eqnarray}
with
\begin{eqnarray*}
	&&\hspace{-1cm} \phi_1^{(3)-}=2\lambda_{3,I}e^{-2{\rm i}\eta_{3,R}}\frac{(\lambda_1-\lambda_3)(\lambda_1-\lambda_3^*)\left(l_{33}-{\rm i}l_{34}\right)^{*}}{(\lambda_1^*-\lambda_3^*)(\lambda_1^*-\lambda_3)\left|l_{33}-{\rm i}l_{34}\right|}{\rm sech}\left(2\eta_{3,I}-a_3^{-}\right),\\
	&&\hspace{-1cm} \phi_2^{(3)-}=2\lambda_{3,I}e^{-2{\rm i}\eta_{3,R}}\frac{(\lambda_1-\lambda_3)(\lambda_1-\lambda_3^*)\left(l_{33}+{\rm i}l_{34}\right)^{*}}{(\lambda_1^*-\lambda_3^*)(\lambda_1^*-\lambda_3)\left|l_{33}+{\rm i}l_{34}\right|}{\rm sech}\left(2\eta_{3,I}-b_3^{-}\right),
\end{eqnarray*}
where $e^{a_3^{-}}=\frac{|\lambda_1^*-\lambda_3|^2\left|l_{33}-{\rm i}l_{34}\right|}{\left|\lambda_1-\lambda_3\right|^2}$ and $e^{b_3^{-}}=\frac{|\lambda_1^*-\lambda_3|^2\left|l_{33}+{\rm i}l_{34}\right|}{\left|\lambda_1-\lambda_3\right|^2}$.

(2) {\em After the interaction} ($t\to  +\infty$)

\hspace{0.65cm}{\em Soliton} $S^{1+}$ $\left(\theta\to -\infty, te^{\theta}\sim\mathcal{O}(1), {\rm i}(\eta_3-\eta_3^*)\to  +\infty \right)$
\begin{eqnarray}
\left(\begin{array}{c}
q_1\\
q_2
\end{array}\right)\to\left(\begin{array}{c}
q_1^{(2)+}\\
q_2^{(2)+}
\end{array}\right)=\frac{1}{2}\left[\left(\begin{array}{c}
\phi_1^{(2)+} \\
-{\rm i} \phi_1^{(2)+}
\end{array}\right)+\left(\begin{array}{c}
\phi_2^{(2)+} \\
{\rm i}\phi_2^{(2)+}
\end{array}\right)\right],\label{AS1p}
\end{eqnarray}
with
\begin{eqnarray*}
	&&\hspace{-1cm} \phi_1^{(2)+}=2{\rm i}\lambda_{1,I}e^{-2{\rm i}\eta_{1,R}}\frac{(\lambda_1^*-\lambda_3^*)(\lambda_1-\lambda_3^*)\left(l_{13}-{\rm i}l_{14}\right)^{*}\chi}{\left|(\lambda_1-\lambda_3)(\lambda_1^*-\lambda_3)\left(l_{13}-{\rm i}l_{14}\right)\chi\right|}{\rm sech}\left(\zeta+a_2^{+}\right),\\
	&&\hspace{-1cm} \phi_2^{(2)+}=2{\rm i}\lambda_{1,I}e^{-2{\rm i}\eta_{1,R}}\frac{(\lambda_1^*-\lambda_3^*)(\lambda_1-\lambda_3^*)\left(l_{13}+{\rm i}l_{14}\right)^{*}\chi}{\left|(\lambda_1-\lambda_3)(\lambda_1^*-\lambda_3)\left(l_{13}+{\rm i}l_{14}\right)\chi\right|}{\rm sech}\left(\zeta+b_2^{+}\right),
\end{eqnarray*}
where $e^{a_2^{+}}=2\lambda_{1,I}\left|\frac{(\lambda_1^*-\lambda_3)\chi}{\left(l_{13}-{\rm i}l_{14}\right)(\lambda_1-\lambda_3)}\right|$ and $e^{b_2^{+}}=2\lambda_{1,I}\left|\frac{(\lambda_1^*-\lambda_3)\chi}{\left(l_{13}+{\rm i}l_{14}\right)(\lambda_1-\lambda_3)}\right|$.

\hspace{0.65cm}{\em Soliton} $S^{2+}$ $\left(\theta\to+\infty, te^{-\theta}\sim\mathcal{O}(1), {\rm i}(\eta_3-\eta_3^*)\to +\infty \right)$
\begin{eqnarray}
\left(\begin{array}{c}
q_1\\
q_2
\end{array}\right)\to\left(\begin{array}{c}
q_1^{(1)+}\\
q_2^{(1)+}
\end{array}\right)=\frac{1}{2}\left[\left(\begin{array}{c}
\phi_1^{(1)+} \\
-{\rm i} \phi_1^{(1)+}
\end{array}\right)+\left(\begin{array}{c}
\phi_2^{(1)+} \\
{\rm i}\phi_2^{(1)+}
\end{array}\right)\right],\label{AS2p}
\end{eqnarray}
with
\begin{eqnarray*}
	&&\hspace{-1cm} \phi_1^{(1)+}=2{\rm i}\lambda_{1,I}e^{-2{\rm i}\eta_{1,R}}\frac{(\lambda_1^*-\lambda_3^*)(\lambda_1-\lambda_3^*)\left(l_{13}-{\rm i}l_{14}\right)^{*}\chi^*}{\left|(\lambda_1-\lambda_3)(\lambda_1^*-\lambda_3)\left(l_{13}-{\rm i}l_{14}\right)\chi\right|}{\rm sech}\left(\xi+a_1^+\right),\\
	&&\hspace{-1cm} \phi_2^{(1)+}=2{\rm i}\lambda_{1,I}e^{-2{\rm i}\eta_{1,R}}\frac{(\lambda_1^*-\lambda_3^*)(\lambda_1-\lambda_3^*)\left(l_{13}+{\rm i}l_{14}\right)^{*}\chi^*}{\left|(\lambda_1-\lambda_3)(\lambda_1^*-\lambda_3)\left(l_{13}+{\rm i}l_{14}\right)\chi\right|}{\rm sech}\left(\xi+b_1^+\right),
\end{eqnarray*}
where $e^{a_1^{+}}=2\lambda_{1,I}\left|\frac{\left(l_{13}-{\rm i}l_{14}\right)\chi(\lambda_1-\lambda_3)}{\lambda_1^*-\lambda_3}\right|$ and $e^{b_1^{+}}=2\lambda_{1,I}\left|\frac{\left(l_{13}+{\rm i}l_{14}\right)\chi(\lambda_1-\lambda_3)}{\lambda_1^*-\lambda_3}\right|$.

\hspace{0.65cm}{\em Soliton} $S^{3+}$ $\left( {\rm i}(\eta_3-\eta_3^*)\sim\mathcal{O}(1), \theta\to -\infty \right)$
\begin{eqnarray}
\left(\begin{array}{c}
q_1\\
q_2
\end{array}\right)\to\left(\begin{array}{c}
q_1^{(3)+}\\
q_2^{(3)+}
\end{array}\right)=\frac{1}{2}\left[\left(\begin{array}{c}
\phi_1^{(3)+} \\
-{\rm i} \phi_1^{(3)+}
\end{array}\right)+\left(\begin{array}{c}
\phi_2^{(3)+} \\
{\rm i}\phi_2^{(3)+}
\end{array}\right)\right],\label{AS3p}
\end{eqnarray}
with
\begin{eqnarray*}
	&&\hspace{-1cm} \phi_1^{(3)+}=2\lambda_{3,I}e^{-2{\rm i}\eta_{3,R}}\frac{(\lambda_1^*-\lambda_3^*)(\lambda_1^*-\lambda_3)\left(l_{33}-{\rm i}l_{34}\right)^{*}}{(\lambda_1-\lambda_3)(\lambda_1-\lambda_3^*)\left|l_{33}-{\rm i}l_{34}\right|}{\rm sech}\left(2\eta_{3,I}-a_3^{+}\right),\\
	&&\hspace{-1cm} \phi_2^{(3)+}=2\lambda_{3,I}e^{-2{\rm i}\eta_{3,R}}\frac{(\lambda_1^*-\lambda_3^*)(\lambda_1^*-\lambda_3)\left(l_{33}+{\rm i}l_{34}\right)^{*}}{(\lambda_1-\lambda_3)(\lambda_1-\lambda_3^*)\left|l_{33}+{\rm i}l_{34}\right|}{\rm sech}\left(2\eta_{3,I}-b_3^{+}\right),
\end{eqnarray*}
where $e^{a_3^{+}}=\frac{|\lambda_1-\lambda_3|^2\left|l_{33}-{\rm i}l_{34}\right|}{\left|\lambda_1^*-\lambda_3\right|^2}$ and $e^{b_3^{+}}=\frac{|\lambda_1-\lambda_3|^2\left|l_{33}+{\rm i}l_{34}\right|}{\left|\lambda_1^*-\lambda_3\right|^2}$.

\vspace{5mm}\noindent{\em ~{\bf B}.~Case 2: $l_{13}^2+l_{14}^2\neq0$ and $l_{33}^2+l_{34}^2=0$}\\

The mixed solitons exhibit the following asymptotic behaviors:

(1) {\em Before the interaction} ($t\to-\infty$)

\hspace{0.65cm}{\em Soliton} $S^{1-}$ $\left(\theta\to+\infty, te^{-\theta}\sim\mathcal{O}(1), {\rm i}(\eta_3-\eta_3^*)\to-\infty \right)$
\begin{eqnarray}
\left(\begin{array}{c}
q_1\\
q_2
\end{array}\right)\to\left(\begin{array}{c}
q_1^{(1)-}\\
q_2^{(1)-}
\end{array}\right)=\frac{1}{2}\left[\left(\begin{array}{c}
\phi_1^{(1)-} \\
-{\rm i} \phi_1^{(1)-}
\end{array}\right)+\left(\begin{array}{c}
\phi_2^{(1)-} \\
{\rm i}\phi_2^{(1)-}
\end{array}\right)\right].\label{BS1n}
\end{eqnarray}

\hspace{0.65cm}{\em Soliton} $S^{2-}$ $\left(\theta\to -\infty, te^{\theta}\sim\mathcal{O}(1), {\rm i}(\eta_3-\eta_3^*)\to-\infty \right)$
\begin{eqnarray}
\left(\begin{array}{c}
q_1\\
q_2
\end{array}\right)\to\left(\begin{array}{c}
q_1^{(2)-}\\
q_2^{(2)-}
\end{array}\right)=\frac{1}{2}\left[\left(\begin{array}{c}
\phi_1^{(2)-} \\
-{\rm i} \phi_1^{(2)-}
\end{array}\right)+\left(\begin{array}{c}
\phi_2^{(2)-} \\
{\rm i}\phi_2^{(2)-}
\end{array}\right)\right].\label{BS2n}
\end{eqnarray}

\hspace{0.65cm}{\em Soliton} $S^{3-}$ $\left( {\rm i}(\eta_3-\eta_3^*)\sim\mathcal{O}(1), \theta\to +\infty \right)$
\begin{eqnarray}
\left(\begin{array}{c}
q_1\\
q_2
\end{array}\right)\to\left(\begin{array}{c}
q_1^{(3)-}\\
q_2^{(3)-}
\end{array}\right)=\lambda_{3,I}e^{-2{\rm i}\eta_{3,R}}\frac{(\lambda_1-\lambda_3)(\lambda_1-\lambda_3^*)l_{33}^{*}}{(\lambda_1^*-\lambda_3^*)(\lambda_1^*-\lambda_3)\left|l_{33}\right|}\left(\begin{array}{c}
{\rm sech}\left(2\eta_{3,I}-\hat{a}_3^{-}\right) \\
\frac{l_{33}}{l_{34}}{\rm sech}\left(2\eta_{3,I}-\hat{a}_3^{-}\right)
\end{array}\right),\label{BS3n}
\end{eqnarray}
with $e^{\hat{a}_3^{-}}=2\frac{|\lambda_1^*-\lambda_3|^2\left|l_{33}\right|}{\left|\lambda_1-\lambda_3\right|^2}$.

(2) {\em After the interaction} ($t\to  +\infty$)

\hspace{0.65cm}{\em Soliton} $S^{1+}$ $\left(\theta\to -\infty, te^{\theta}\sim\mathcal{O}(1), {\rm i}(\eta_3-\eta_3^*)\to  +\infty \right)$
\begin{eqnarray}
\left(\begin{array}{c}
q_1\\
q_2
\end{array}\right)\to\left(\begin{array}{c}
q_1^{(2)+}\\
q_2^{(2)+}
\end{array}\right)=\frac{1}{2}\left[\left(\begin{array}{c}
\phi_{1B}^{(2)+} \\
-{\rm i} \phi_{1B}^{(2)+}
\end{array}\right)+\left(\begin{array}{c}
\phi_{2B}^{(2)+} \\
{\rm i}\phi_{2B}^{(2)+}
\end{array}\right)\right],\label{BS1p}
\end{eqnarray}
with
\begin{eqnarray*}
	&&\hspace{-1cm} \phi_{1B}^{(2)+}=2{\rm i}\lambda_{1,I}e^{-2{\rm i}\eta_{1,R}}\frac{\left(l_{13}-{\rm i}l_{14}\right)^{*}\chi}{\left|\left(l_{13}-{\rm i}l_{14}\right)\chi\right|}{\rm sech}\left(\zeta+a_{2B}^{+}\right),\quad \phi_{2B}^{(2)+}=\phi_{2}^{(2)+},
\end{eqnarray*}
where $e^{a_{2B}^{+}}=2\lambda_{1,I}\left|\frac{\chi}{l_{13}-{\rm i}l_{14}}\right|$.

\hspace{0.65cm}{\em Soliton} $S^{2+}$ $\left(\theta\to+\infty, te^{-\theta}\sim\mathcal{O}(1), {\rm i}(\eta_3-\eta_3^*)\to +\infty \right)$
\begin{eqnarray}
\left(\begin{array}{c}
q_1\\
q_2
\end{array}\right)\to\left(\begin{array}{c}
q_1^{(1)+}\\
q_2^{(1)+}
\end{array}\right)=\frac{1}{2}\left[\left(\begin{array}{c}
\phi_{1B}^{(1)+} \\
-{\rm i} \phi_{1B}^{(1)+}
\end{array}\right)+\left(\begin{array}{c}
\phi_{2B}^{(1)+} \\
{\rm i}\phi_{2B}^{(1)+}
\end{array}\right)\right],\label{BS2p}
\end{eqnarray}
with
\begin{eqnarray*}
	&&\hspace{-1cm} \phi_{1B}^{(1)+}=2{\rm i}\lambda_{1,I}e^{-2{\rm i}\eta_{1,R}}\frac{\left(l_{13}-{\rm i}l_{14}\right)^{*}\chi^*}{\left|\left(l_{13}-{\rm i}l_{14}\right)\chi\right|}{\rm sech}\left(\xi+a_{1B}^+\right),\quad  \phi_{2B}^{(1)+}=\phi_2^{(1)+},
\end{eqnarray*}
where $e^{a_{1B}^{+}}=2\lambda_{1,I}\left|\left(l_{13}-{\rm i}l_{14}\right)\chi\right|$.

\hspace{0.65cm}{\em Soliton} $S^{3+}$ $\left( {\rm i}(\eta_3-\eta_3^*)\sim\mathcal{O}(1), \theta\to -\infty \right)$
\begin{eqnarray}
\left(\begin{array}{c}
q_1\\
q_2
\end{array}\right)\to\left(\begin{array}{c}
q_1^{(3)+}\\
q_2^{(3)+}
\end{array}\right)=\lambda_{3,I}e^{-2{\rm i}\eta_{3,R}}\frac{(\lambda_1^*-\lambda_3^*)(\lambda_1^*-\lambda_3)l_{33}^{*}}{(\lambda_1-\lambda_3)(\lambda_1-\lambda_3^*)\left|l_{33}\right|}\left(\begin{array}{c}
{\rm sech}\left(2\eta_{3,I}-\hat{a}_3^{+}\right) \\
\frac{l_{33}}{l_{34}}{\rm sech}\left(2\eta_{3,I}-\hat{a}_3^{+}\right)
\end{array}\right),\label{BS3p}
\end{eqnarray}
with $e^{\hat{a}_3^{+}}=2\frac{|\lambda_1-\lambda_3|^2\left|l_{33}\right|}{\left|\lambda_1^*-\lambda_3\right|^2}$.

\vspace{5mm}\noindent{\em ~{\bf C}.~Case 3: $l_{13}^2+l_{14}^2=0$ and $l_{33}^2+l_{34}^2\neq0$}\\

The mixed solitons admit the following asymptotic behaviors:

(1) {\em Before the interaction} ($t\to-\infty$)

\hspace{0.65cm}{\em Soliton} $S^{1-}$ $\left(\theta\to+\infty, te^{-\theta}\sim\mathcal{O}(1), {\rm i}(\eta_3-\eta_3^*)\to-\infty \right)$
\begin{eqnarray}
\left(\begin{array}{c}
q_1\\
q_2
\end{array}\right)\to\left(\begin{array}{c}
q_1^{(1)-}\\
q_2^{(1)-}
\end{array}\right)={\rm i}\lambda_{1,I}e^{-2{\rm i}\eta_{1,R}}\frac{(\lambda_1-\lambda_3)(\lambda_1^*-\lambda_3)l_{13}^{*}\chi^*}{\left|(\lambda_1-\lambda_3)(\lambda_1^*-\lambda_3)l_{13}\chi\right|}\left(\begin{array}{c}
{\rm sech}\left(\xi+\hat{a}_1^-\right) \\
\frac{l_{13}}{l_{14}}{\rm sech}\left(\xi+\hat{a}_1^-\right)
\end{array}\right),\label{CS1n}
\end{eqnarray}
where $e^{\hat{a}_1^{-}}=4\lambda_{1,I}\left|\frac{l_{13}\chi(\lambda_1^*-\lambda_3)}{\lambda_1-\lambda_3}\right|$.

\hspace{0.65cm}{\em Line soliton} $S^{{\rm line}-}$ $\left(\theta \sim\mathcal{O}(1), {\rm i}(\eta_3-\eta_3^*)\to-\infty, m_{11}-\frac{l_{13}}{l_{14}} m_{12}\neq0 \right)$

\begin{eqnarray}
&&\hspace{-1.7cm}\left(\begin{array}{c}
q_1\\
q_2
\end{array}\right)\to\left(\begin{array}{c}
q_1^{{\rm line}-}\\
q_2^{{\rm line}-}
\end{array}\right)\nonumber\\
&&\hspace{-0.3cm}={\rm i}\lambda_{1,I}e^{-2{\rm i}\eta_{1,R}}\frac{(\lambda_1-\lambda_3)(\lambda_1^*-\lambda_3)\left(m_{11}-\frac{l_{13}}{l_{14}}m_{12}\right)}{\left|(\lambda_1-\lambda_3)(\lambda_1^*-\lambda_3)\left(m_{11}-\frac{l_{13}}{l_{14}}m_{12}\right)\right|}\left(\begin{array}{c}
{\rm sech}\left(\theta-\frac{\hat{\rho}^-}{2}\right) \\
-\frac{l_{13}}{l_{14}}{\rm sech}\left(\theta-\frac{\hat{\rho}^-}{2}\right)
\end{array}\right),\label{CSlinen}
\end{eqnarray}
where $e^{\frac{\hat{\rho}^-}{2}}=2\lambda_{1,I}\left|\frac{\left(m_{11}-\frac{l_{13}}{l_{14}}m_{12}\right)\left(\lambda_1^*-\lambda_3\right)}{\lambda_1-\lambda_3}\right|$.

\hspace{0.65cm}{\em Soliton} $S^{2-}$ $\left(\theta\to -\infty, te^{\theta}\sim\mathcal{O}(1), {\rm i}(\eta_3-\eta_3^*)\to-\infty \right)$
\begin{eqnarray}
\left(\begin{array}{c}
q_1\\
q_2
\end{array}\right)\to\left(\begin{array}{c}
q_1^{(2)-}\\
q_2^{(2)-}
\end{array}\right)={\rm i}\lambda_{1,I}e^{-2{\rm i}\eta_{1,R}}\frac{(\lambda_1-\lambda_3)(\lambda_1^*-\lambda_3)l_{13}^{*}\chi}{\left|(\lambda_1-\lambda_3)(\lambda_1^*-\lambda_3)l_{13}\chi\right|}\left(\begin{array}{c}
{\rm sech}\left(\zeta+\hat{a}_2^{-}\right) \\
\frac{l_{13}}{l_{14}}{\rm sech}\left(\zeta+\hat{a}_2^{-}\right)
\end{array}\right),\label{CS2n}
\end{eqnarray}
with  $e^{\hat{a}_2^{-}}=\lambda_{1,I}\left|\frac{(\lambda_1-\lambda_3)\chi}{l_{13}(\lambda_1^*-\lambda_3)}\right|$.

\hspace{0.65cm}{\em Soliton} $S^{3-}$ $\left( {\rm i}(\eta_3-\eta_3^*)\sim\mathcal{O}(1), \theta\to +\infty \right)$
\begin{eqnarray}
\left(\begin{array}{c}
q_1\\
q_2
\end{array}\right)\to\left(\begin{array}{c}
q_1^{(3)-}\\
q_2^{(3)-}
\end{array}\right)=\frac{1}{2}\left[\left(\begin{array}{c}
\phi_1^{(3)-} \\
-{\rm i} \phi_1^{(3)-}
\end{array}\right)+\left(\begin{array}{c}
\phi_2^{(3)-} \\
{\rm i}\phi_2^{(3)-}
\end{array}\right)\right].\label{CS3n}
\end{eqnarray}

(2) {\em After the interaction} ($t\to  +\infty$)

\hspace{0.65cm}{\em Soliton} $S^{1+}$ $\left(\theta\to -\infty, te^{\theta}\sim\mathcal{O}(1), {\rm i}(\eta_3-\eta_3^*)\to  +\infty \right)$
\begin{eqnarray}
\left(\begin{array}{c}
q_1\\
q_2
\end{array}\right)\to\left(\begin{array}{c}
q_1^{(2)+}\\
q_2^{(2)+}
\end{array}\right)={\rm i}\lambda_{1,I}e^{-2{\rm i}\eta_{1,R}}\frac{(\lambda_1^*-\lambda_3^*)(\lambda_1-\lambda_3^*)l_{13}^{*}\chi}{\left|(\lambda_1-\lambda_3)(\lambda_1^*-\lambda_3)l_{13}\chi\right|}\left(\begin{array}{c}
{\rm sech}\left(\zeta+\hat{a}_2^{+}\right) \\
\frac{l_{13}}{l_{14}}{\rm sech}\left(\zeta+\hat{a}_2^{+}\right)
\end{array}\right),\label{CS1p}
\end{eqnarray}
where $e^{\hat{a}_2^{+}}=\lambda_{1,I}\left|\frac{(\lambda_1^*-\lambda_3)\chi}{l_{13}(\lambda_1-\lambda_3)}\right|$.

\hspace{0.65cm}{\em Line soliton} $S^{{\rm line}+}$ $\left(\theta \sim\mathcal{O}(1), {\rm i}(\eta_3-\eta_3^*)\to+\infty, m_{11}-\frac{l_{13}}{l_{14}} m_{12}\neq0 \right)$

\begin{eqnarray}
&&\hspace{-1.7cm}\left(\begin{array}{c}
q_1\\
q_2
\end{array}\right)\to\left(\begin{array}{c}
q_1^{{\rm line}+}\\
q_2^{{\rm line}+}
\end{array}\right)\nonumber\\
&&\hspace{-0.3cm}={\rm i}\lambda_{1,I}e^{-2{\rm i}\eta_{1,R}}\frac{(\lambda_1^*-\lambda_3^*)(\lambda_1-\lambda_3^*)\left(m_{11}-\frac{l_{13}}{l_{14}}m_{12}\right)}{\left|(\lambda_1-\lambda_3)(\lambda_1^*-\lambda_3)\left(m_{11}-\frac{l_{13}}{l_{14}}m_{12}\right)\right|}\left(\begin{array}{c}
{\rm sech}\left(\theta-\frac{\hat{\rho}^+}{2}\right) \\
-\frac{l_{13}}{l_{14}}{\rm sech}\left(\theta-\frac{\hat{\rho}^+}{2}\right)
\end{array}\right),\label{CSlinep}
\end{eqnarray}
where $e^{\frac{\hat{\rho}^+}{2}}=2\lambda_{1,I}\left|\frac{\left(m_{11}-\frac{l_{13}}{l_{14}}m_{12}\right)\left(\lambda_1-\lambda_3\right)}{\lambda_1^*-\lambda_3}\right|$.

\hspace{0.65cm}{\em Soliton} $S^{2+}$ $\left(\theta\to+\infty, te^{-\theta}\sim\mathcal{O}(1), {\rm i}(\eta_3-\eta_3^*)\to +\infty \right)$
\begin{eqnarray}
\left(\begin{array}{c}
q_1\\
q_2
\end{array}\right)\to\left(\begin{array}{c}
q_1^{(1)+}\\
q_2^{(1)+}
\end{array}\right)={\rm i}\lambda_{1,I}e^{-2{\rm i}\eta_{1,R}}\frac{(\lambda_1^*-\lambda_3^*)(\lambda_1-\lambda_3^*)l_{13}^{*}\chi^*}{\left|(\lambda_1-\lambda_3)(\lambda_1^*-\lambda_3)l_{13}\chi\right|}\left(\begin{array}{c}
{\rm sech}\left(\xi+\hat{a}_1^+\right)\\
\frac{l_{13}}{l_{14}}{\rm sech}\left(\xi+\hat{a}_1^+\right)
\end{array}\right),\label{CS2p}
\end{eqnarray}
with $e^{\hat{a}_1^{+}}=4\lambda_{1,I}\left|\frac{l_{13}\chi(\lambda_1-\lambda_3)}{\lambda_1^*-\lambda_3}\right|$.

\hspace{0.65cm}{\em Soliton} $S^{3+}$ $\left( {\rm i}(\eta_3-\eta_3^*)\sim\mathcal{O}(1), \theta\to -\infty \right)$
\begin{eqnarray}
\left(\begin{array}{c}
q_1\\
q_2
\end{array}\right)\to\left(\begin{array}{c}
q_1^{(3)+}\\
q_2^{(3)+}
\end{array}\right)=\frac{1}{2}\left[\left(\begin{array}{c}
\phi_{1C}^{(3)+} \\
-{\rm i} \phi_{1C}^{(3)+}
\end{array}\right)+\left(\begin{array}{c}
\phi_{2C}^{(3)+} \\
{\rm i}\phi_{2C}^{(3)+}
\end{array}\right)\right],\label{CS3p}
\end{eqnarray}
with
\begin{eqnarray*}
	&&\hspace{-1cm} \phi_{1C}^{(3)+}=2\lambda_{3,I}e^{-2{\rm i}\eta_{3,R}}\frac{\left(l_{33}-{\rm i}l_{34}\right)^{*}}{\left|l_{33}-{\rm i}l_{34}\right|}{\rm sech}\left(2\eta_{3,I}-a_{3C}^{+}\right), \quad \phi_{2C}^{(3)+}=\phi_2^{(3)+},
\end{eqnarray*}
where $e^{a_{3C}^{+}}=\left|l_{33}-{\rm i}l_{34}\right|$.

\vspace{5mm}\noindent{\em ~{\bf D}.~Case 4: $l_{13}^2+l_{14}^2=0$ and $l_{33}^2+l_{34}^2=0$}\\

In this case, the asymptotic behaviors of the mixed solitons are as follows:

(1) {\em Before the interaction} ($t\to-\infty$)

\hspace{0.65cm}{\em Soliton} $S^{1-}$ $\left(\theta\to+\infty, te^{-\theta}\sim\mathcal{O}(1), {\rm i}(\eta_3-\eta_3^*)\to-\infty \right)$
\begin{eqnarray}
\left(\begin{array}{c}
q_1\\
q_2
\end{array}\right)\to\left(\begin{array}{c}
q_1^{(1)-}\\
q_2^{(1)-}
\end{array}\right)={\rm i}\lambda_{1,I}e^{-2{\rm i}\eta_{1,R}}\frac{(\lambda_1-\lambda_3)(\lambda_1^*-\lambda_3)l_{13}^{*}\chi^*}{\left|(\lambda_1-\lambda_3)(\lambda_1^*-\lambda_3)l_{13}\chi\right|}\left(\begin{array}{c}
{\rm sech}\left(\xi+\hat{a}_1^-\right) \\
\frac{l_{13}}{l_{14}}{\rm sech}\left(\xi+\hat{a}_1^-\right)
\end{array}\right).\label{DS1n}
\end{eqnarray}

\hspace{0.65cm}{\em Line soliton} $S^{{\rm line}-}$ $\left(\theta \sim\mathcal{O}(1), {\rm i}(\eta_3-\eta_3^*)\to-\infty, m_{11}-\frac{l_{13}}{l_{14}} m_{12}\neq0 \right)$

\begin{eqnarray}
&&\hspace{-1.7cm}\left(\begin{array}{c}
q_1\\
q_2
\end{array}\right)\to\left(\begin{array}{c}
q_1^{{\rm line}-}\\
q_2^{{\rm line}-}
\end{array}\right)\nonumber\\
&&\hspace{-0.3cm}={\rm i}\lambda_{1,I}e^{-2{\rm i}\eta_{1,R}}\frac{(\lambda_1-\lambda_3)(\lambda_1^*-\lambda_3)\left(m_{11}-\frac{l_{13}}{l_{14}}m_{12}\right)}{\left|(\lambda_1-\lambda_3)(\lambda_1^*-\lambda_3)\left(m_{11}-\frac{l_{13}}{l_{14}}m_{12}\right)\right|}\left(\begin{array}{c}
{\rm sech}\left(\theta-\frac{\hat{\rho}^-}{2}\right) \\
-\frac{l_{13}}{l_{14}}{\rm sech}\left(\theta-\frac{\hat{\rho}^-}{2}\right)
\end{array}\right).\label{DSlinen}
\end{eqnarray}

\hspace{0.65cm}{\em Soliton} $S^{2-}$ $\left(\theta\to -\infty, te^{\theta}\sim\mathcal{O}(1), {\rm i}(\eta_3-\eta_3^*)\to-\infty \right)$
\begin{eqnarray}
\left(\begin{array}{c}
q_1\\
q_2
\end{array}\right)\to\left(\begin{array}{c}
q_1^{(2)-}\\
q_2^{(2)-}
\end{array}\right)={\rm i}\lambda_{1,I}e^{-2{\rm i}\eta_{1,R}}\frac{(\lambda_1-\lambda_3)(\lambda_1^*-\lambda_3)l_{13}^{*}\chi}{\left|(\lambda_1-\lambda_3)(\lambda_1^*-\lambda_3)l_{13}\chi\right|}\left(\begin{array}{c}
{\rm sech}\left(\zeta+\hat{a}_2^{-}\right) \\
\frac{l_{13}}{l_{14}}{\rm sech}\left(\zeta+\hat{a}_2^{-}\right)
\end{array}\right).\label{DS2n}
\end{eqnarray}

\hspace{0.65cm}{\em Soliton} $S^{3-}$ $\left( {\rm i}(\eta_3-\eta_3^*)\sim\mathcal{O}(1), \theta\to +\infty \right)$
\begin{eqnarray}
\left(\begin{array}{c}
q_1\\
q_2
\end{array}\right)\to\left(\begin{array}{c}
q_1^{(3)-}\\
q_2^{(3)-}
\end{array}\right)=\lambda_{3,I}e^{-2{\rm i}\eta_{3,R}}\frac{(\lambda_1-\lambda_3)(\lambda_1-\lambda_3^*)l_{33}^{*}}{(\lambda_1^*-\lambda_3^*)(\lambda_1^*-\lambda_3)\left|l_{33}\right|}\left(\begin{array}{c}
{\rm sech}\left(2\eta_{3,I}-\hat{a}_3^{-}\right) \\
\frac{l_{33}}{l_{34}}{\rm sech}\left(2\eta_{3,I}-\hat{a}_3^{-}\right)
\end{array}\right).\label{DS3n}
\end{eqnarray}

(2) {\em After the interaction} ($t\to  +\infty$)

\hspace{0.65cm}{\em Soliton} $S^{1+}$ $\left(\theta\to -\infty, te^{\theta}\sim\mathcal{O}(1), {\rm i}(\eta_3-\eta_3^*)\to  +\infty \right)$
\begin{eqnarray}
\left(\begin{array}{c}
q_1\\
q_2
\end{array}\right)\to\left(\begin{array}{c}
q_1^{(2)+}\\
q_2^{(2)+}
\end{array}\right)={\rm i}\lambda_{1,I}e^{-2{\rm i}\eta_{1,R}}\frac{(\lambda_1^*-\lambda_3^*)(\lambda_1-\lambda_3^*)l_{13}^{*}\chi}{\left|(\lambda_1-\lambda_3)(\lambda_1^*-\lambda_3)l_{13}\chi\right|}\left(\begin{array}{c}
{\rm sech}\left(\zeta+\hat{a}_2^{+}\right) \\
\frac{l_{13}}{l_{14}}{\rm sech}\left(\zeta+\hat{a}_2^{+}\right)
\end{array}\right).\label{DS1p}
\end{eqnarray}

\hspace{0.65cm}{\em Line soliton} $S^{{\rm line}+}$ $\left(\theta \sim\mathcal{O}(1), {\rm i}(\eta_3-\eta_3^*)\to+\infty, m_{11}-\frac{l_{13}}{l_{14}} m_{12}\neq0 \right)$

\begin{eqnarray}
&&\hspace{-1.7cm}\left(\begin{array}{c}
q_1\\
q_2
\end{array}\right)\to\left(\begin{array}{c}
q_1^{{\rm line}+}\\
q_2^{{\rm line}+}
\end{array}\right)={\rm i}\lambda_{1,I}e^{-2{\rm i}\eta_{1,R}}\frac{m_{11}-\frac{l_{13}}{l_{14}}m_{12}}{\left|m_{11}-\frac{l_{13}}{l_{14}}m_{12}\right|}\left(\begin{array}{c}
{\rm sech}\left(\theta-\frac{{\rho}}{2}\right) \\
-\frac{l_{13}}{l_{14}}{\rm sech}\left(\theta-\frac{{\rho}}{2}\right)
\end{array}\right).\label{DSlinep}
\end{eqnarray}

\hspace{0.65cm}{\em Soliton} $S^{2+}$ $\left(\theta\to+\infty, te^{-\theta}\sim\mathcal{O}(1), {\rm i}(\eta_3-\eta_3^*)\to +\infty \right)$
\begin{eqnarray}
\left(\begin{array}{c}
q_1\\
q_2
\end{array}\right)\to\left(\begin{array}{c}
q_1^{(1)+}\\
q_2^{(1)+}
\end{array}\right)={\rm i}\lambda_{1,I}e^{-2{\rm i}\eta_{1,R}}\frac{(\lambda_1^*-\lambda_3^*)(\lambda_1-\lambda_3^*)l_{13}^{*}\chi^*}{\left|(\lambda_1-\lambda_3)(\lambda_1^*-\lambda_3)l_{13}\chi\right|}\left(\begin{array}{c}
{\rm sech}\left(\xi+\hat{a}_1^+\right)\\
\frac{l_{13}}{l_{14}}{\rm sech}\left(\xi+\hat{a}_1^+\right)
\end{array}\right).\label{DS2p}
\end{eqnarray}

\hspace{0.65cm}{\em Soliton} $S^{3+}$ $\left( {\rm i}(\eta_3-\eta_3^*)\sim\mathcal{O}(1), \theta\to -\infty \right)$
\begin{eqnarray}
\left(\begin{array}{c}
q_1\\
q_2
\end{array}\right)\to\left(\begin{array}{c}
q_1^{(3)+}\\
q_2^{(3)+}
\end{array}\right)=\lambda_{3,I}e^{-2{\rm i}\eta_{3,R}}\frac{(\lambda_1^*-\lambda_3^*)(\lambda_1^*-\lambda_3)l_{33}^{*}}{(\lambda_1-\lambda_3)(\lambda_1-\lambda_3^*)\left|l_{33}\right|}\left(\begin{array}{c}
{\rm sech}\left(2\eta_{3,I}-\hat{a}_3^{+}\right) \\
\frac{l_{33}}{l_{34}}{\rm sech}\left(2\eta_{3,I}-\hat{a}_3^{+}\right)
\end{array}\right).\label{DS3p}
\end{eqnarray}

\vspace{5mm}\noindent\textbf{~5.~Interactions of the mixed solitons}\\

By means of the asymptotic behaviors obtained in Section~4, we will discuss the interaction properties of the mixed solitons for System~(\ref{equations}) in this section.

{\bf A}. In virtue of Expressions~(\ref{AS1n})-(\ref{AS3p}), we derive the following interaction properties of the mixed solitons for the case of $l_{13}^2+l_{14}^2\neq0$ and $l_{33}^2+l_{34}^2\neq0$: (\textrm{i}) The amplitudes and shapes of solitons $S^1$, $S^2$ and $S^3$
keep invariant before and after the interaction. This reveals that the interaction among $S^1$, $S^2$ and $S^3$ is elastic. (\textrm{ii}) Position shifts of solitons $S^1$ and $S^2$ are both $-\frac{1}{\lambda_{1,I}}\ln\left|\frac{\lambda_1-\lambda_3}{\lambda_1^*-\lambda_3}\right|$, while the position shift of the soliton $S^3$ is obtained as $\frac{2}{\lambda_{3,I}}\ln\left|\frac{\lambda_1-\lambda_3}{\lambda_1^*-\lambda_3}\right|$.
(\textrm{iii}) Velocities of the soliton $S^1$ and soliton $S^2$ are derived as
\begin{subequations}
\begin{eqnarray}
v_{(1)}^-(t)=v_{(2)}^+(t)=-4\left(\lambda_{1,R}+3\lambda_{1,R}^2\varepsilon-\lambda_{1,I}^2\varepsilon\right)-\frac{1}{2\lambda_{1,I}t},\\
v_{(1)}^+(t)=v_{(2)}^-(t)=-4\left(\lambda_{1,R}+3\lambda_{1,R}^2\varepsilon-\lambda_{1,I}^2\varepsilon\right)+\frac{1}{2\lambda_{1,I}t},
\end{eqnarray}\label{velocitiess1s2}
\end{subequations}
\hspace{-0.1cm}while the soliton $S^3$ has an invariant velocity of $v_{3}=-4\left(\lambda_{3,R}+3\lambda_{3,R}^2\varepsilon-\lambda_{3,I}^2\varepsilon\right)$ before and after the interaction.

In this case, the interactions between a degenerate soliton and a bell-shaped soliton is elastic except for a little position shifts at the interaction regions are displayed in Figs.~9.
When $l_{13}=l_{33}={\rm i}$, Fig.~8(a$_1$) presents the elastic interaction between a single-hump degenerate soliton and a single-hump bell-shaped soliton in the $q_1$ component, while
Fig.~8(a$_2$) illustrates the interaction between a double-hump degenerate soliton and a double-hump bell-shaped soliton in the $q_2$ component. When $l_{13}=1$, $l_{33}={\rm i}$, Figs.~8(b$_1$-b$_2$) display the elastic interaction between a single-hump degenerate soliton and a double-hump bell-shaped soliton in both the $q_1$ and $q_2$ components. When $l_{13}={\rm i}$, $l_{33}=1$, Figs.~8(c$_1$-c$_2$) depict the elastic interaction between a single-hump degenerate soliton and a single-hump bell-shaped soliton in the $q_1$
component, and between a double-hump degenerate soliton and a single-hump bell-shaped soliton in the $q_2$ component.
\begin{center} \vspace{0cm}
	\hspace{0.0cm}{\includegraphics[scale =0.34]{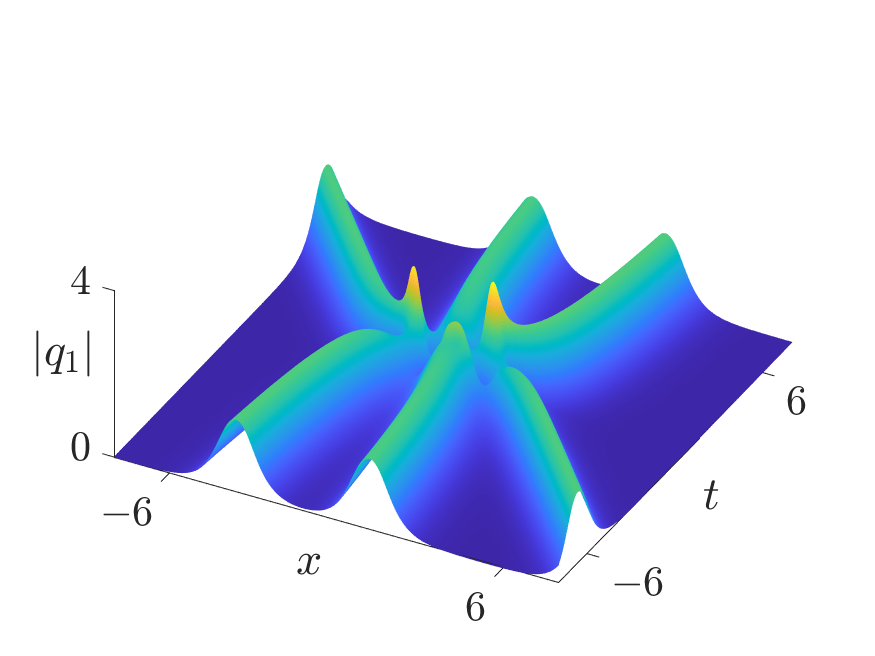}}
	\hspace{0.2cm}{\includegraphics[scale =0.34]{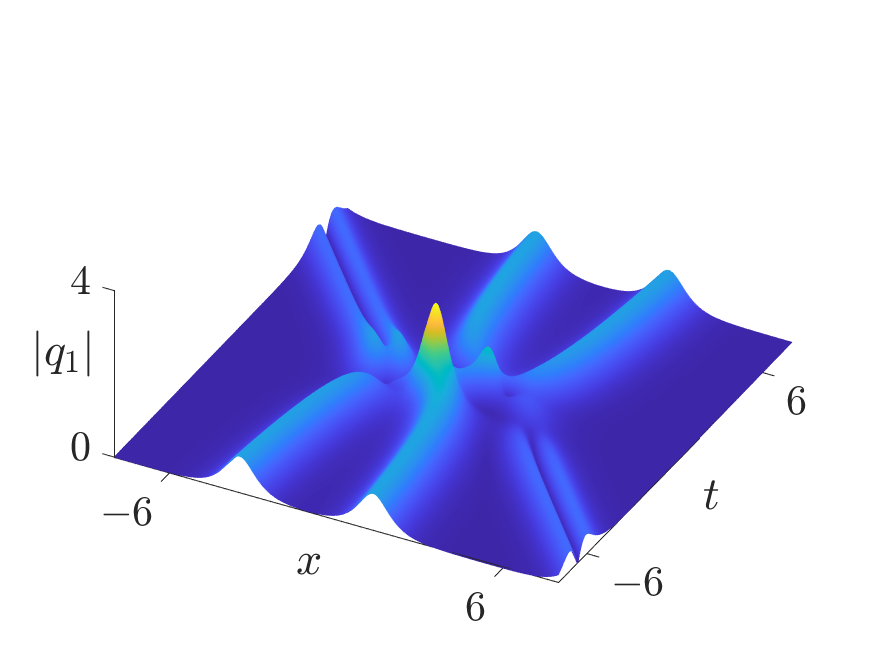}}
	\hspace{0.2cm}{\includegraphics[scale =0.34]{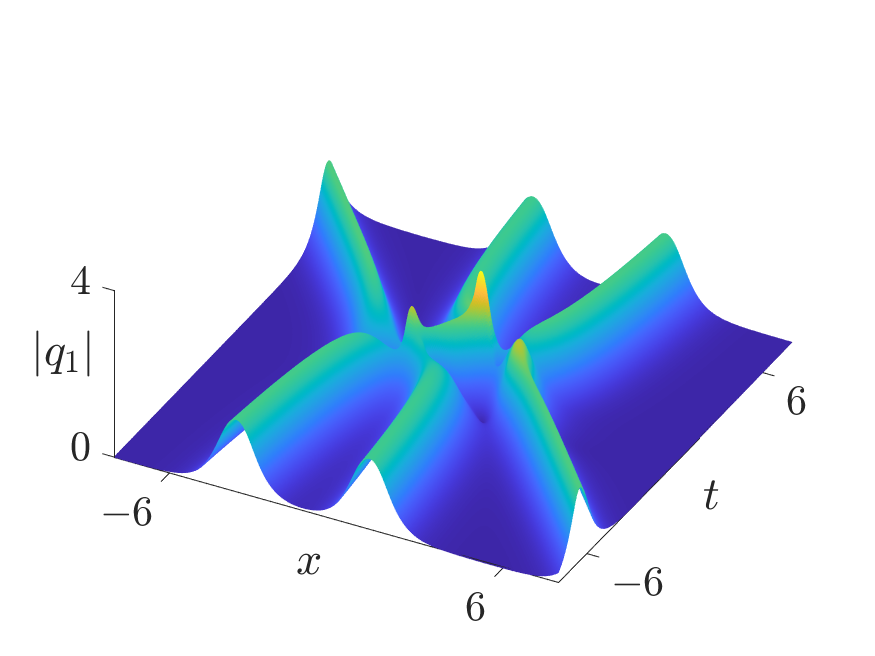}}\\
	\vspace{-0.2cm}{\footnotesize\hspace{-0.5cm}(a$_1$) \hspace{4.2cm}(b$_1$)  \hspace{4.0cm}(c$_1$)}\\
	\hspace{-1cm}\\
	\hspace{0.0cm}{\includegraphics[scale =0.34]{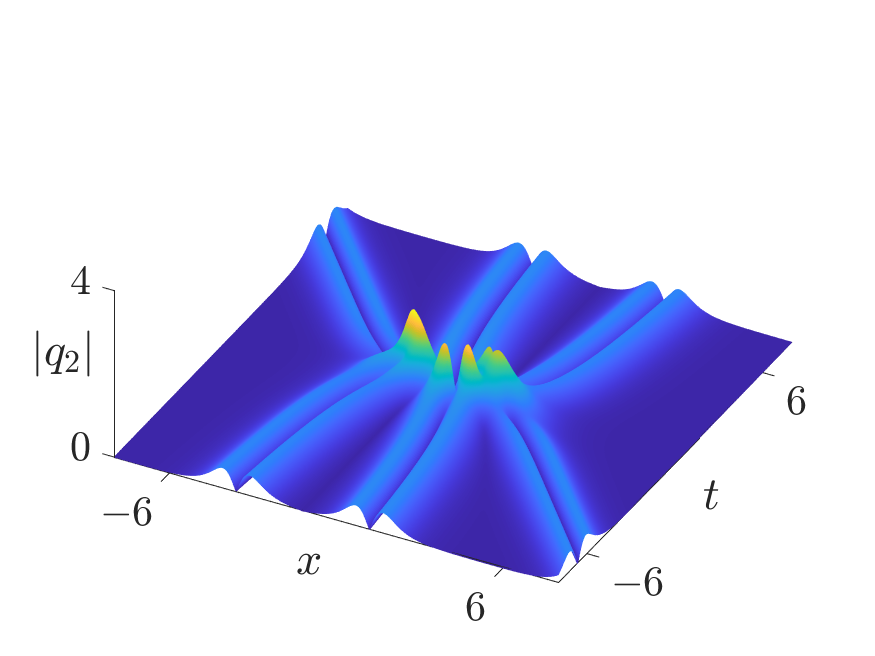}}
	\hspace{0.2cm}{\includegraphics[scale =0.34]{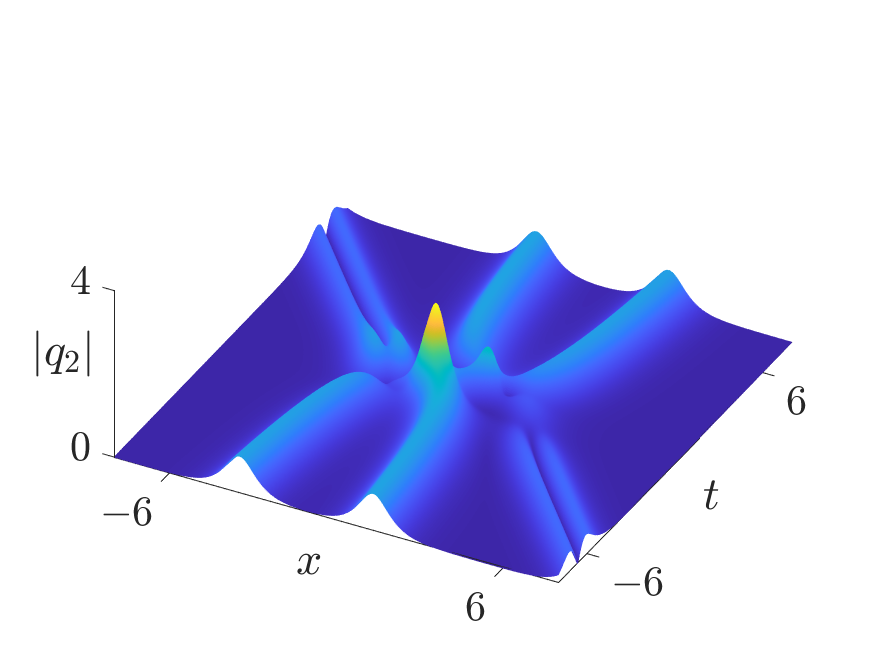}}
	\hspace{0.2cm}{\includegraphics[scale =0.34]{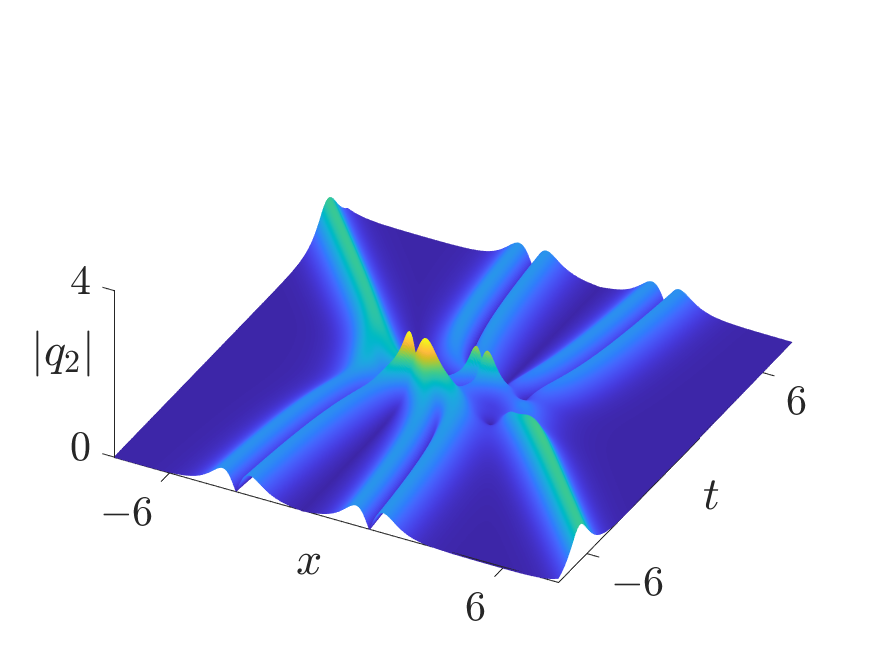}}\\
	\vspace{-0.2cm}{\footnotesize\hspace{-0.5cm}(a$_2$) \hspace{4.2cm}(b$_2$)  \hspace{4.0cm}(c$_2$)}\\
	\hspace{-1cm}\\
	\vspace{-0.3cm}\flushleft{\footnotesize{\bf Figs.}~8\,
		Elastic interactions between a degenerate soliton and a bell-shaped soliton via Solutions~(\ref{mixedsoliton}) with $m_{11}=-1+{\rm i}$, $m_{12}=1$, $\lambda_{1,R}=\frac{-1+\sqrt{1+12\lambda_{1,I}^2\varepsilon^2}}{6\varepsilon}$, $\lambda_{1,I}=1$, $\lambda_3=\frac{1}{4}+{\rm i}$, $\varepsilon=\frac{1}{25}$,  $l_{14}=l_{34}=\frac{1}{2}$, (a$_1$-a$_2$) $l_{13}=l_{33}={\rm i}$; (b$_1$-b$_2$) $l_{13}=1$, $l_{33}={\rm i}$; (c$_1$-c$_2$) $l_{13}={\rm i}$, $l_{33}=1$.}
\end{center}

{\bf B}. By means of Expressions~(\ref{BS1n})-(\ref{BS3p}), we find that the mixed solitons for the case of $l_{13}^2+l_{14}^2\neq0$ and $l_{33}^2+l_{34}^2=0$ admit the following interaction properties: (\textrm{i}) Both solitons $S^1$ and $S^2$ experience changes in amplitudes and shapes in both $q_1$ and $q_2$ components before and after the interaction, while the soliton $S^3$ maintains its amplitude and shape with only minor position shifts at the interaction regions during the propagation. This reflects that the intensities of $S^1$ and $S^2$ are redistributed, while the intensity distribution of $S^3$ remains unchanged before and after the interaction. (\textrm{ii}) The solitons $S^1$ and $S^2$ share the same velocity given by Expressions~(\ref{velocitiess1s2}).
The soliton $S^3$ keeps the constant velocity of $v_{3}=-4\left(\lambda_{3,R}+3\lambda_{3,R}^2\varepsilon-\lambda_{3,I}^2\varepsilon\right)$ before and after the interaction.

Therefore, in such case, the interaction is inelastic for the degenerate soliton composed of $S^1$ and $S^2$ in both $q_1$ and $q_2$ components, while it is elastic for the bell-shaped soliton $S^3$, as displayed in Figs.~9. In Figs.~9(a$_1$-a$_2$), the degenerate soliton transforms from a lower double-hump to a slightly higher single-hump profile in both $q_1$ and $q_2$ components. Figs.~9(b$_1$-b$_2$) show that in the $q_1$ component, the degenerate soliton changes from a lower double-hump to a notably higher single-hump profile, while in the $q_2$ component, it shifts from a double-hump to an almost invisible single-hump profile. This indicates a significant increase in wave energy in the $q_1$ component and a notable decrease in the $q_2$
component. This intriguing property have potential applications in optical switching.

\begin{center} \vspace{0cm}
	\hspace{-1cm}\\
	\hspace{0.0cm}{\includegraphics[scale =0.36]{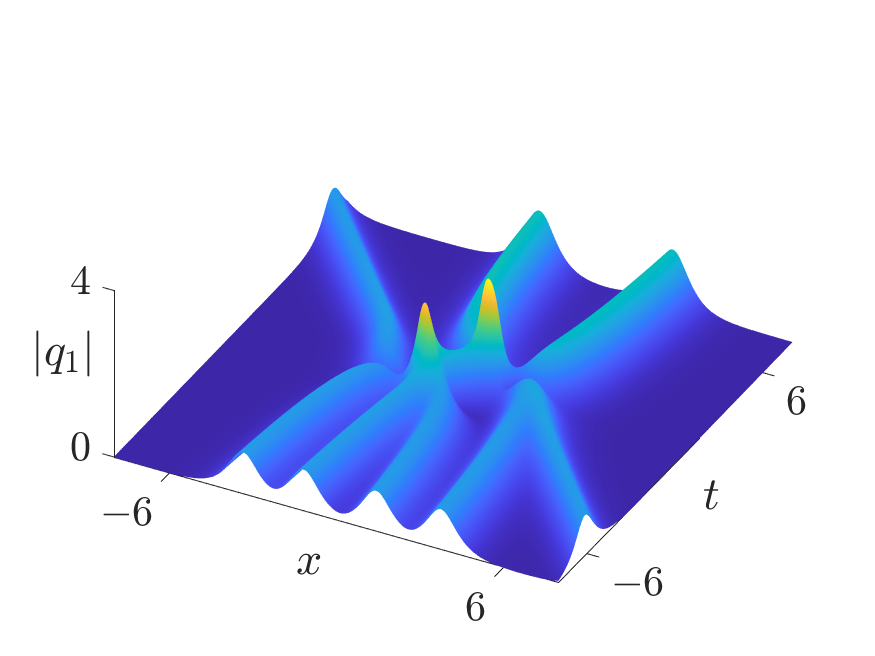}}
\hspace{0.5cm}{\includegraphics[scale =0.36]{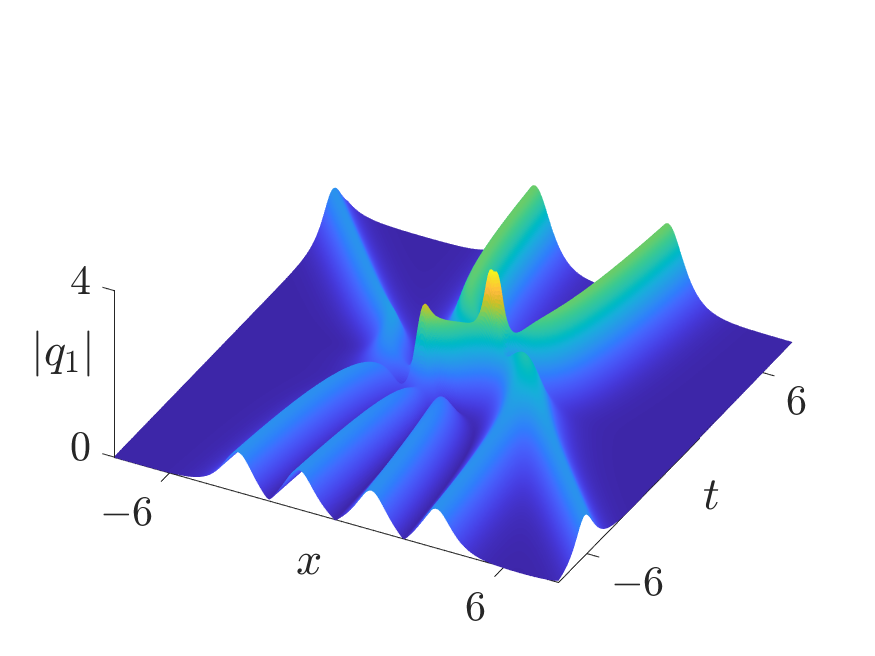}}\\
\vspace{-0.2cm}{\footnotesize\hspace{-0.2cm}(a$_1$)\hspace{5.3cm}(b$_1$)}\\
\hspace{-1cm}\\
\hspace{0.0cm}{\includegraphics[scale =0.36]{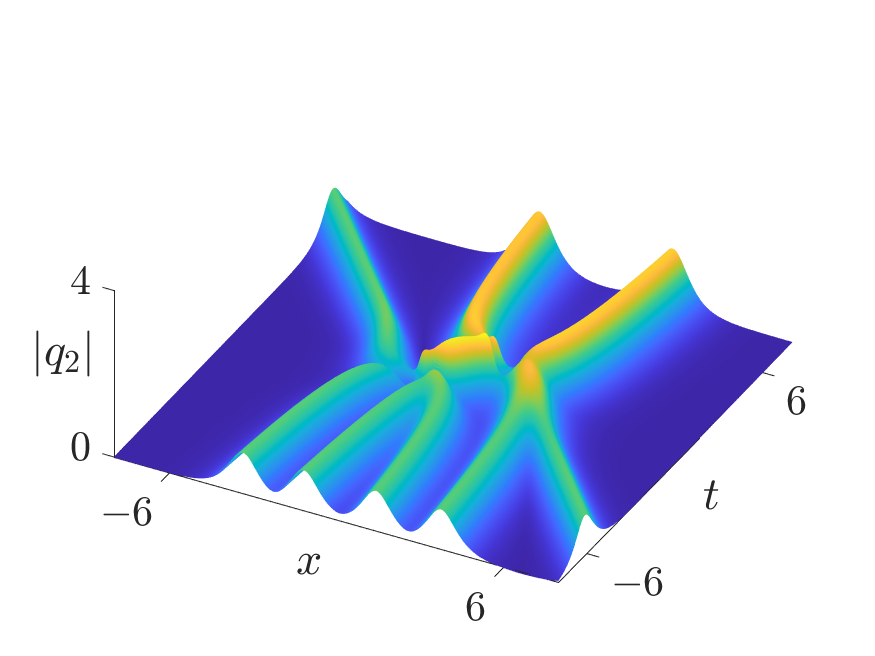}}
\hspace{0.5cm}{\includegraphics[scale =0.36]{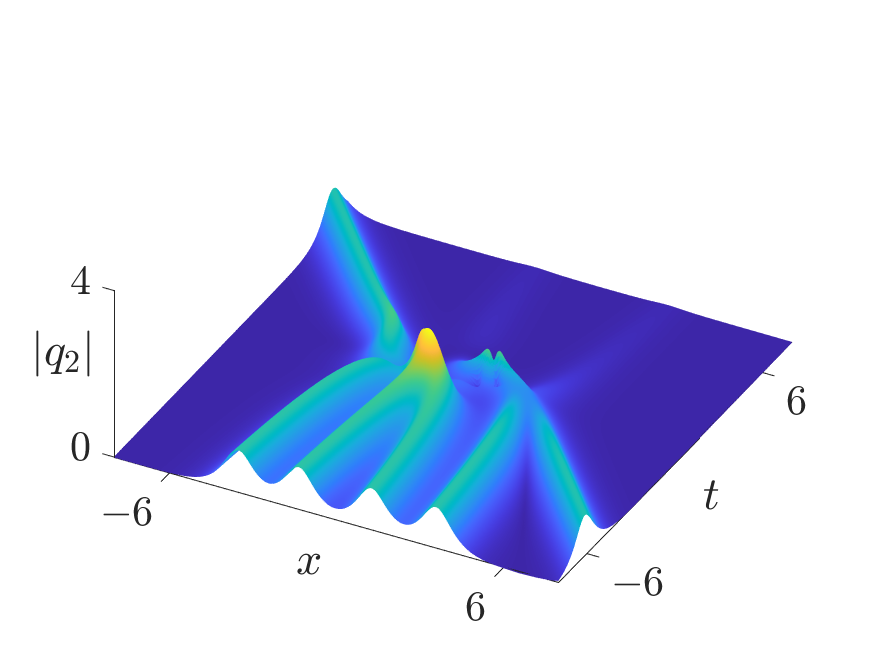}}\\
\vspace{-0.2cm}{\footnotesize\hspace{-0.2cm}(a$_2$)\hspace{5.3cm}(b$_2$)}\\
\hspace{-1cm}\\
	\vspace{-0.3cm}\flushleft{\footnotesize{\bf Figs.}~9\,
	Inelastic interactions between a degenerate soliton and a bell-shaped soliton via Solutions~(\ref{mixedsoliton}) with the same parameters as those in Figs.~9(b$_1$-b$_2$) except for $m_{11}=0$, $l_{34}=1$, (a$_1$-a$_2$) $l_{14}=-1$; (b$_1$-b$_2$) $l_{14}=0$.}
\end{center}

{\bf C}. Based on Expressions~(\ref{CS1n})-(\ref{CS3p}), we obtain the following results for the case of $l_{13}^2+l_{14}^2=0$ and $l_{33}^2+l_{34}^2\neq0$ ($m_{11}-\frac{l_{13}}{l_{14}} m_{12}\neq0$): (\textrm{i}) The degenerate soliton consists of three branches: solitons $S^1$, $S^2$ and $S^{\rm line}$ in both $q_1$ and $q_2$ components. The amplitudes and shapes of solitons $S^1$, $S^2$ and $S^{\rm line}$ remain unchanged, except for minor position shifts during the interaction. In contrast, the soliton $S^3$ experiences changes in both amplitude and shape in the $q_1$ and $q_2$ components before and after the interaction. This indicates that the intensities of solitons $S^1$, $S^2$ and $S^{\rm line}$ remain constant, while the soliton $S^3$ undergoes a redistribution of its intensities in the internal states. (\textrm{ii}) Position shifts for solitons $S^1$, $S^2$ and $S^{\rm line}$ are given by $-\frac{1}{\lambda_{1,I}}\ln\left|\frac{\lambda_1-\lambda_3}{\lambda_1^*-\lambda_3}\right|$, while the position shift for the soliton $S^3$ is $\frac{2}{\lambda_{3,I}}\ln\left|\frac{\lambda_1-\lambda_3}{\lambda_1^*-\lambda_3}\right|$. The soliton $S^{\rm line}$ maintains a constant velocity of $v_{\rm line}=-4\left(\lambda_{1,R}+3\lambda_{1,R}^2\varepsilon-\lambda_{1,I}^2\varepsilon\right)$ during the interaction.

Thus, in the case of $l_{13}^2+l_{14}^2=0$ and $l_{33}^2+l_{34}^2\neq0$ $\left(m_{11}-\frac{l_{13}}{l_{14}} m_{12}\neq0\right)$, the interaction is elastic for the degenerate soliton consisting of $S^1$, $S^2$ and $S^{\rm line}$ in both $q_1$ and $q_2$ components, but inelastic for the bell-shaped soliton $S^3$, as illustrated in Figs.~10(a$_1$-a$_2$) and 10(b$_1$-b$_2$).  Figs.~10(a$_1$-a$_2$) display that the bell-shaped soliton $S^3$  transitions from a slightly higher single-hump profile to a lower double-hump profile in both $q_1$ and $q_2$ components. Figs.~10(b$_1$-b$_2$) reveal that in the $q_1$ component, the soliton changes from a notably higher single-hump to a lower double-hump profile, while in the $q_2$ component, it shifts from an invisible hump to a double-hump profile, indicating a wave energy exchange between the $q_1$ and $q_2$ components. Comparing Figs.~10(a$_1$-a$_2$) with Figs.~10(c$_1$-c$_2$), we observe that the line soliton in the degenerate soliton $S^{\rm line}$ disappears in both $q_1$ and $q_2$ components when $m_{11}-\frac{l_{13}}{l_{14}} m_{12}=0$.

\begin{center} \vspace{0cm}
	\hspace{0.0cm}{\includegraphics[scale =0.34]{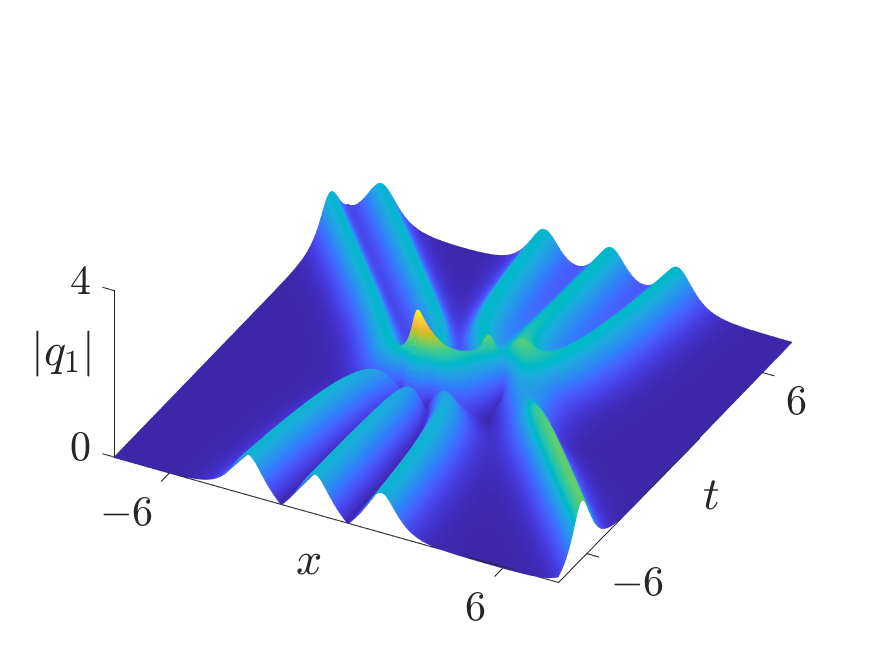}}
	\hspace{0.2cm}{\includegraphics[scale =0.34]{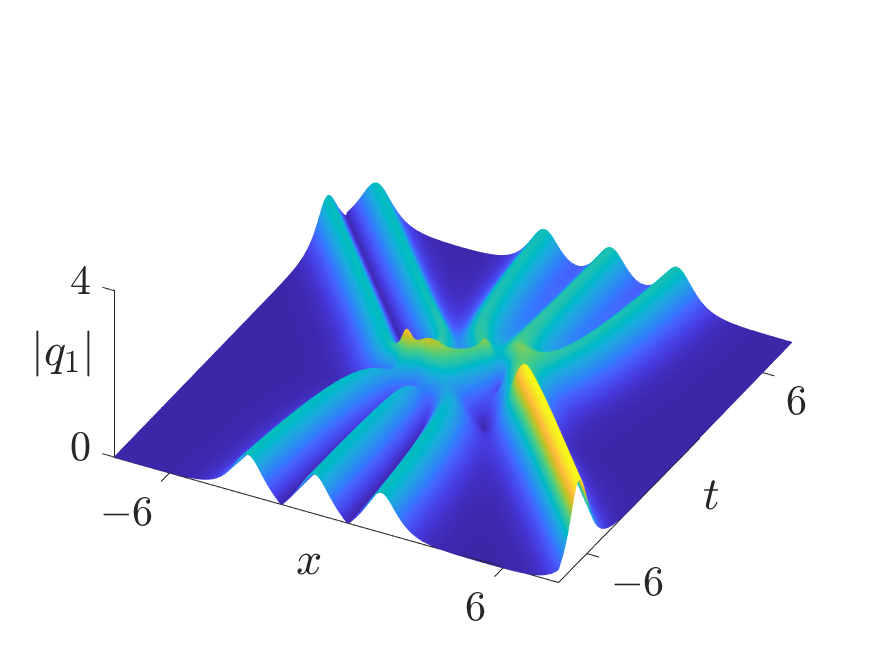}}
	\hspace{0.2cm}{\includegraphics[scale =0.34]{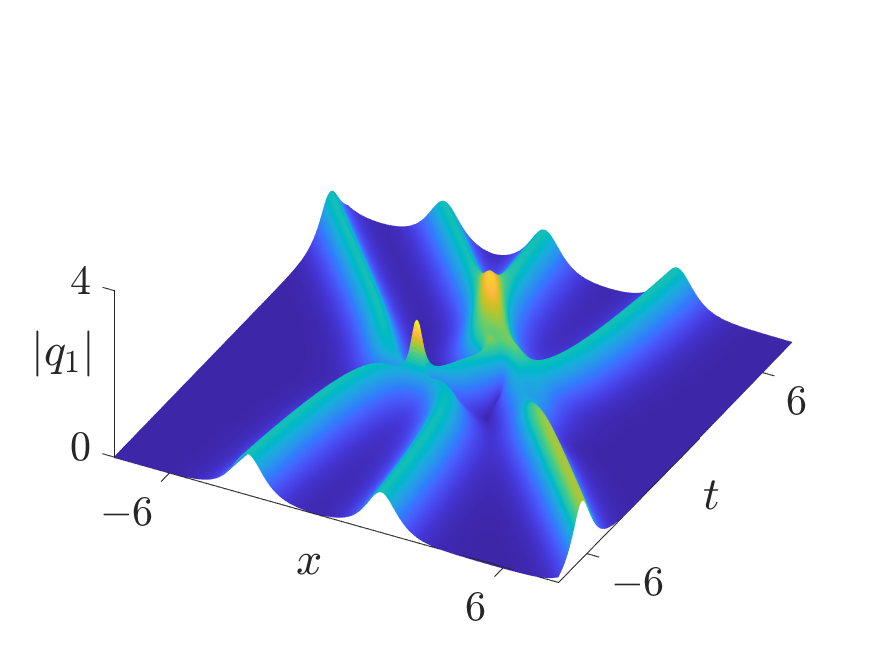}}\\
	\vspace{-0.2cm}{\footnotesize\hspace{-0.5cm}(a$_1$) \hspace{4.2cm}(b$_1$)  \hspace{4.0cm}(c$_1$)}\\
	\hspace{-1cm}\\
	\hspace{0.0cm}{\includegraphics[scale =0.34]{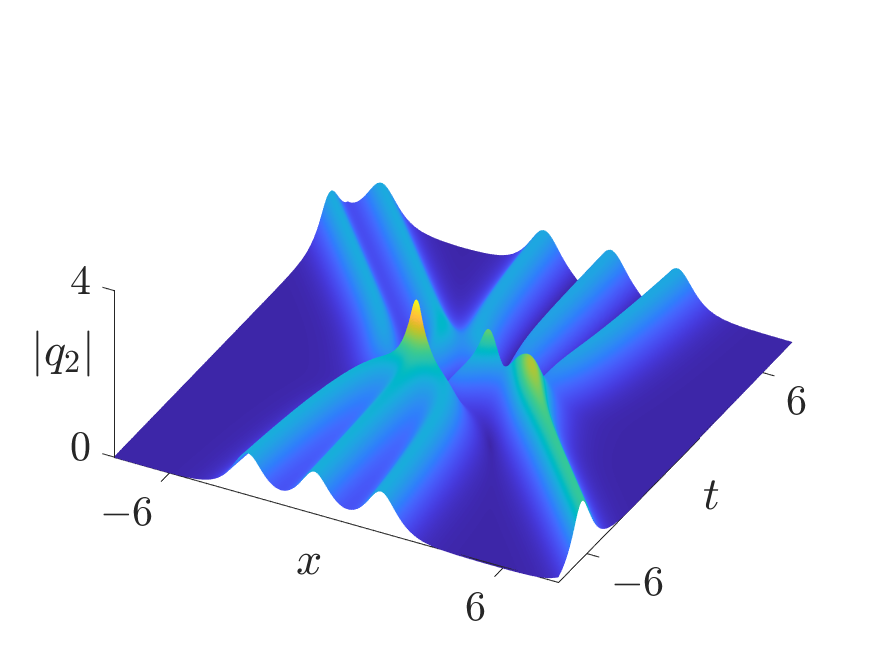}}
	\hspace{0.2cm}{\includegraphics[scale =0.34]{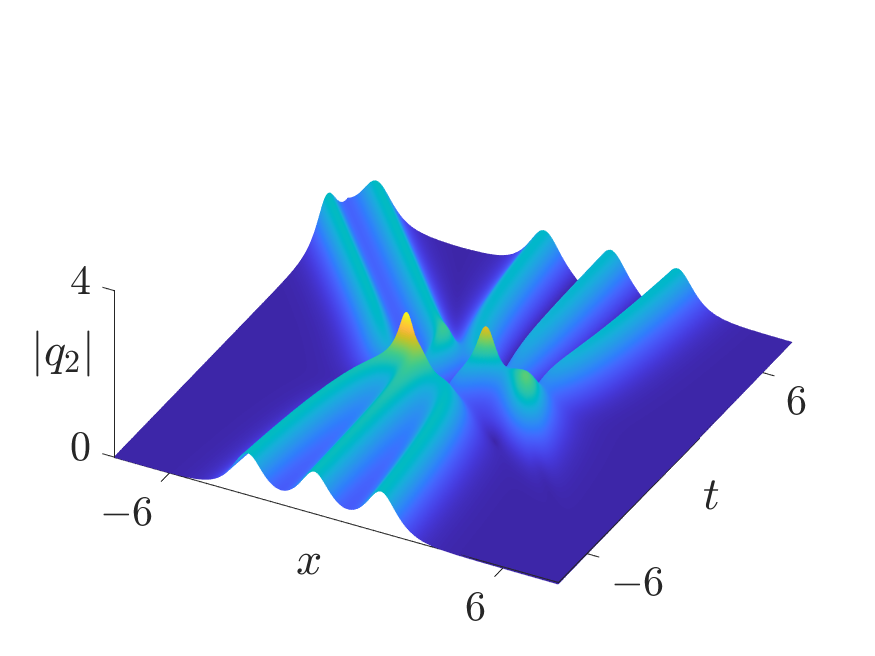}}
	\hspace{0.2cm}{\includegraphics[scale =0.34]{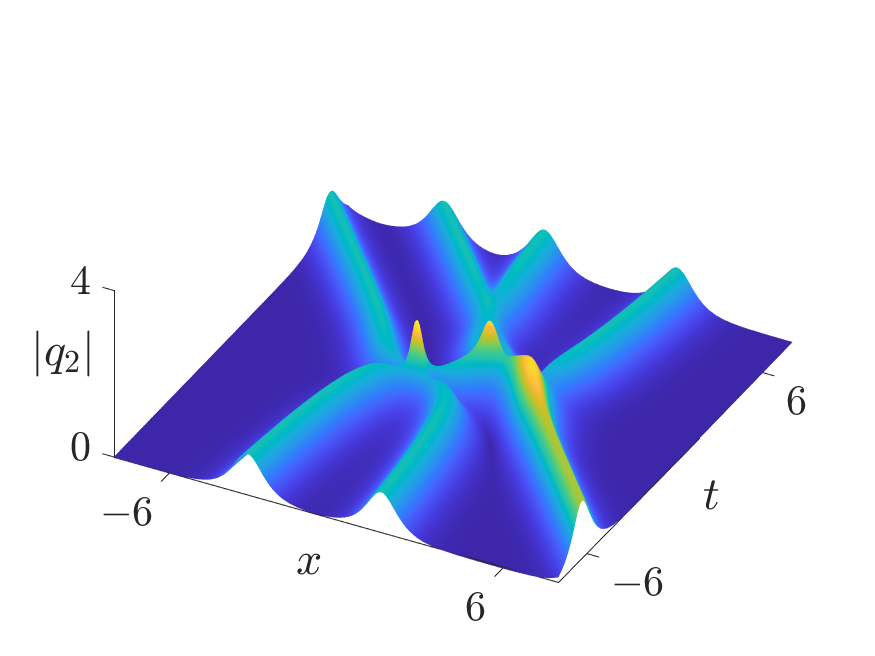}}\\
	\vspace{-0.2cm}{\footnotesize\hspace{-0.5cm}(a$_2$) \hspace{4.2cm}(b$_2$)  \hspace{4.0cm}(c$_2$)}\\
	\hspace{-1cm}\\
	\vspace{-0.3cm}\flushleft{\footnotesize{\bf Figs.}~10.\,
			Inelastic interactions between a degenerate soliton and a bell-shaped soliton via Solutions~(\ref{mixedsoliton}) with the same parameters as those in Figs.~9(c$_1$-c$_2$) except for $l_{14}=1$, (a$_1$-a$_2$) $m_{11}=0$, $l_{34}=-1$; (b$_1$-b$_2$) $m_{11}=0$, $l_{34}=0$; (c$_1$-c$_2$) $m_{11}={\rm i}$, $l_{34}=-1$.}
\end{center}

{\bf D}. According to Expressions~(\ref{DS1n})-(\ref{DS3p}), for the case of $l_{13}^2+l_{14}^2\neq0$ and $l_{33}^2+l_{34}^2\neq0$ $\left(m_{11}-\frac{l_{13}}{l_{14}}m_{12}\neq0\right)$, the mixed solitons exhibit the following interaction properties: The degenerate soliton also comprises three branches: solitons $S^1$, $S^2$ and $S^{\rm line}$ in both $q_1$ and $q_2$ components. The maximum asymptotic amplitudes of solitons $S^1$, $S^2$ and $S^{\rm line}$ before and after the interaction are the same as $|\lambda_{1,I}|$. The soliton $S^3$ maintains an unchanged amplitude of $|\lambda_{3,I}|$ during the interaction.

In this context, it is noted that the intensity redistribution between soliton components cannot be conclusively determined from asymptotic amplitudes alone. Degenerate soliton branches may exhibit some coherent interaction phenomena during a longer interaction region, as illustrated in Figs.~11(a$_1$-a$_2$) and 11(b$_1$-b$_2$). With $m_{11}-\frac{l_{13}}{l_{14}}m_{12}\neq0$, Figs.~11(a$_1$-a$_2$) and 11(b$_1$-b$_2$) reveal that degenerate soliton branches $S^1$ and $S^{\rm line}$ experience coherent interaction before interacting with $S^3$ in the $q_1$ and $q_2$ components. However, when $m_{11}-\frac{l_{13}}{l_{14}}m_{12}=0$, the line soliton in the degenerate soliton $S^{\rm line}$ disappears, resulting in an elastic interaction without coherent effects in both $q_1$ and $q_2$ components, as depicted in Figs.~11(c$_1$-c$_2$).

\begin{center} \vspace{1cm}
	\hspace{0.0cm}{\includegraphics[scale =0.34]{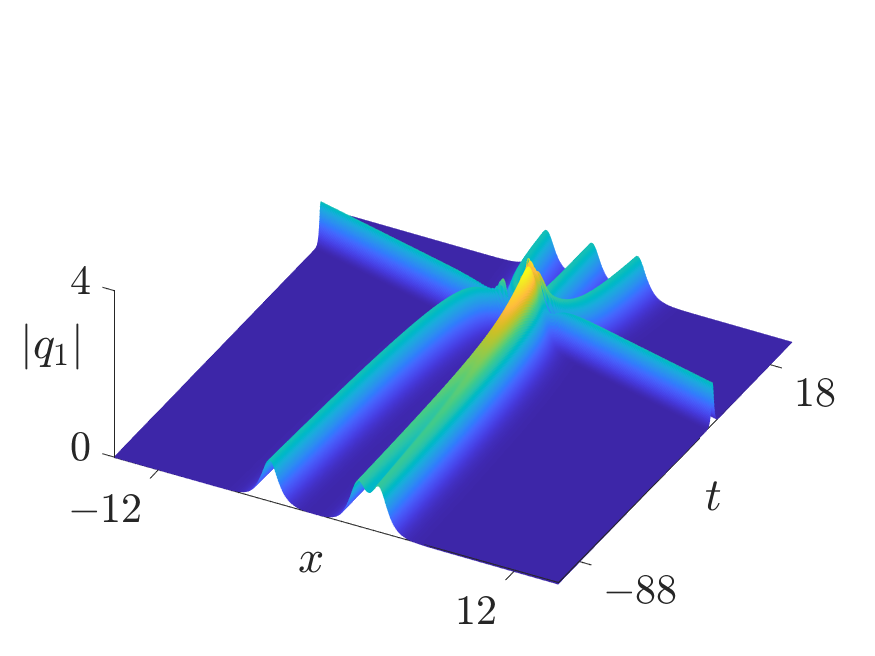}}
	\hspace{0.2cm}{\includegraphics[scale =0.32]{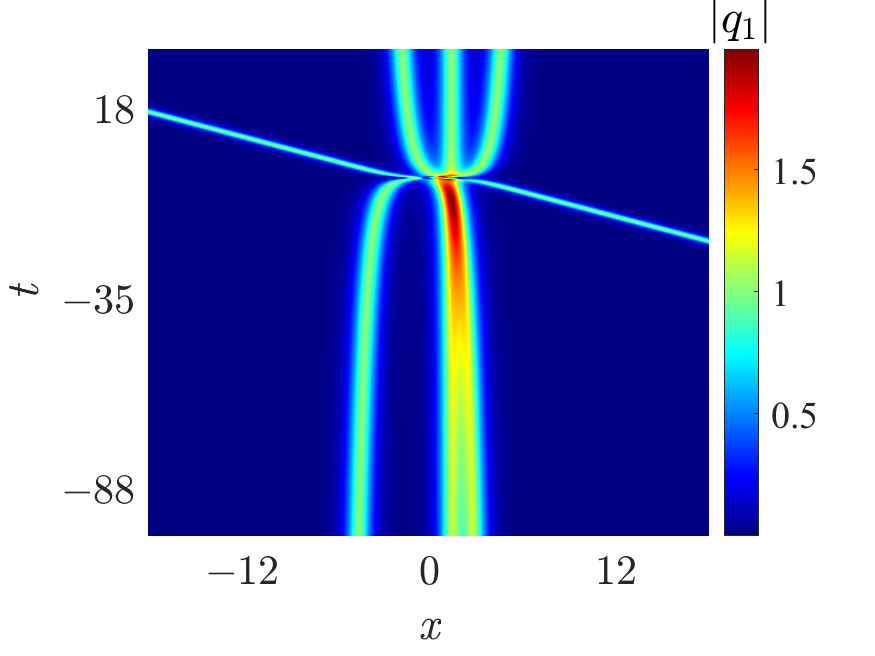}}
	\hspace{0.2cm}{\includegraphics[scale =0.34]{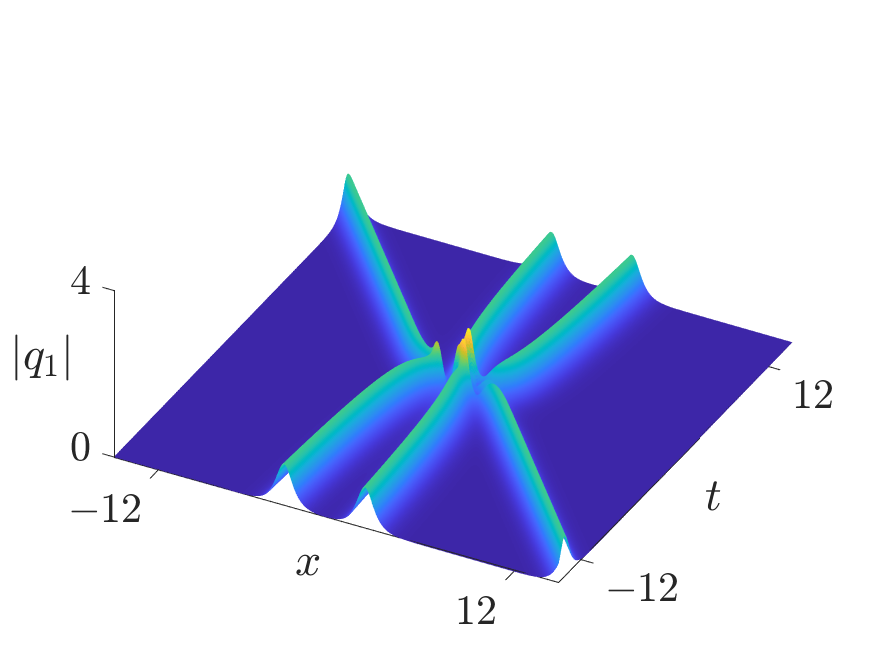}}\\
	\vspace{-0.2cm}{\footnotesize\hspace{-0.5cm}(a$_1$) \hspace{4.2cm}(b$_1$)  \hspace{4.0cm}(c$_1$)}\\
	\hspace{-1cm}\\
	\hspace{0.0cm}{\includegraphics[scale =0.34]{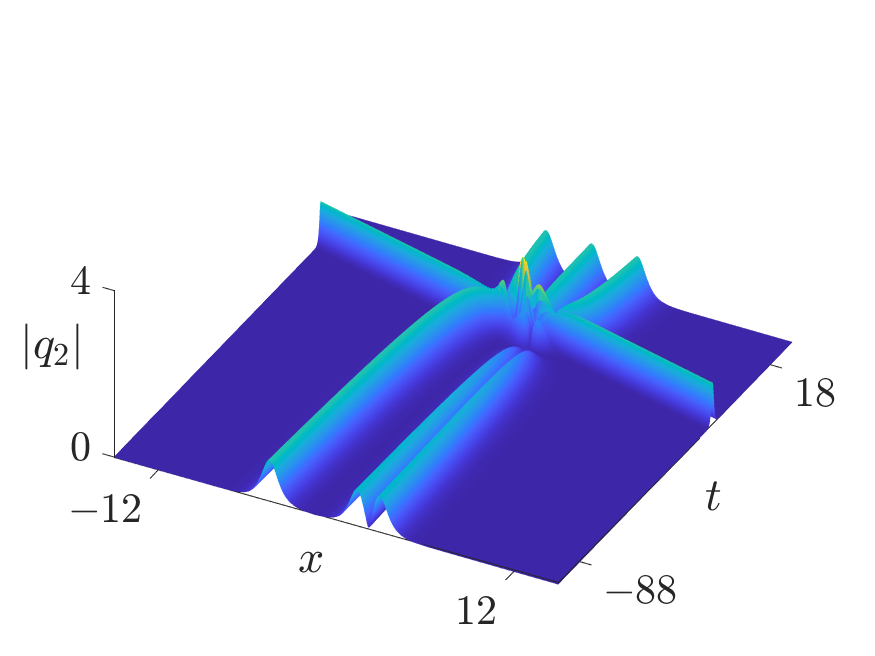}}
	\hspace{0.2cm}{\includegraphics[scale =0.32]{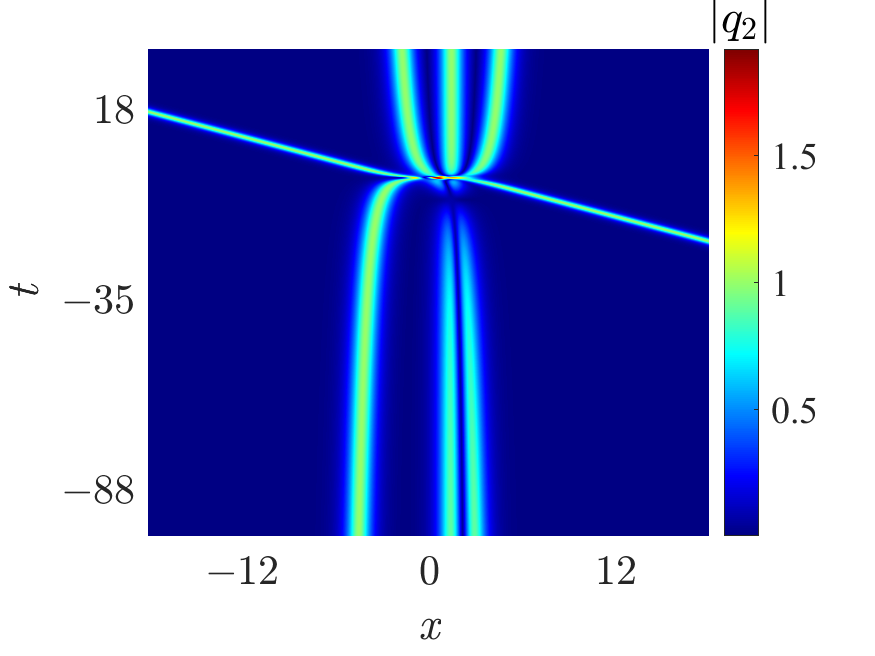}}
	\hspace{0.2cm}{\includegraphics[scale =0.34]{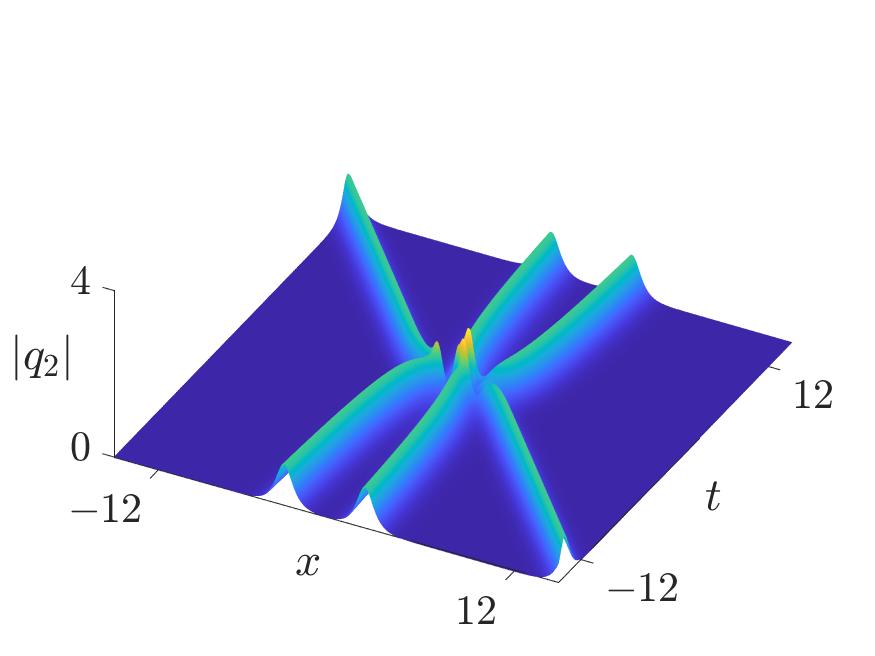}}\\
	\vspace{-0.2cm}{\footnotesize\hspace{-0.5cm}(a$_2$) \hspace{4.2cm}(b$_2$)  \hspace{4.0cm}(c$_2$)}\\
	\hspace{-1cm}\\
	\vspace{-0.3cm}\flushleft{\footnotesize{\bf Figs.}~11.\,
		Interactions between the degenerate solitons and the bell-shaped solitons via Solutions~(\ref{mixedsoliton}). Relevant parameters are the same as those in Figs.~9(a$_1$-a$_2$) except for $l_{14}=l_{34}=1$, (a$_1$-a$_2$) and (b$_1$-b$_2$) $m_{11}=0$; (c$_1$-c$_2$) $m_{11}={\rm i}$.}
\end{center}

\vspace{5mm}\noindent\textbf{~6.~Strong coherence phenomena and robust analysis}\\

In the above soliton interactions, we find several degenerate vector solitons which show significant coherence effects.
In this section, we discuss a special vector soliton in Figs.~2 for simplicity. Without loss of generality, we always consider this soliton with a fixed zero-asymptotic velocity when $t\to\pm\infty$ by controlling the real part
of eigenvalue and higher-order perturbation parameter $\varepsilon$ to satisfy the relation $\lambda_{1,R}=\frac{-1+\sqrt{1+12\lambda_{1,I}^2\varepsilon^2}}{6\varepsilon}$ (If $\varepsilon=0$, we set $\lambda_{1,R}=0$).

Based on Solutions~(\ref{desoliton}), we can know that parameter $\varepsilon$ is independent of the initial intensities $|q_1(x,0)|$  and $|q_2(x,0)|$ of degenerate solitons, which are shown in Figs.~13(a$_1$-a$_2$). Figs.~13(b$_1$-b$_2$) show the intensity profiles of degenerate solitons at $t=2$.  We can see that the wave energy increases significantly in the $q_1$ component, but decays significantly in the $q_2$ component. Such phenomena can only occur in a coherent systems, which are different from the solitons in some incoherent systems, e.g., the conventional coupled NLS system or coupled Hirota system.
More importantly,  when we increase $\varepsilon$, the wave energy changes more sharp. Particularly, when we choose $\varepsilon=2$, the soliton intensities in the $q_1$ component expand about $7$ times to maximum from the initial state, but decrease to only $1/7$ of their initial intensities in the $q_2$ component,
see the blue line in Figs.~12(b$_1$-b$_2$). Figs.~13 show the strong coherence degenerate solitons.
In other words, the higher-order perturbation parameter $\varepsilon$  distinctly
affects the coherence of degenerate solitons. The larger $\varepsilon$ leads to the stronger coherence.

The above conclusion can be confirmed based on System~(\ref{equations}). After the calculations, we have
\begin{subequations}
\begin{eqnarray}
&&\hspace{-1.5cm}(|q_{1}|^2)_t={\rm i}(q_{1,x}q_1^*-q_{1,x}^*q_1)_x-2{\rm i}(q_1^{*2}q_2^{2}-q_1^{2}q_2^{*2})-\varepsilon
(q_{1,xx}q_1^*+q_{1,xx}^*q_1-|q_{1,x}|^2)_x\nonumber\\
&&\hspace{-1cm}-6\varepsilon\left[(|q_1|^2+|q_2|^2)|q_1|^2-\frac{1}{2}|q_1|^4\right]_x+
3\varepsilon\left[q_1^{*2}(q_2^{2})_x+q_1^{2}(q_2^{*2})_x\right],\\
&&\hspace{-1.5cm}(|q_{2}|^2)_t={\rm i}(q_{2,x}q_2^*-q_{2,x}^*q_2)_x+2{\rm i}(q_1^{*2}q_2^{2}-q_1^{2}q_2^{*2})-\varepsilon
(q_{2,xx}q_2^*+q_{2,xx}^*q_2-|q_{2,x}|^2)_x\nonumber\\
&&\hspace{-1cm}-6\varepsilon\left[(|q_1|^2+|q_2|^2)|q_2|^2-\frac{1}{2}|q_2|^4\right]_x+
3\varepsilon\left[q_2^{*2}(q_1^{2})_x+q_2^{2}(q_1^{*2})_x\right],
\end{eqnarray}
\end{subequations}
Because $q_{1}, q_{2}, q_{1,x}, q_{2,x}\rightarrow0$ when $x\rightarrow\pm\infty$, we obtain
\begin{subequations}
\begin{eqnarray}
&&\hspace{-1.5cm}\left(\int_{-\infty}^{+\infty}|q_{1}|^2d x\right)_t=-2\int_{-\infty}^{+\infty}{\rm i}(q_1^{*2}q_2^{2}-q_1^{2}q_2^{*2})dx+3\varepsilon
\int_{-\infty}^{+\infty}\left[q_1^{*2}(q_2^{2})_x+q_1^{2}(q_2^{*2})_x\right]dx,\\
&&\hspace{-1.5cm}\left(\int_{-\infty}^{+\infty}|q_{2}|^2d x\right)_t=2\int_{-\infty}^{+\infty}{\rm i}(q_1^{*2}q_2^{2}-q_1^{2}q_2^{*2})dx+3\varepsilon
\int_{-\infty}^{+\infty}\left[q_2^{*2}(q_1^{2})_x+q_2^{2}(q_1^{*2})_x\right]dx,
\end{eqnarray}
\end{subequations}
Thus the coherent coupling terms of System~(\ref{equations}) actually have two parts, and the higher-order perturbation parameter $\varepsilon$ play a significant role on the coherence effect. The energy exchange between $q_1$ and $q_2$ components may become stronger when $|\varepsilon|$ increases. Such phenomena of System~(\ref{equations}) are completely different from the matrix NLS system and the vector Hirota system without coherent coupling~\cite{LLPRL,LLcnsns,LLpla}.

\begin{center} \vspace{0cm}
	\hspace{0.0cm}{\includegraphics[scale =0.36]{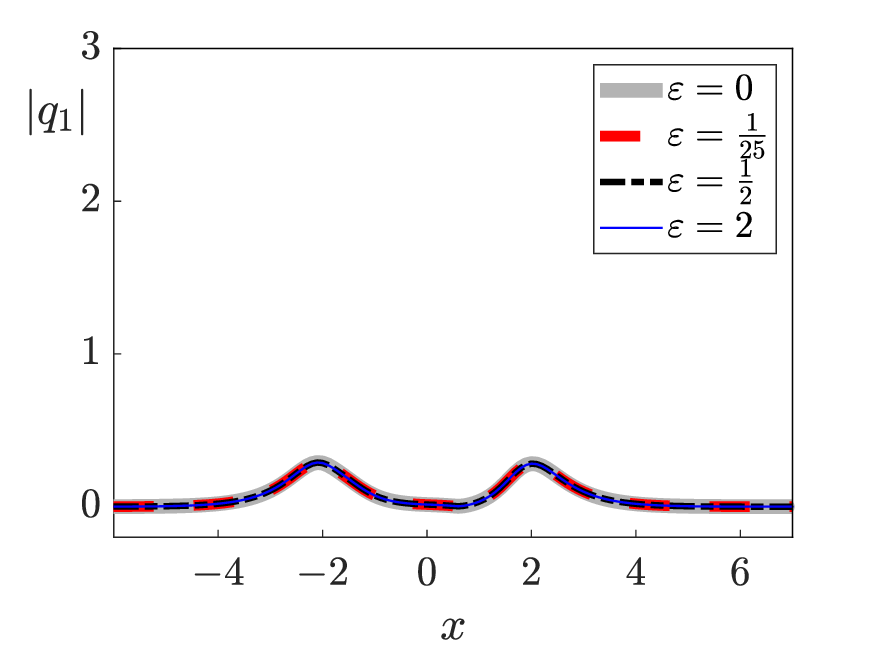}}
	\hspace{0.5cm}{\includegraphics[scale =0.36]{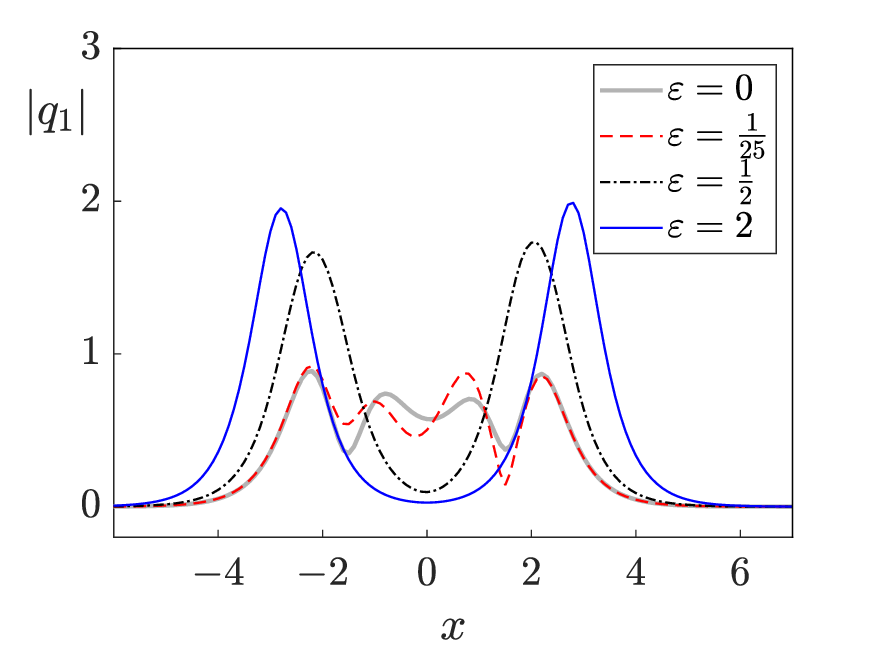}}\\
	\vspace{-0.2cm}{\footnotesize\hspace{-0.2cm}(a$_1$)\hspace{5.3cm}(b$_1$)}\\
	\hspace{-1cm}\\
	\hspace{0.0cm}{\includegraphics[scale =0.36]{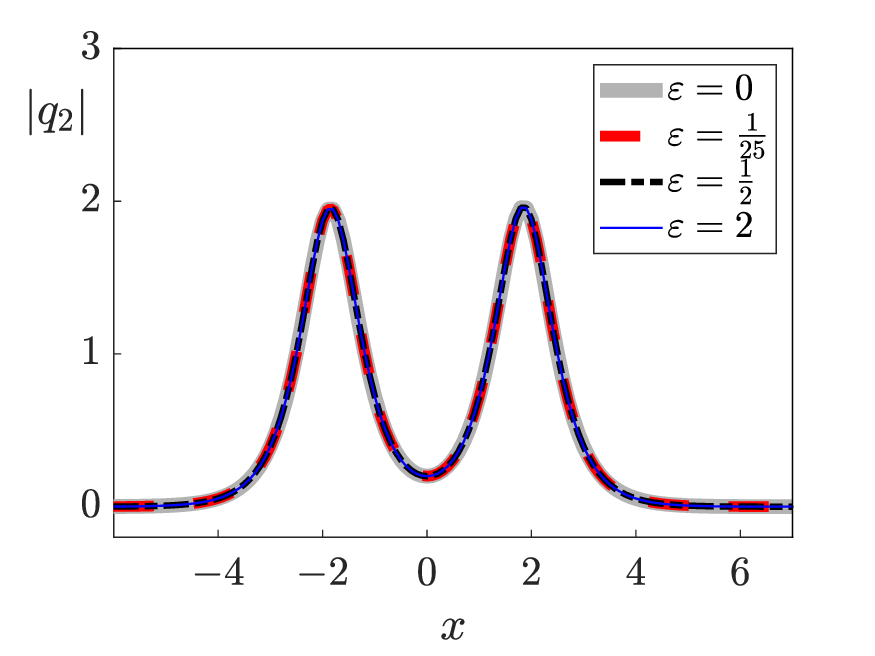}}
	\hspace{0.5cm}{\includegraphics[scale =0.36]{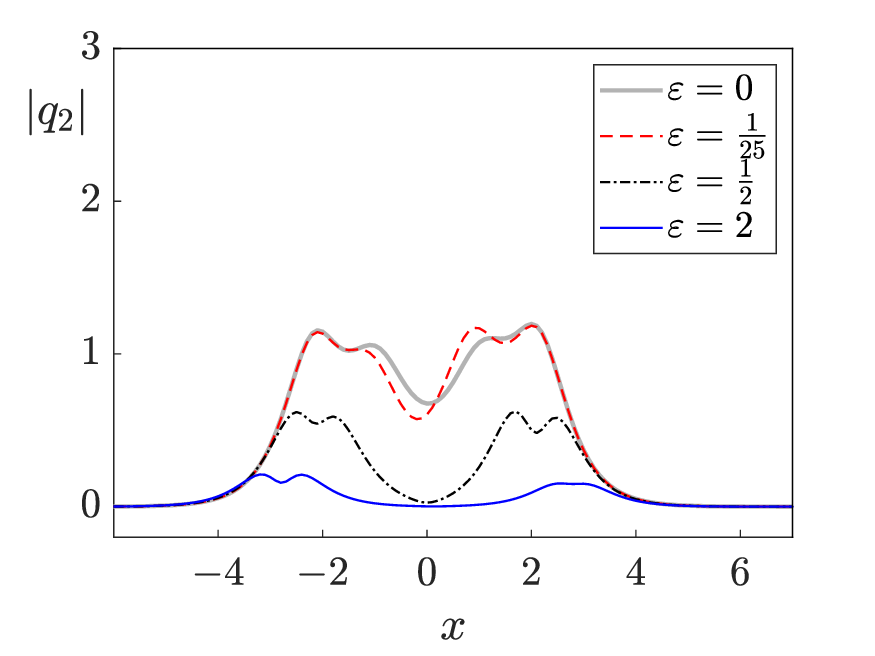}}\\
	\vspace{-0.2cm}{\footnotesize\hspace{-0.2cm}(a$_2$)\hspace{5.3cm}(b$_2$)}\\
	\hspace{-1cm}\\
	\vspace{-0.3cm}\flushleft{\footnotesize{\bf Figs.}~12.\,
		Intensity profiles of degenerate solitons for some different $\varepsilon$ settings at (a$_1$-a$_2$) $t=0$; (b$_1$-b$_2$) $t=2$. Other relevant parameters are the same as those in Figs.~2.}
\end{center}

\begin{center} \vspace{0cm}
	\hspace{0.0cm}{\includegraphics[scale =0.36]{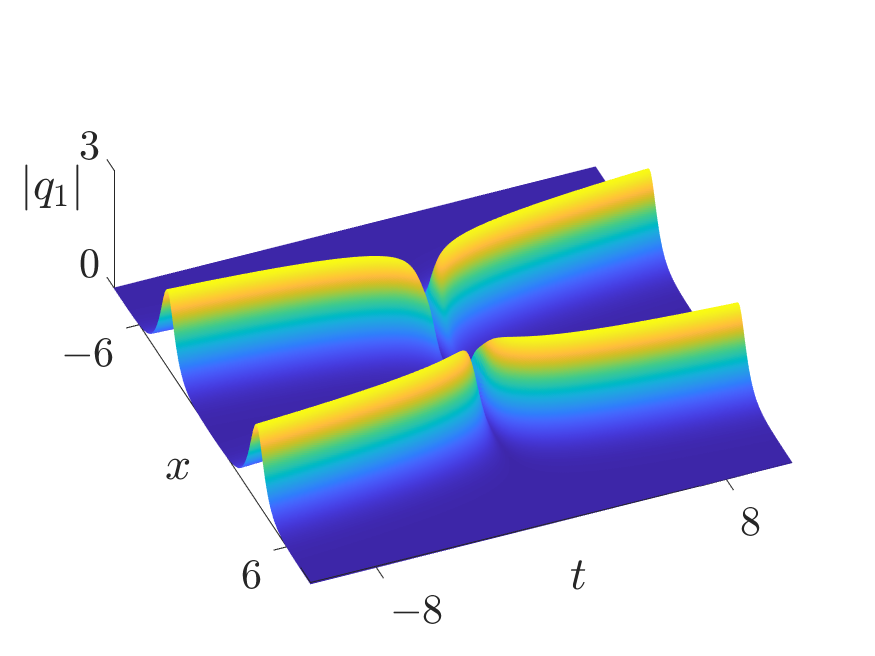}}
	\hspace{0.5cm}{\includegraphics[scale =0.36]{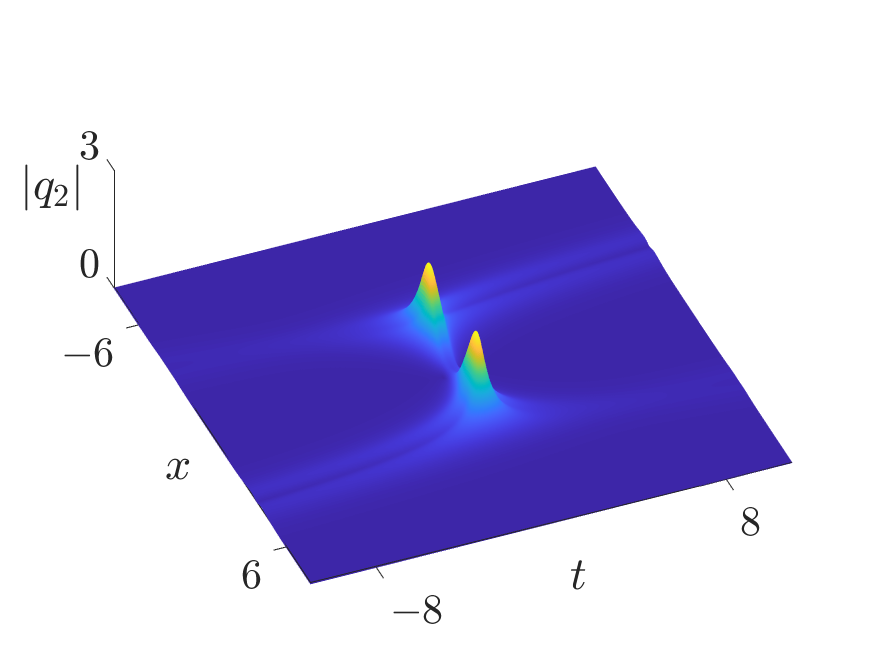}}\\
	\vspace{-0.2cm}{\footnotesize\hspace{-0.2cm}(a)\hspace{5.3cm}(b)}\\
	\vspace{-0.3cm}\flushleft{\footnotesize{\bf Figs.}~13.\,
		The degenerate solitons via Solutions~(\ref{desoliton}) with $\varepsilon=2$. Other relevant parameters are the same as those in Figs.~2.}
\end{center}

Finally, we numerically analyze the relationship between the robustness of such solitons and $\varepsilon$.  Here, a small rapidly decaying perturbation to the initial condition is considered, i.e.,
$q_{1,2}^\delta(x,t=0)=q_{1,2}(x,0)+\delta f(x)$, where $\delta$ is a small real constant and $f(x)=e^{-x^2}$. Here, to obtain a high resolution, we use the Fourier pseudospectral discretization with at least $N=2^9$ Fourier modes to deal with variable $x$ and the fourth-order Runge-Kutta scheme with a step size less than $10^{-4}$ is used to the time discretization.
The computation interval is taken as $[-15,15]$, which is large enough to ignore the effect of periodic boundary errors. The corresponding simulations for each of the cases presented only took a few minutes of computer time on a standard desktop computer.

For a fixed small perturbation and different $\varepsilon$, the initial intensities $q_1$ and $q_2$ are almost the same. We can see that the soliton evolution keep basically stable in a short time $t=4$ for $\varepsilon=\frac{1}{25}$, as seen in Figs.~14(a$_1$-b$_1$). However, the robustness of such degenerate solitons become weaker with increasing $\varepsilon$, as displayed in Figs.~14(a$_2$-b$_2$) and (a$_3$-b$_3$). The coherence effects have a negative effect on the dynamics stability of the degenerate solitons. It is noted that such the degenerate solitons are generally dynamic unstable under a perturbation in a large time.
\begin{center} \vspace{0cm}
	\hspace{0.0cm}{\includegraphics[scale =0.36]{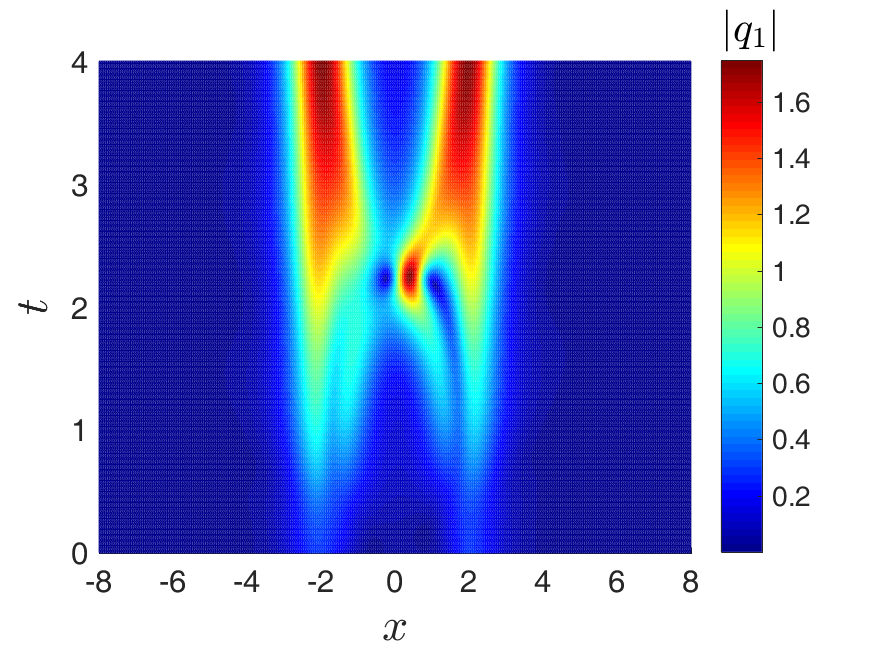}}
	\hspace{0.5cm}{\includegraphics[scale =0.36]{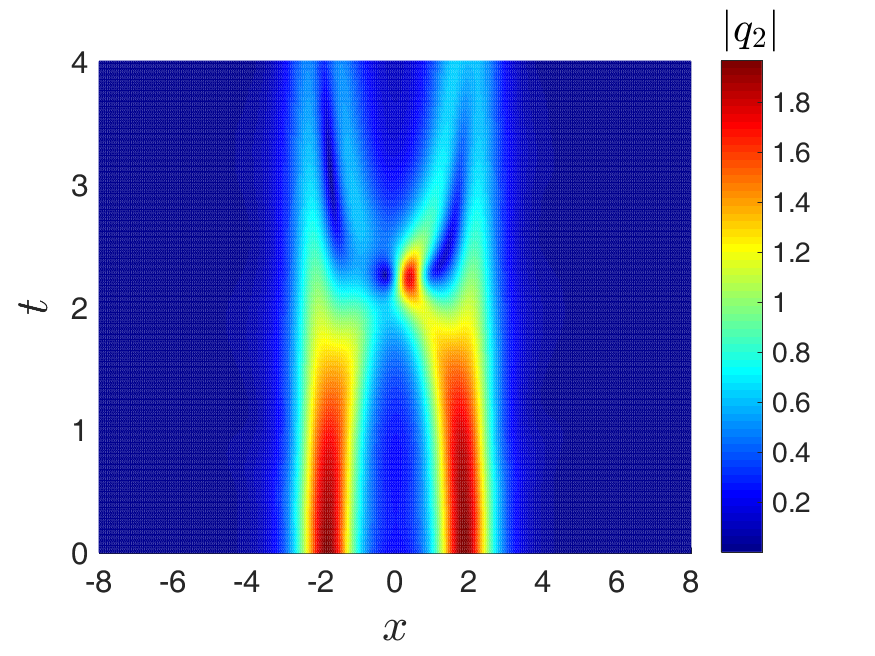}}\\
	\vspace{-0.2cm}{\footnotesize\hspace{-0.2cm}(a$_1$)\hspace{5.3cm}(b$_1$)}\\
	\hspace{0.0cm}{\includegraphics[scale =0.36]{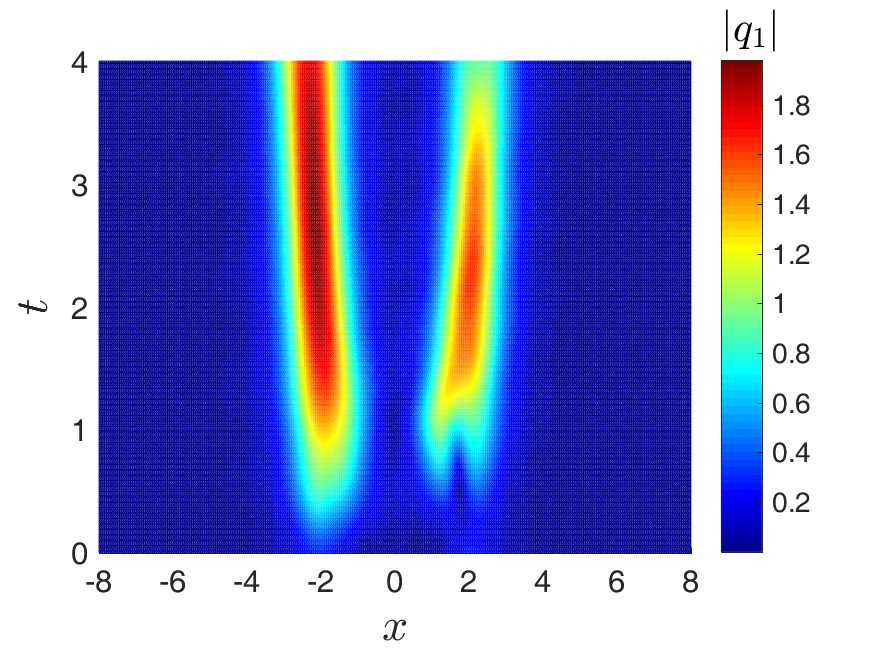}}
	\hspace{0.5cm}{\includegraphics[scale =0.36]{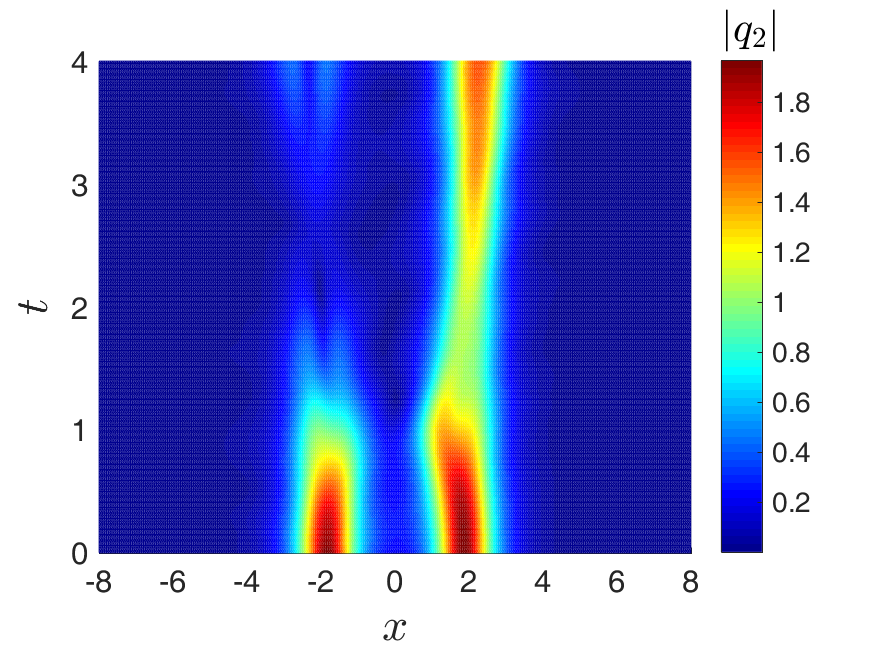}}\\
	\vspace{-0.2cm}{\footnotesize\hspace{-0.2cm}(a$_2$)\hspace{5.3cm}(b$_2$)}\\
	\hspace{0.0cm}{\includegraphics[scale =0.36]{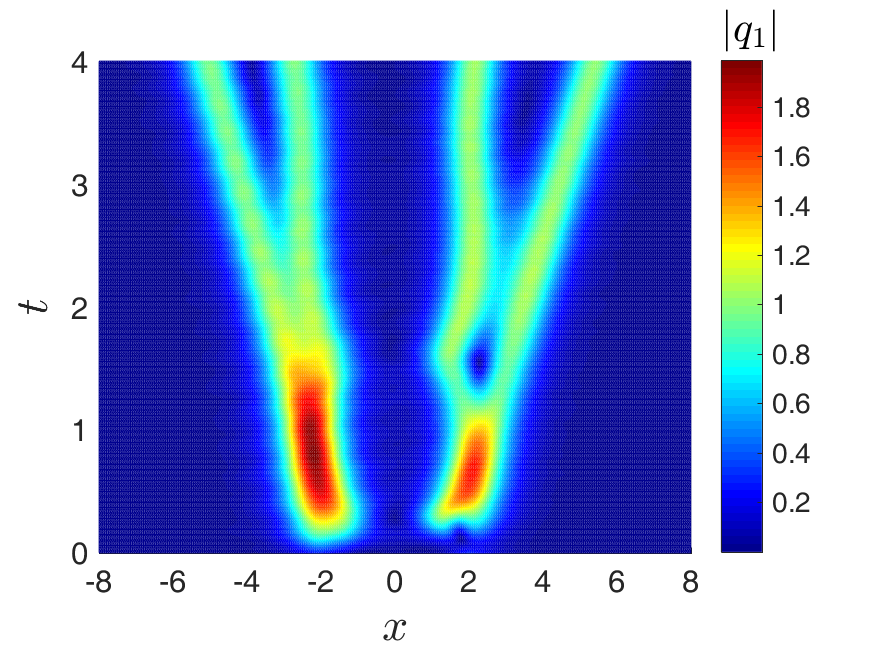}}
	\hspace{0.5cm}{\includegraphics[scale =0.36]{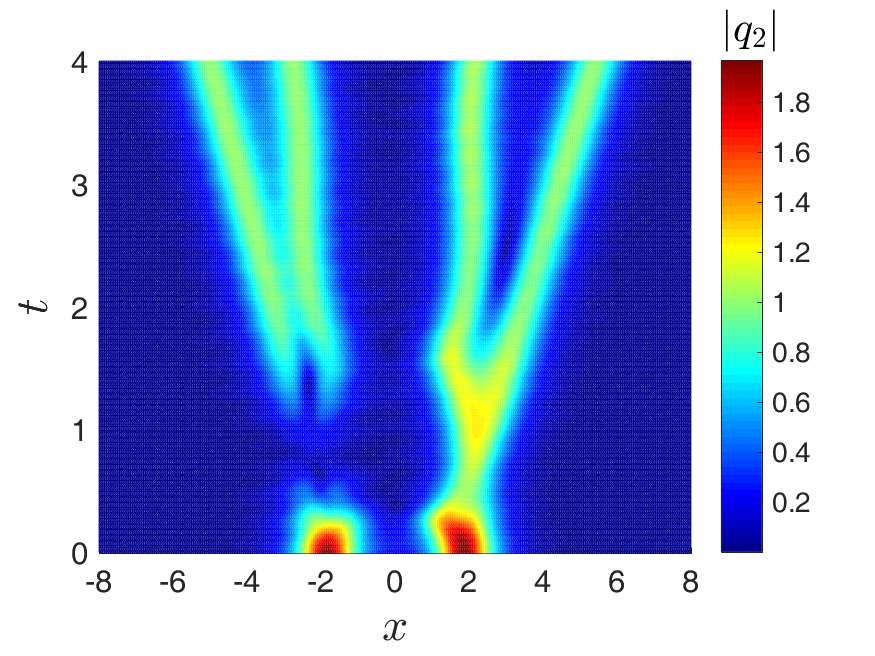}}\\
	\vspace{-0.2cm}{\footnotesize\hspace{-0.2cm}(a$_3$)\hspace{5.3cm}(b$_3$)}\\
	\vspace{-0.3cm}\flushleft{\footnotesize{\bf Figs.}~14.\,
		Time evolution of degenerate solitons under a local initial perturbation with $\delta=0.05$ for three different $\varepsilon$ by numerical simulation. (a$_1$-b$_1$) $\varepsilon=\frac{1}{25}$; (a$_2$-b$_2$) $\varepsilon=\frac{1}{2}$; (a$_3$-b$_3$) $\varepsilon=2$. Other relevant parameters are the same as those in Figs.~2.}
\end{center}

\vspace{5mm}\noindent\textbf{~7.~Conclusions and discussions}\\

In this paper, we have investigated the asymptotic behaviors and dynamics of degenerate and mixed solitons for the coupled Hirota system~(\ref{equations}), which can describe the optical pulse propagation in
isotropic nonlinear medium.
Firstly, using Binary DT (\ref{BDT}), we have derived Solutions~(\ref{desoliton}) to represent the degenerate soliton solutions with two eigenvalues that are conjugate to each other. We have obtained three types of degenerate solitons, with their asymptotic expressions detailed in Expressions~(\ref{asymptotic2}) and (\ref{asymptotic3}). Notably, these degenerate solitons exhibit time-dependent velocities, as illustrated in Expressions~(\ref{velocity1}) and (\ref{velocity2}). Based on those solutons and expressions, we have obtained the following prominent results:
\begin{itemize}
	
	\item We have revealed that higher-order perturbation parameter $\varepsilon$ has a significant impact on the coherence and robustness of solitons. Specifically, the relative distance between asymptotic solitons increases logarithmically with the increase of the high-order perturbation parameter $|\varepsilon|$. 
	
	\item We have asymptotically and graphically discovered four different interaction mechanisms between degenerate solitons and bell-shaped solitons, including elastic interactions with position shifts, interactions where degenerate solitons are inelastic but bell-shaped solitons are elastic, interactions where degenerate solitons are elastic but bell-shaped solitons are inelastic, and coherent interactions that occur over longer interaction regions.
		
	\item By means of numerical simulation, we have shown that the coherence of the degenerate soliton is strongly affected by the parameter $\varepsilon$, and its robustness decreases with the increase of $|\varepsilon|$. This discovery highlights the importance of considering higher-order effects in practical applications.
\end{itemize}



Besides, the findings in this paper have several physical implications, which are described in detail as follows: (1) Detailed analysis of degenerate solitons and mixed solitons provides new insights into the behavior of optical pulses in isotropic nonlinear media. (2) The research results show that the high-order effect significantly affects the coherence and robustness of solitons. (3) The research results on energy redistribution and coherence effects indicate that these solitons have potential application value in fields such as nonlinear optics and fluid mechanics.

Moreover, although the binary Darboux transformation method adopted in this paper successfully derive degenerate and mixed soliton solutions for the coupled Hirota system with strong coherent coupling effects, it also has some limitations. For instance, its solutions are sensitive to initial conditions and parameter choices, and the calculation process is complex, which may introduce numerical instability. Although theoretical analysis and numerical simulations provide important insights into soliton dynamics, the lack of direct experimental verification and the potential impact of higher-order perturbation parameters on the robustness and stability of solitons in practical applications are issues. Therefore, future research needs to further explore these limitations and verify the applicability of the solutions in a wider range of systems.

While the primary focus of our study is on the coupled Hirota system in the context of nonlinear optics, the methodology and findings have broader implications, particularly in the field of fluid mechanics, such as studying the wave dynamics in fluids, turbulence phenomena, and designing fluid systems through the analysis of soliton propagation and interaction. The higher-order effects considered in our model, such as third-order dispersion and self-steepening, are particularly important for accurately describing the propagation of ultrashort pulses in the marine environment. Our research can enhance the understanding of these phenomena and their potential impacts on ocean dynamics, coastal processes and fluid-structure interactions such as the response of offshore structures to extreme wave events. Meanwhile, future research directions and scope should include: (1) Experimentally verifying theoretical results; (2) Extending the method to other fluid systems; (3) Using numerical simulations to study more complex fluid environments; (4) Studying practical applications in engineering and technology, such as designing fluid devices and wave energy converters. These efforts will help to realize the potential application of this research in fluid mechanics.



\vspace{5mm}\noindent\textbf{Acknowledgments}\\

This work has been supported by the National Natural Science Foundation of China under Grant Nos.~12305001 and 12205029, by the Fundamental Research Funds for the Central Universities (Grant Nos. 2024CDJXY018 and 2025MS183), and by the China Scholarship Council (No. 202306050162). 

\end{document}